\documentclass{aa}  

\usepackage{graphicx}
\usepackage{xcolor}
\usepackage{amsmath}
\usepackage[switch]{lineno}
\newcommand{\dg}{$^{\circ}$}
\usepackage{txfonts}
\begin{document}

   \title{Composition and thermal properties of Ganymede's surface from JWST/NIRSpec and MIRI observations}

   \author{ 
   D. Bockel\'ee-Morvan
         \inst{1}
          \and
          E. Lellouch\inst{1}
          \and
          O. Poch \inst{2}
          \and
          E. Quirico \inst{2}
           \and
          S. Cazaux \inst{3,4}
           \and
          I. de Pater \inst{5,6}
          \and
          T. Fouchet \inst{1}
           \and
         P. M. Fry \inst{7}
          \and
          P. Rodriguez-Ovalle \inst{1}
          \and 
          F. Tosi \inst{8}
          \and
          M. H. Wong \inst{5}
           \and
          I. Boshuizen \inst{3}
         \and
          K. de Kleer \inst{9}
          \and
          L. N. Fletcher \inst{10}
          \and
          L. Meunier \inst{2}
          \and
          A. Mura \inst{8}
         \and
          L. Roth \inst{11}
          \and 
          J. Saur \inst{12} 
          \and
          B. Schmitt \inst{2}
          \and
          S. K. Trumbo \inst{13} 
          \and
          M. E. Brown \inst{9}
          \and
          J. O'Donoghue \inst{14}
%          Department of Solar System Science, JAXA Institute of Space and Astronautical Science, Japan
          \and
          G. S. Orton \inst{15}
          \and         
          M. R. Showalter \inst{16}
          %SETI Institute, Mountain View, CA 94043 
%          \fnmsep\thanks{Just to show the usage          of the elements in the author field}
          }

% 1
   \institute{LESIA, Observatoire de Paris, Université PSL, Sorbonne Universit\'e, Universit\'e Paris Cité, CNRS, 92195, Meudon, France\\
%              \email{Dominique.Bockelee@obspm.fr}
         \and
% 2
             Univ. Grenoble Alpes, CNRS, IPAG, 38000 Grenoble, France%Université Grenoble Alpes, Centre National de la Recherche Scientifique (CNRS), 33 Institut de Planétologie et d’Astrophysique de Grenoble (IPAG), 38000 Grenoble, France.
% 3
             \and
             Faculty of Aerospace Engineering, Delft University of Technology, Delft, The Netherlands
             \and
% 4
             Leiden Observatory, Leiden University, P.O. Box 9513, NL 2300 RA Leiden, The 28 Netherlands.
             \and
% 5
             Department of Astronomy, University of California, 22 Berkeley, CA 94720, USA.
             \and
% 6             
             Department of Earth and Planetary Science, University of California, 22 Berkeley, CA 94720, USA.             
% 7 
            \and
            University of Wisconsin, Madison, WI, 53706
% 8
             \and
             Istituto Nazionale di AstroFisica – Istituto di Astrofisica e Planetologia Spaziali (INAF-IAPS), 00133 Rome, Italy
%9
            \and
            Division of Geological and Planetary Sciences, Caltech, Pasadena, CA 91125 USA
%10
            \and
            School of Physics and Astronomy, University of Leicester, University Road, Leicester, LE1 7RH, UK
%11 
             \and
             Space and Plasma Physics, KTH Royal Institute of Technology, Stockholm, Sweden
%12             
             \and
             Institute of Geophysics and Meteorology, University of Cologne, Albertus Magnus Platz, 50923 Cologne, Germany
%13
             \and
             Cornell Center for Astrophysics and Planetary Science, Cornell University, Ithaca, NY 14853, USA.
% 14 O'Donoghue
             \and
            Department of Solar System Science, JAXA Institute of Space and Astronautical Science, Japan
% 15 Orton
             \and
             Jet Propulsion Laboratory, California Institute of Technology, Pasadena, California 91109, USA
% 16 Showalter
             \and
             SETI Institute, Mountain View, CA 94043
             %\thanks{The university of heaven temporarily does not accept e-mails}
             }

% Leigh Fletcher: School of Physics and Astronomy, University of Leicester, University Road, Leicester, LE1 7RH.

   \date{Received; accepted}

% \abstract{}{}{}{}{} 
% 5 {} token are mandatory

\abstract
   {We present the first spectroscopic observations of Ganymede by the James Webb Space Telescope undertaken in August 2022 as part of the proposal "ERS observations of the Jovian System as a demonstration of JWST's capabilities for Solar System science".}
   {We aimed to investigate the composition and thermal properties of the surface, and to study the relationships of ice and non-water-ice materials and their distribution.  }
   {NIRSpec IFU (2.9--5.3 $\mu$m) and MIRI MRS (4.9--28.5 $\mu$m) observations were performed on both the leading and trailing hemispheres of Ganymede, with a spectral resolution of $\sim$ 2700 and a spatial sampling of 0.1 to 0.17'' (while Ganymede size was $\sim$1.68''). We characterized the spectral signatures and their spatial distribution on the surface. The distribution of brightness temperatures was analyzed with standard thermophysical modelling including surface roughness. }
   {Reflectance spectra show signatures of water ice, CO$_2$ and H$_2$O$_2$. An absorption feature at 5.9 $\mu$m, with a shoulder at 6.5 $\mu$m, is revealed, and is tentatively assigned to sulfuric acid hydrates. The CO$_2$ 4.26-$\mu$m band shows latitudinal and longitudinal variations in depth, shape and position over the two hemispheres, unveiling different CO$_2$ physical states. In the ice-rich polar regions, which are the most exposed to Jupiter's plasma irradiation, the CO$_2$ band is redshifted with respect to other terrains. In the boreal region of the leading hemisphere, the CO$_2$ band is dominated by a high wavelength component at $\sim$ 4.27 $\mu$m, consistent with CO$_2$ trapped in amorphous water ice. At equatorial latitudes (and especially on dark terrains) the observed band is broader and shifted towards the blue, suggesting CO$_2$ adsorbed on non-icy materials, such as minerals or salts. Maps of the H$_2$O Fresnel peak area correlate with Bond albedo maps and follow the distribution of water ice inferred from H$_2$O absorption bands. Amorphous ice is detected in the ice-rich polar regions, and is especially abundant on the northern polar cap of the leading hemisphere. Leading and trailing polar regions exhibit different H$_2$O, CO$_2$ and H$_2$O$_2$ spectral properties. However in both hemispheres the north polar cap ice appears to be more processed than the south polar cap. A longitudinal modification of the H$_2$O ice molecular structure and/or nano/micrometre-scale texture, of diurnal or geographic origin, is observed in both hemispheres. Ice frost is tentatively observed on the morning limb of the trailing hemisphere, possibly formed during the night from the recondensation of water subliming from the warmer subsurface. Reflectance spectra of the dark terrains are compatible with the presence of Na-/Mg-sulfate salts, sulfuric acid hydrates, and possibly phyllosilicates mixed with fine-grained opaque minerals, having an highly porous texture. Latitude and local time variations of the brightness temperatures indicate a rough surface with mean slope angles of 15\dg--25\dg and a low thermal inertia $\Gamma$ = 20--40 J m$^{-2}$ s$^{-0.5}$ K$^{-1}$, consistent with a porous surface, with no obvious difference between the leading and trailing sides. }
   {}

   \keywords{Planets and satellites: individual: Ganymede --  Planets and satellites: surfaces -- Planets and satellites: composition}
   %Planets and satellites: individual: Ganymede -- Planets and satellites: surfaces -- Planets and satellites: composition -- Infrared: planetary systems}               }

\titlerunning{Ganymede from JWST}
\authorrunning{Bockel\'ee-Morvan et al.}
\maketitle
%

%-------------------------------------------------------------------
%\clearpage

\section{Introduction}
Ganymede is the largest Galilean satellite and the largest natural satellite in our Solar System. Ganymede's internal structure reveals a differentiated body with a molten core producing an intrinsic magnetic field, a silicate mantle, and a complex icy crust which hides a deep ocean, making it an archetype of water worlds. Ganymede also possesses a thin oxygen atmosphere \citep{Hall98} and two auroral ovals \citep{Feldman2000}.

Ganymede’s main provinces are the dark "\textit{Regiones}" and bright, ice-rich, grooved terrains. Based on photo-interpretation and crater counting, the \textit{Regiones} are older than the bright terrains, holding clues about Ganymede's geologic evolution. Like other Galilean moons immersed in Jupiter's magnetosphere, the icy surface of Ganymede undergoes space weathering processes due to the impact of energetic particles, solar (UV) flux, thermal cycling, and bombardment by micro-meteoroids. Optical images returned by previous space missions revealed that Ganymede's polar caps are brighter than the equatorial region, and that the trailing hemisphere is darker than the leading hemisphere in the equatorial region \citep[e.g.,][]{Khurana2007-og}. Ganymede's magnetosphere is thought to play an important role in shielding the equatorial region at latitudes below 40\dg~from the incident plasma, while the radiolytic flux is much higher at polar latitudes \citep[e.g.,][]{Fatemi2016-uy, Poppe2018-wo, Liuzzo2020-ww, Plainaki2020-hc}.

Current knowledge of Ganymede’s surface composition comes from close exploration carried out by the NASA Galileo and Juno spacecrafts, and ground-based telescopic observations. Near infrared reflectance spectra of Ganymede, returned by the Near Infrared Mapping Spectrometer (NIMS) onboard Galileo \citep{Carlson1992-tq} 
 revealed the presence of amorphous ice, especially in the polar regions, probably formed by the more intense energetic bombardment in these regions \citep{Hansen2004}. They also revealed  
spectral features at 3.4, 3.88, 4.05, 4.25 and 4.57 $\mu$m - weaker than those on Callisto - attributable to sulfur dioxide (SO$_2$), carbon dioxide (CO$_2$), and organic compounds \citep{McCord1997-bi, McCord1998, McCord2001-pi}. Observations at shorter wavelengths returned by the Very Large Telescope (VLT) in recent years  
suggested the contributions of chlorinated and sulfate salts as well as sulfuric acid hydrates \citep{Ligier2019, King2022-is}.

%--------------------------------------------------------------------
\begin{table*}[ht]
\caption{JWST Observations of Ganymede. \label{tab:observations}  } 
\begin{tabular}{|l|c|c|c|c|c|c|c|}
\hline
Observation$^a$ & Side & UT Date  & Sub. Obs. $^b$ & Sub. solar $^b$   & Helio.    & Diameter  & Exp. time$^c$  \\
& & (begin/end) & coord. &  coord. &distance(au)  & (arcsec)  &   (s)\\
\hline
19 NIRSpec & Leading & 2022/08/07 19:32/20:13  & 2.55\dg N / 72.3\dg W & 2.02\dg / 82.0\dg W &  4.9557& 1.695 & 1718 \\
28 NIRSpec & Trailing & 2022/08/03 01:10/01:50 & 2.54 \dg N / 269.7\dg W & 2.00\dg N /279.7\dg W  & 4.9557 & 1.672 & 1718 \\

18 MIRI & Leading  & 2022/08/07 02:14/05:18   & 2.55\dg N / 77.1\dg W & 2.02\dg N / 86.7\dg W  & 4.9597   & 1.690  & 8758  \\
27 MIRI & Trailing & 2022/08/03 20:36/23:40   & 2.54\dg N / 274.4\dg W & 2.01\dg N / 284.4\dg W & 4.9608  & 1.675 & 8758 \\
\hline
\end{tabular}
\footnotesize{$^a$ Observation number of the ERS program and instrument.}
\footnotesize{$^b$ At mid-time.}
\footnotesize{$^c$ Effective exposure time.}
\end{table*}
%---------------------------------------------------------------------

The most abundant non-ice compound, CO$_2$, does not display any clear leading/trailing asymmetry on large regional scales, whereas such asymmetry was observed on Callisto \citep{Hibbitts2002-ps, Hibbitts2003}. Also, unlike Callisto, there are no systematic correlations at local scale \citep[e.g. CO$_2$-enriched impact craters,][]{Hibbitts2003}, even though some CO$_2$-rich impact craters may not be ruled out \citep{Tosi2023}. CO$_2$ appears to be mostly correlated with moderately hydrated non-ice material primarily associated with the dark \textit{Regiones} \citep{Hibbitts2009}. Observations of Ganymede by Juno/JIRAM \citep{Adriani2017-nm} revealed a latitudinal trend in the strength of the CO$_2$ feature, with a band depth slightly higher at low latitudes \citep{Mura2020}, consistent with previous NIMS results \citep{Hibbitts2003}. While in the 3–5 $\mu$m range the strongest signature of free CO$_2$ ice is centered at 4.27 $\mu$m, on Ganymede this signature is slightly shifted to shorter wavelengths at 4.25-4.26 $\mu$m, in average, similar to what is observed on Saturn's icy satellites. This reveals that the CO$_2$ molecule must be trapped in another host material rather than present as pure ice \citep{Chaban2007, Cruikshank2010-pz}.

Radiolytic H$_2$O$_2$ has a diagnostic, weak absorption at 3.5 $\mu$m and was first observed on Europa by NIMS \citep{Carlson1999-gt}, providing the rationale for investigating the presence of this compound also on Ganymede. Newly detected and mapped on Ganymede with JWST, this species exists primarily at the polar caps, consistent with production by radiolysis driven by particles directed by Ganymede's magnetic field \citep{Trumbo2023}. Ganymede's intrinsic magnetic field and its effects on the charged particle precipitation are also relevant for the generation of the sputtered atmosphere. The global O$_2$ atmosphere is thought to be sourced by ice radiolysis from sputtering in the polar regions \citep{Marconi2007}. Around the sub-solar point, H$_2$O from surface sublimation (not affected by the magnetic field) is suggested to be more abundant than O$_2$ \citep{2021NatAs...5.1043R, Leblanc2023}.

Following previous observations, there remain several open questions on the nature, the origin and the processes making up Ganymede’s current surface composition. Here, we discuss the first spectroscopic observations of Ganymede by the James Webb Space Telescope (JWST) as part of the proposal "ERS observations of the Jovian System as a demonstration of JWST's capabilities for Solar System science", submitted in response to the ERS call and selected in November 2017 \citep{2022DPS....5430607D}. The program's objectives on Ganymede were the investigation of both the surface and the exosphere. 
JWST observations of Ganymede were performed in August 2022 using MIRI MRS (4.9–28.5 $\mu$m), and NIRSpec IFU High Res (2.9–5.3 $\mu$m, grating G395H). This paper focuses on the main results obtained on the surface, with an emphasis on: i) the mapping of the CO$_2$ molecule to derive constraints on the properties of the material in which it is trapped; ii) the distribution and properties of water ice mainly from the Fresnel reflection peaks, iii) the origin of a newly detected 5.9-$\mu$m absorption band, and iv) the physical properties of the surface as derived from its thermal emission.

%%%%%%%%%%%%%%%%%% SECTION 2 - MIRI & NIRSpec observations
\section{Observations}
\subsection{NIRSpec observations and data reduction}
\label{sec:NIRSPEC-obs}
Ganymede Trailing and Leading hemispheres were observed with NIRSpec/IFU onboard JWST on 3 August (Obs. 28) and 7 August 2022 (Obs. 19), respectively. NIRSpec IFU observations provide spatially resolved imaging spectroscopy over a 3\arcsec~$\times$ 3\arcsec~field-of-view with 0.1 \arcsec $\times$ 0.1\arcsec (310 $\times$ 310 km at Ganymede) spatial elements (spaxels). They were taken with the G395H/F290LP grating/filter pair, and cover the spectral range 2.86--5.28 $\mu$m (with gaps) with an average spectral resolution of $\sim$ 2700. In this high-spectral resolution configuration, the dispersion direction in the focal plane is covered by two NIRSpec detector arrays (NRS1 \& NRS2) separated by a physical gap which is responsible for missing wavelengths in the $\sim$ 4.0-4.1 $\mu$m range. The observations were taken with a four-point dither (offset of about 0.4 \arcsec~between dithers), with 4 integrations constituted of 10 groups (i.e., single exposures) per dither position. The data were acquired with the NRSRAPID readout pattern appropriate for bright sources. Details concerning the dates, geometries, and exposure times are given in Table~\ref{tab:observations}.

The data reduction was performed with the JWST pipeline version 1.9.0., using the context file version $jwst\_1039.pmap$. The individual raw images were first processed for detector-level corrections using the $Detector1Pipeline$ module of the pipeline (Stage 1). The resulting count-rate images were corrected from the 1/$\it f$ noise introduced during the detector readout using the method described by \cite{Trumbo2023}. The cleaned count-rate images were then calibrated using the $Calwebb\_spec2$ module (Stage 2). At this stage, WCS-correction, flat-fielding, and flux calibrations were applied to convert the data from units of count rate to flux density. The individual Stage 2 dither images were then resampled and combined onto a final data cube through the $Calwebb\_spec3$ processing (Stage 3). In this processing stage, the $outlier\_detection$ step was ultimately not applied, as it rejected pixels on the Ganymede disk. The $cube\_build$ step, which produces the final Level-3 3-D spectral cubes, was run with the drizzle-weighting and ifualign-geometry options. In this geometry, the IFU cube is aligned with the instrument IFU plane, so that North is not along the y axis, contrary to the default skyalign option. The ifualign option is recommended to minimize artifacts in the cube-build step. 

The observations were affected by some pointing issues that were taken into account to compute the geographical coordinates of each spaxel, and recenter the images. The central position of the NRS1 and NRS2 images were calculated as the (signal-unweighted) mean RA, DEC value of all spaxels where the measured radiance (at 2.89 $\mu$m for NRS1 and 4.20 $\mu$m for NRS2) is above some threshold (ensuring that the spaxel sees signal from the target; a threshold of 5 \% of the maximum signal was found optimal). The measured offsets with respect to the position given by the header are in the range $\sim$ 0.43--0.46'' in RA and less than 0.04'' in DEC.

Essentially all the detected signal  from Ganymede with NIRSpec/IFU is reflected sunlight. Indeed, the relative contribution of thermal emission is typically 12\% at the longest wavelengths (5.2 $\mu$m), but less than 5\% in average at wavelengths shorter than 4.9 $\mu$m based on MIRI data and associated thermal modelling (Sect.~\ref{sec:MIRI-an}). The measured radiances were therefore converted into the radiance factor $I/F$ (i.e., uncorrected from solar incidence and emission angles), where $F$ = $I_{\odot}$/$\pi$, and $I_{\odot}$ is the solar radiance at the heliocentric distance of Ganymede. At the  spectral resolution of these observations (spectral resolution element size of 1.49 nm), the detailed spectrum of the Sun is critical for properly calibrating the data in $I/F$ and correcting for the solar lines. We used the ACE-FTS atlas of \cite{Hase2010}, which provides the depth of the infrared solar lines from 2.6 to 5.4 $\mu$m with a step of 0.005 cm$^{-1}$, and the infrared solar continuum from R.L. Kurucz \footnote{\url{http://kurucz.harvard.edu/sun.html}}. The solar spectrum was convolved to the NIRSpec G395H spectral resolution, and resampled at the wavelengths of the spectral elements in the Ganymede velocity rest frame, taking into account the heliocentric velocity of Ganymede and Ganymede's velocity with respect to JWST. It was found that solar lines are best removed by shifting the wavelength of the spectel elements\footnote{a spectel is a spectral element, to not be confused with a spaxel which is a spatial element of the reconstructed IFU data cube.} by a fraction (0.2 and 0.4, for NRS1 and NRS2, respectively) of the dispersion element (of 0.66 nm). Applying these shifts, solar lines residuals are however observed in the $I/F$ spectra at the 0.5\% level. Spikes present in the $I/F$ spectra were flagged using sigma-clipping (which identifies spectels that are above a specified number of standard deviations; we set a 3-$\sigma$ threshold), and replaced by the mean $I/F$ of the ten nearest spectels. Inspecting the spectrum associated with each spaxel, those at the limb present mid-frequency wave-like structures, and so may be less reliable. 

\subsection{MIRI observations and data reduction}
   \begin{figure}
   \includegraphics[angle=90,width=9cm]{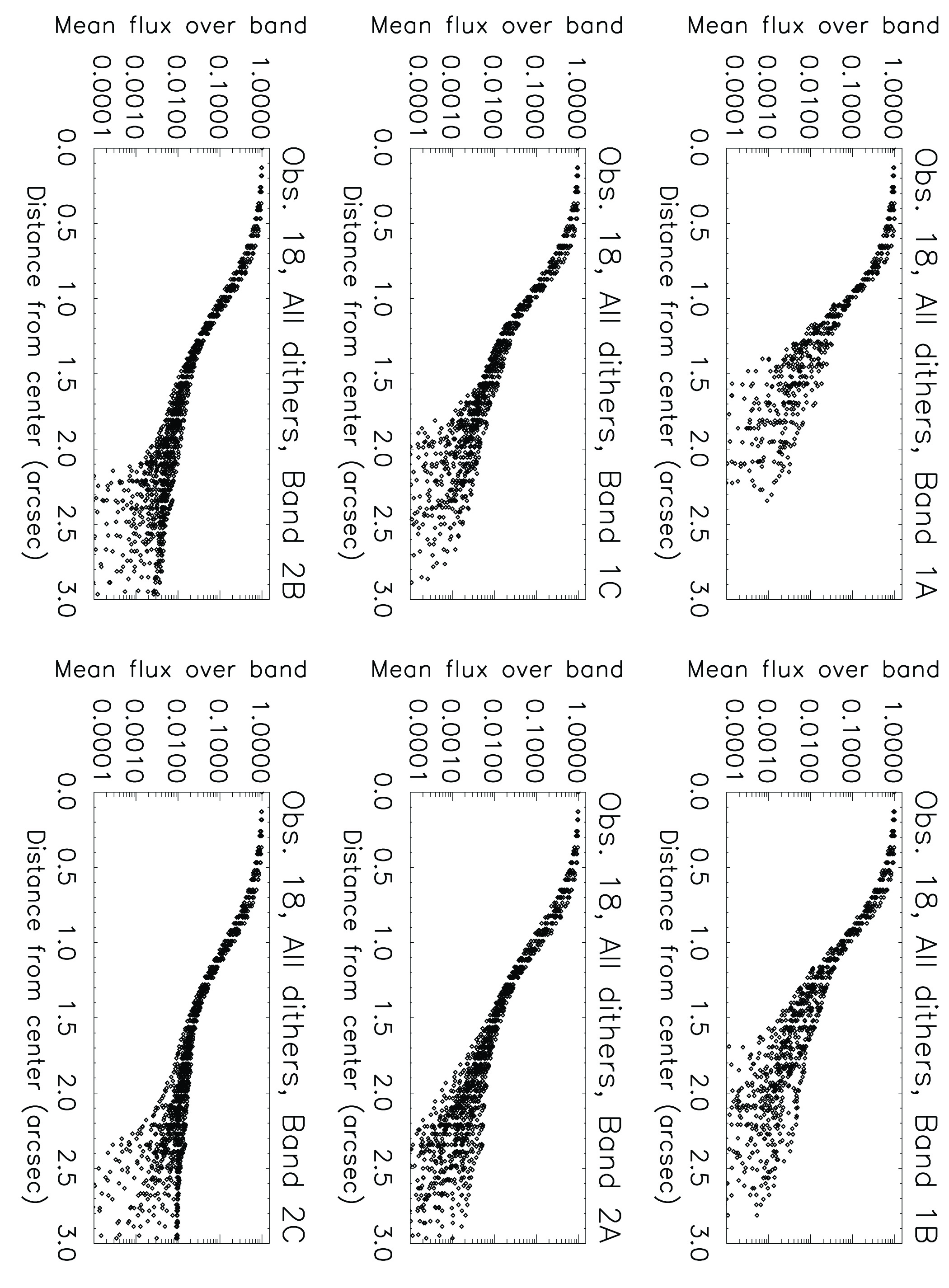}
   \caption{Radial dependence of band-averaged radiances for Obs. 18. Note that because the RA,DEC grid extends up to $\pm$2.34", distances beyond this distance are only partially sampled.}
   \label{fig:straylight}
    \end{figure}

MIRI/MRS observations of Ganymede were taken on Aug. 3, 2022 (trailing side, Obs. 28) and Aug. 7, 2022 (leading side, Obs. 17). The instrument has 4 separate IFUs (channels 1 through 4), each covering a separate wavelength range between 4.9 and 27.9 $\mu$m. All 4 channels are observed simultaneously, but covering the entire spectral range requires the use of three grating settings: SHORT (A), MEDIUM (B), and LONG (C). Combining the 4 channels and 3 grating settings, this yields 12 different bands increasing in wavelength from 1A to 4C. Observations were acquired in the form of a 4-point dither optimized for extended targets. We used the FASTR1 readout mode, with 5 groups (Ngroups = 5) per integration to optimize the slope-fitting algorithm. For each dither and grating position, 44 integrations were acquired, 
for a total 8758 sec exposure time for each visit. Given the way observations were conducted (grating A, dither positions 1-4; then grating B, 1-4; then grating C, 1-4), and the 7.15455-day rotation period of Ganymede, spectra in gratings A, B, C do not exactly correspond to the same geographical locations on Ganymede, with a shift of 4\dg~longitude between grating A and grating C; this small effect was ignored upon spectra reconstruction\footnote{For example, for channel 1, a spaxel size of 0.13" corresponds to a 14\dg~longitude excursion at equator and central meridian.}. 
Table 1 summarizes the observational and some relevant ephemeris parameters. The distance of Ganymede from Jupiter center was 340$\pm$1" for Obs. 27 and 331$\pm$5" for Obs. 18.

Background observations were also acquired. The goal of these observations was both to evaluate the straylight from Jupiter at Ganymede's distance and to mitigate the effects of bad pixels %and cosmic rays 
in the Ganymede data. These observations were obtained on July 26 (Obs. 29), Aug. 4 (Obs. 30), and Aug. 6 (Obs. 22),  but it turned out that the specified background position (20" North of Ganymede) was too close to Ganymede. As a consequence, the background data actually show scattered light from Ganymede itself. Therefore, these data were not used.

Observations were initially reduced using the standard JWST pipeline (version 1.9.2, context file $jwst\_1041.pmap$), with the {\em skyalign} option. MIRI observations are known to be affected by fringing issues, caused by interferences between the reflective layers of the detectors.
Those were handled at the data reduction level, using the alternative ''residual fringe correction step'' patch while running the Stage 3 processing in the pipeline. As expected, the use of Ngroups=5 in data acquisition led to partial or total
saturation of the signals at least over some parts of the Ganymede disk, especially near disk center, longwards of 8.5 $\mu$m (from bands 2B to 4C). Therefore, data were also reduced restricting the ramps to Ngroups = 2 and Ngroups=1. For Ngroups = 1, the calibration was performed by applying a band-dependent correction factor to the uncalibrated \verb|UNCAL| data. The factor was determined by comparing, on non-saturated spaxels, the Level-2 calibrated radiances obtained when restricting the ramp to a single group to those obtained using the full ramp (Ngroups = 5). The correction factor ranges from 1.02 to 1.14 depending on the band. Doing so considerably alleviates saturation issues, but at the expense of S/N. With an appropriate choice of Ngroups for each band, we could recover unsaturated and high S/N data at all wavelengths from 4.9 to 11.7 $\mu$m (bands 1A to 2C).  The pipeline output consists of 3D cubes (2D spaxels $\times$ 1D spectels) in fits format, calibrated in radiance (MJy/sr), with 
spaxel (x,y) coordinates aligned with sky RA, DEC. 

Starting from the fits files, we applied the following steps to construct cubes usable for scientific analysis. All of steps (1-6) described below were applied independently to
data corresponding to the different dither positions.
\begin{enumerate}
\item Bad pixel correction. Spectra at one given spatial position appear affected by spikes, presumably due to bad pixels and cosmic rays. These were removed spectrum-by-spectrum  and band-by-band using sigma-clipping at 5-$\sigma$ threshold and subsequent reinterpolation. 
\item Resampling of the cubes to a common RA,DEC grid. The 2D spaxel grid delivered by the pipeline has a NX $\times$ NY size that depends on band (e.g. 37 $\times$ 37 for band 1A, 35 $\times$ 33 for band 2C, etc.) and the size of the spaxels is also band-dependent (0.13", 0.17", 0.20", and 0.35" in channels 1, 2, 3, 4). All cubes were resampled from interpolation to a common RA, DEC grid (that of band 1A).
\item Merging of data with different Ngroups. By combining data with different values of Ngroups, we optimized both S/N and usable spectral coverage. Specifically, for trailing side (Obs. 27) data, we kept Ngroups=5 for bands 1A to 1C, Ngroups=2 for band 2A and Ngroups=1 for band 2B, 2C. (A single Ngroup value, independent of spaxel and spectel, was adopted for a given band). For the leading side (Obs. 18) where signals are slightly lower due to lower surface temperatures (see below), we could use Ngroups=2 for band 2B. Data in channels 3 and 4 remained unusable in all cases, being saturated even at Ngroups=1.
\item Cube recentering. As for NIRSpec, spectral images are not centered, i.e. Ganymede is shifted, typically by 0.05"-0.25", with respect to the sky position indicated by the MT-RA, MT-DEC header keywords. For each spectel of each band, we computed the central position of the image using the method adopted for NIRSpec data, taking a threshold value of 10\% of the maximum signal over the image, found to be optimum (see Sect.~\ref{sec:NIRSPEC-obs}). For each band, we then adopted the median of the values determined from each spectel. If the approach was perfect, the so-determined central position should be exactly the same for the different bands associated with the same grating position (i.e. 1A and 2A, 1B and 2B, 1C and 2C), given that the 4 channels are observed simultaneously. In reality, residual differences (mostly in DEC) between channels at the 0.03--0.05" level occurred, indicating that the recentering method is accurate to only $\sim$1/3 of a spaxel. The cubes were then spatially shifted according to the determined central position, and 
resampled to a common RA, DEC grid (that of band 1A).
\item Calibration adjustment. Spectra extracted at a variety of positions on the common RA, DEC grid show slight flux discontinuities at band edges (typically a few percent within the disk, and somewhat more near the limb, a likely result of the pointing corrections being imperfect). To minimize flux discontinuities, we proceeded as follows. For each pair of adjacent bands (e.g. 1C/2A), the longer-wavelength channel was scaled in flux to
match the shorter-wavelength channel in the overlapping region.  In doing so, the flux scale in Band 1A was taken (arbitrarily) as the reference. Although there is some arbitrariness in selecting this particular band, the calibration of bands 1A to 1C may be somewhat more trustworthy due to the larger number of retained groups. This led to correction factors usually of 0.97--1.03, occasionally reaching 1.08. 
Absolute flux uncertainties of $\sim$ 5 \% are inconsequential for the analysis.
\item Wavelength calibration. Similarly to NIRSpec, we used the solar line spectrum from \citet{Hase2010} to calibrate the wavelength scale in bands 1A and 1B, and found it to be accurate to within $\sim$0.001 $\mu$m, i.e. about 1 spectel. At longer wavelengths, where solar lines are more rare and weak and where the spectrum becomes progressively dominated by thermal emission, this approach could not be applied, so the default wavelengths from the pipeline were adopted.

\item Finally, for most of the scientific analyses, we co-added the recentered and recalibrated data from the different dither positions to enhance signal-to-noise. In doing so, we admittedly lose one of the features enabled by dithering, i.e a better sampling of the spatial point spread function. Nonetheless, this possibility is anyways compromised by the pointing inaccuracies described above.
\end{enumerate}

%---------------------NIRSPEC RADIANCE AND REFLECTANCE MAPS  --- 
\begin{figure}[ht!]
\begin{minipage}{8cm}
\includegraphics[width=4.cm]{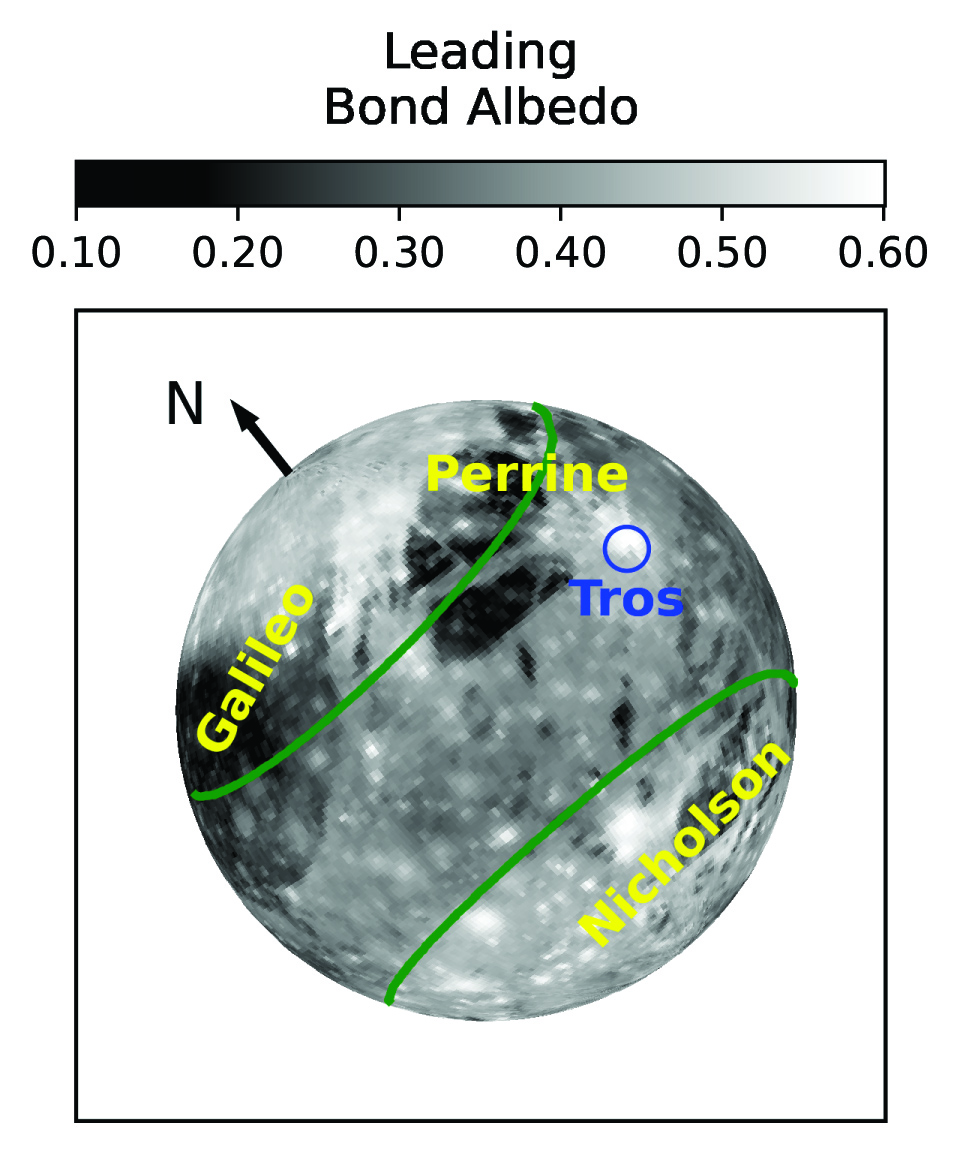}\hfill
\includegraphics[width=4.cm]{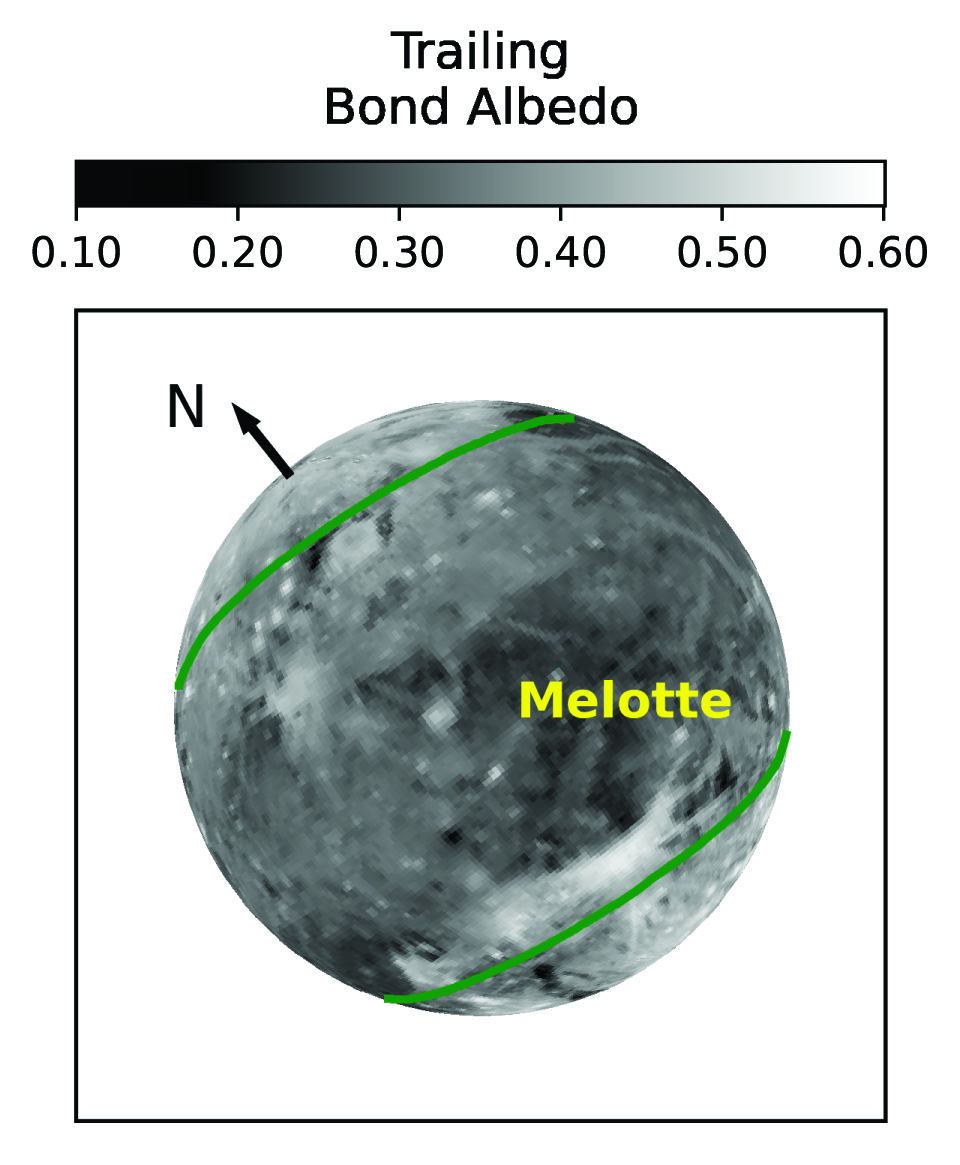}
\end{minipage}
\begin{minipage}{8cm}
\includegraphics[width=4.cm]{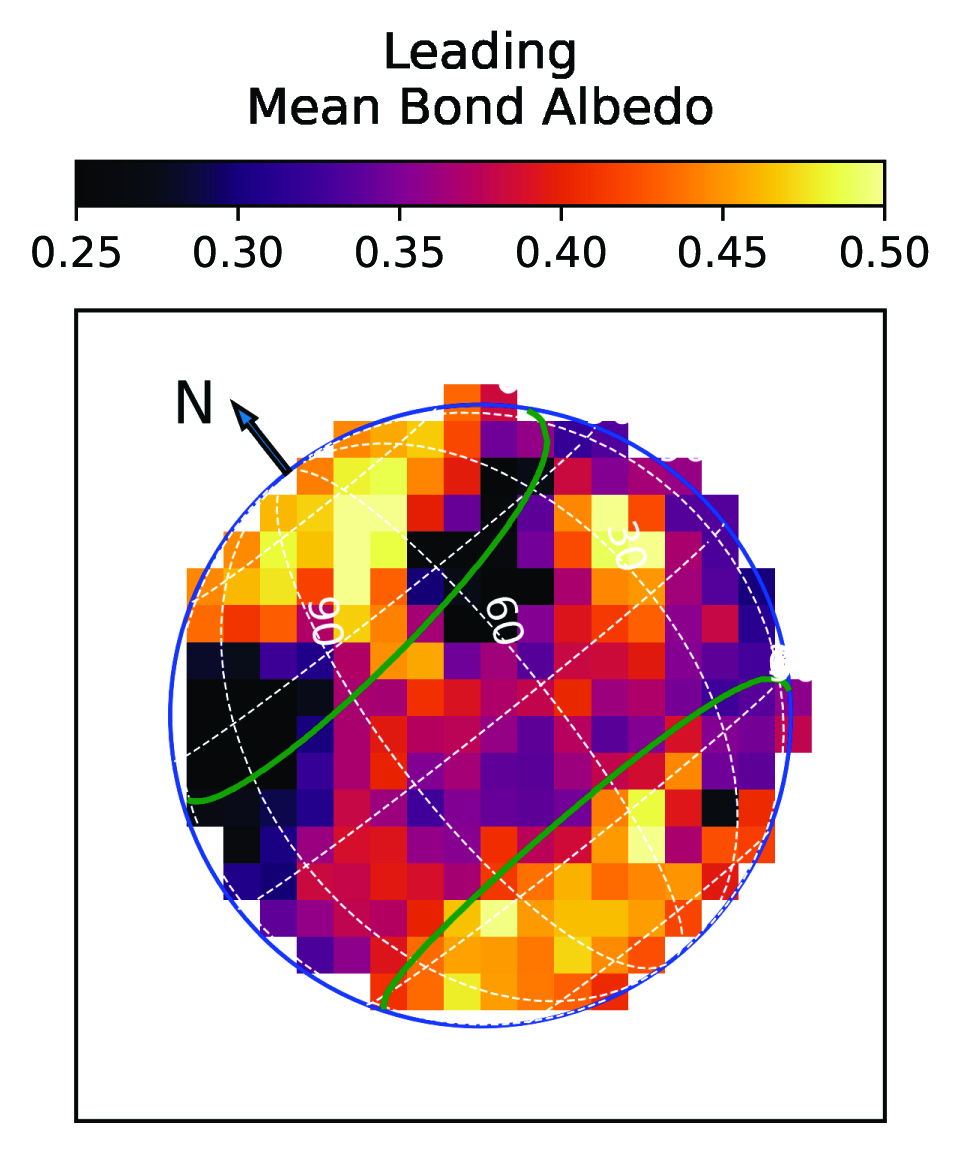}\hfill
\includegraphics[width=4.cm]{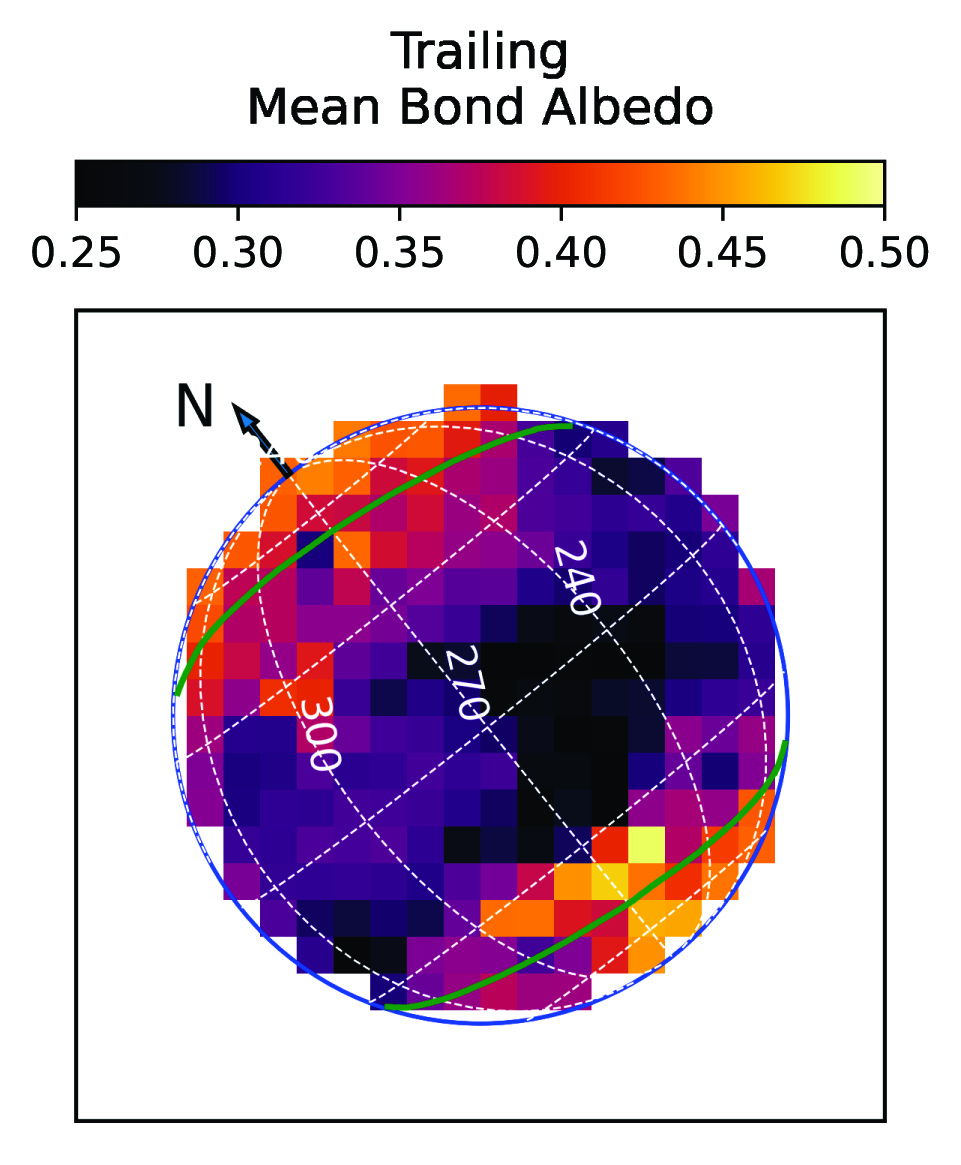}
\end{minipage}
\begin{minipage}{8cm} %4.36/4.5
\includegraphics[width=3.88cm]{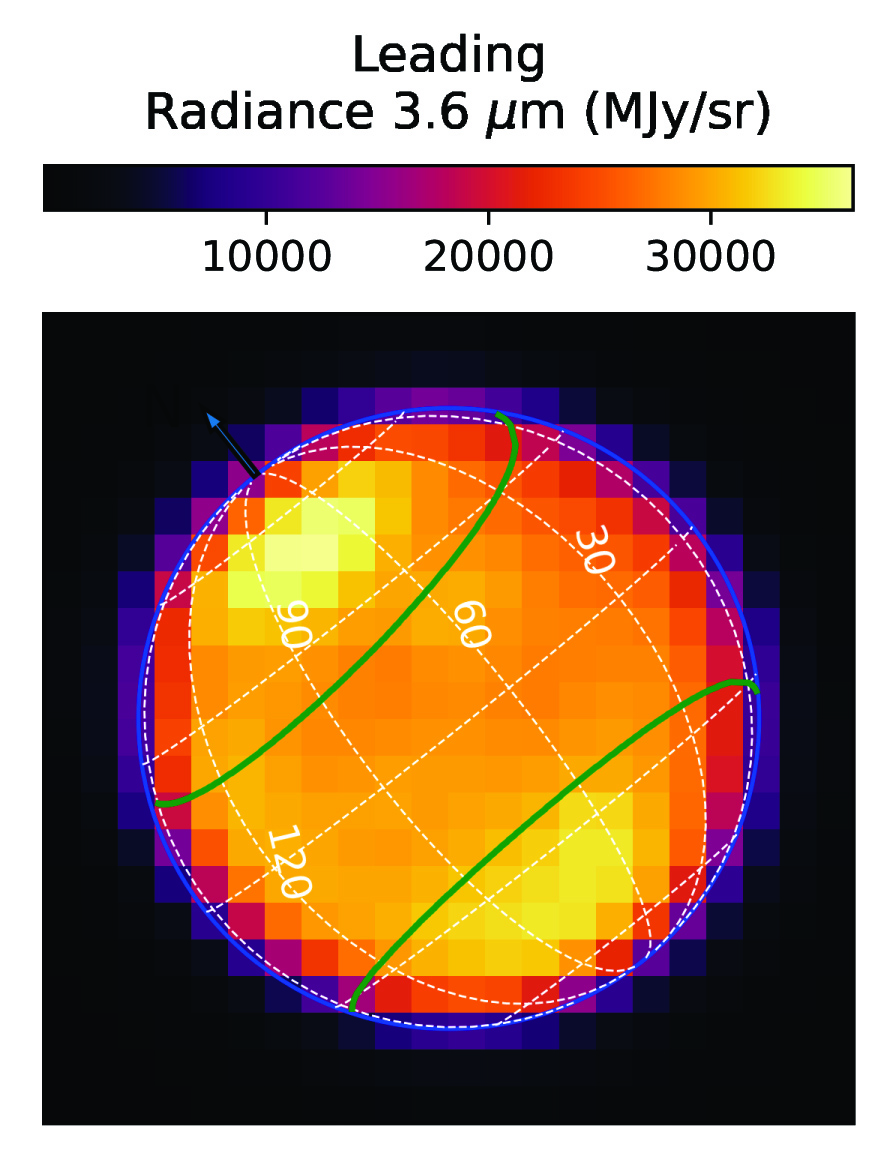}\hfill
\includegraphics[width=4.0cm]{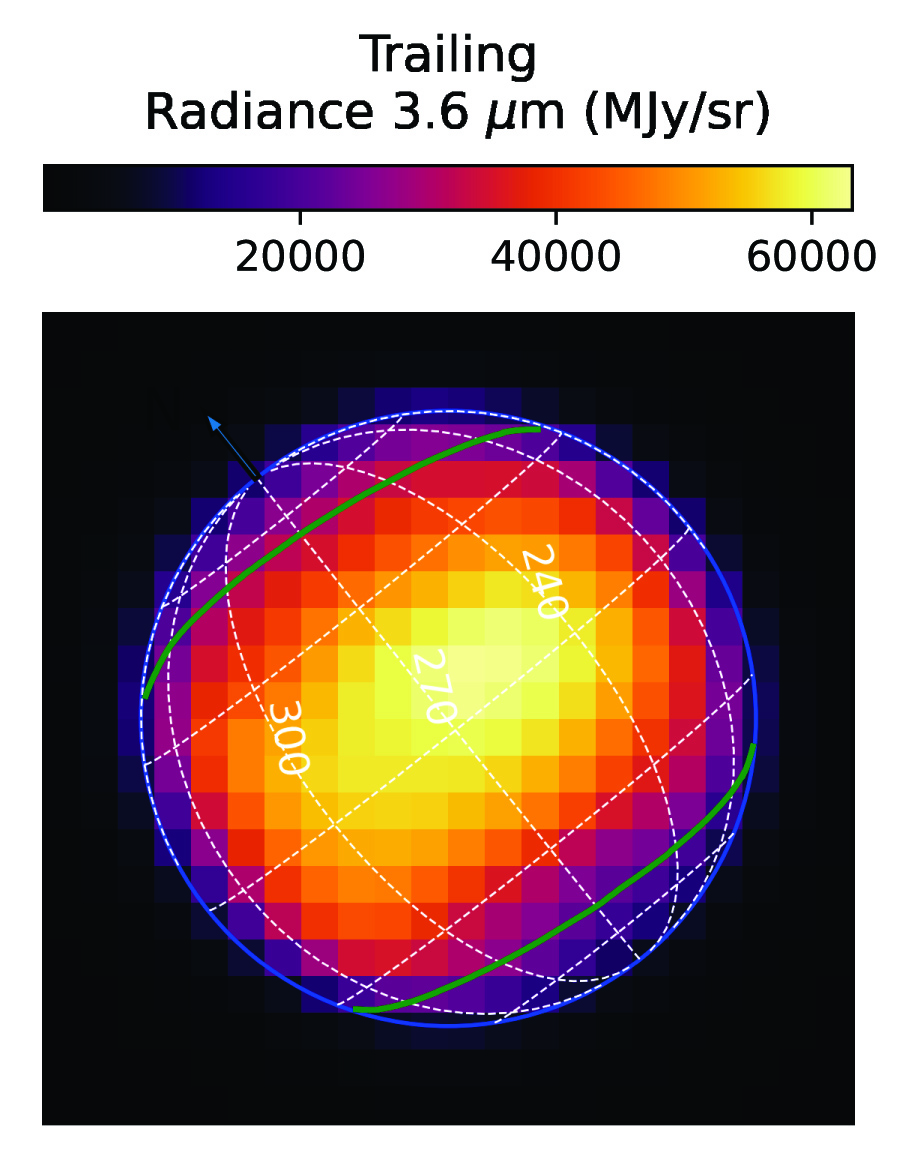}
\end{minipage}
\begin{minipage}{8cm}
\includegraphics[width=3.88cm]{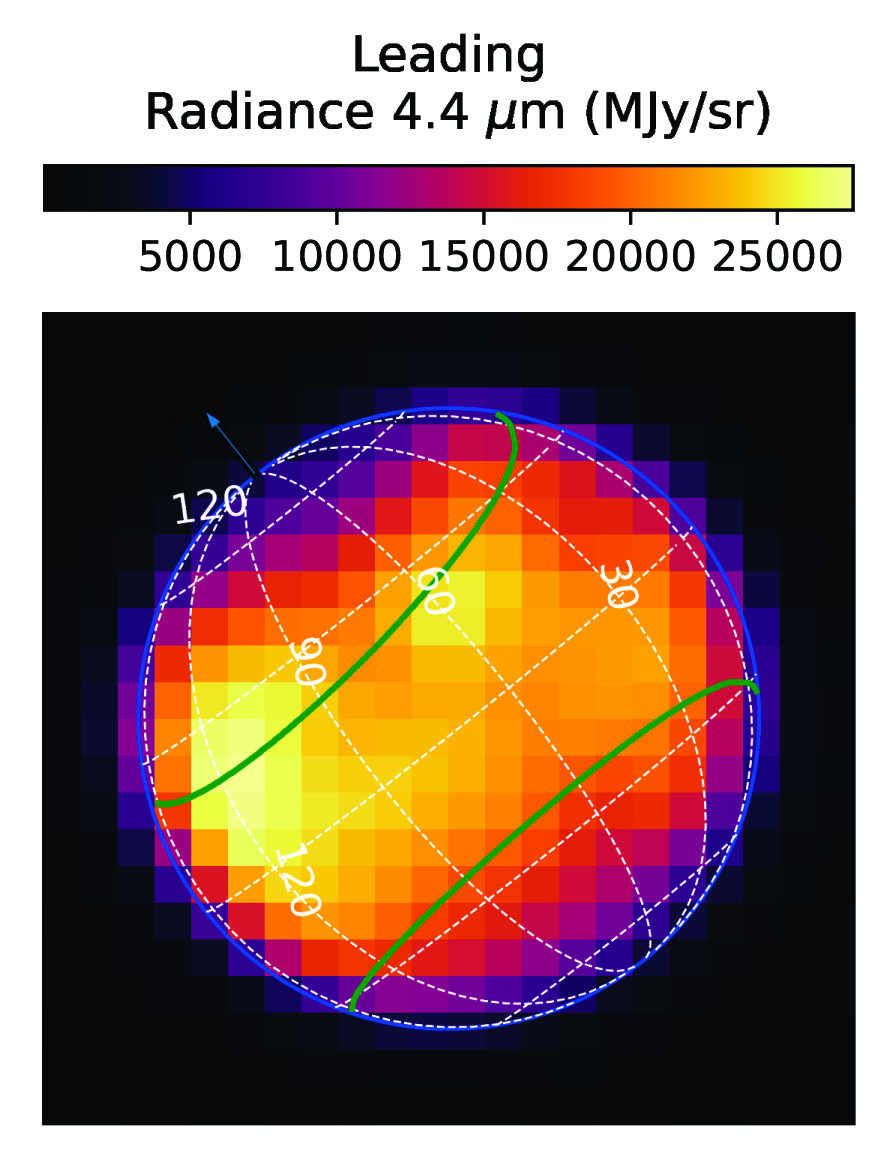}\hfill
\includegraphics[width=4.0cm]{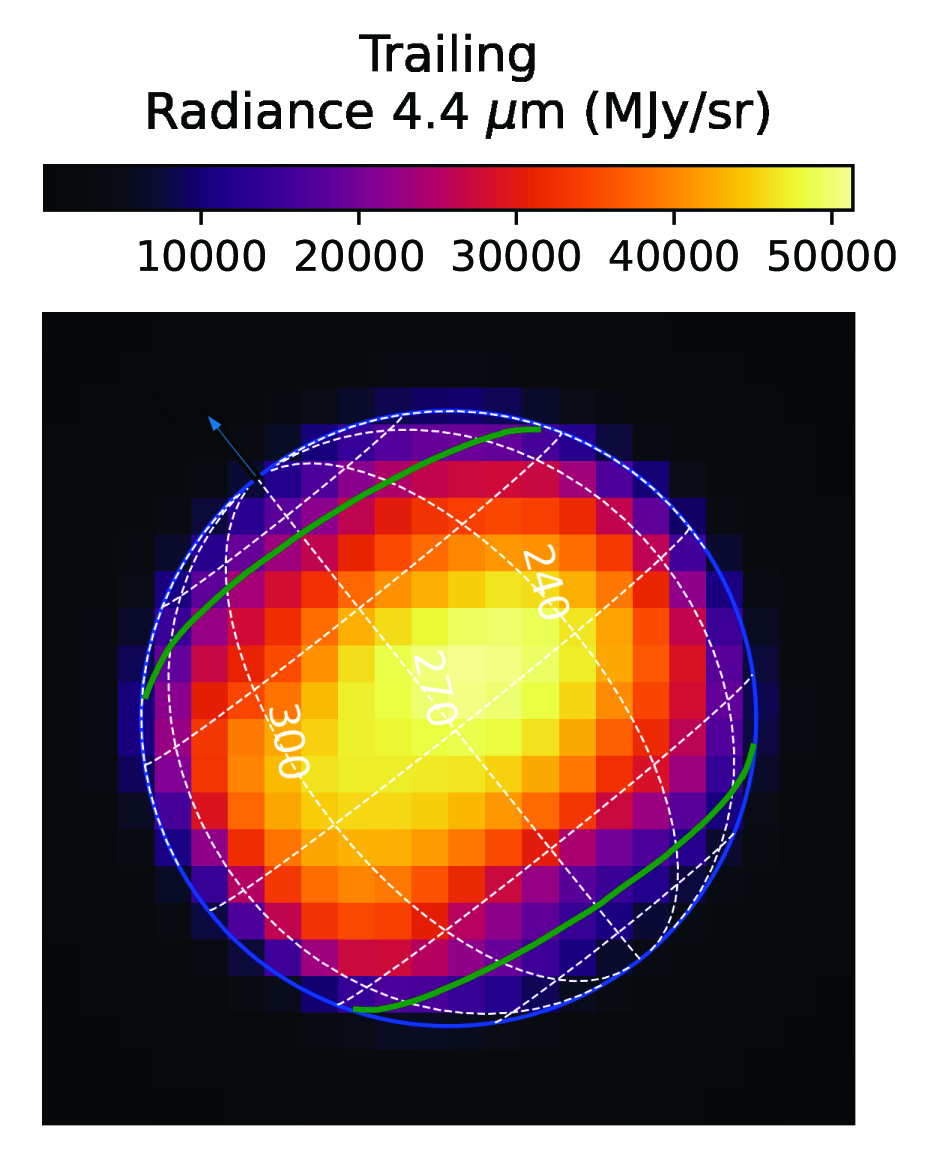}
\end{minipage}
\caption{From top to bottom: context Bond albedo maps derived by \citet{2021PSJ.....2....5D} from {\it Voyager-Galileo } mosaic with the dashed green lines showing the open-closed-field line-boundary  \citep{2022GeoRL..4901688D}, resulting Bond albedo maps after projection in the NIRSpec IFU spaxels (see text), NIRSpec radiance images at 3.6 (NRS1) and 4.4 (NRS2) $\mu$m. In the context maps, dark {\it Regiones} are indicated by their names in yellow, and the blue circle shows the position of the Tros crater.  Iso-latitude and longitude lines are drawn by increments of 30$^{\circ}$. The color scale is different for each map, except for the Bond albedo maps. Left side: Leading hemisphere; Right side: Trailing hemisphere.  }
\label{fig:albedo}
\end{figure}

\begin{figure}[ht]
\centering
\begin{minipage}{8cm} %4.36/4.5
\includegraphics[width=4.0cm]{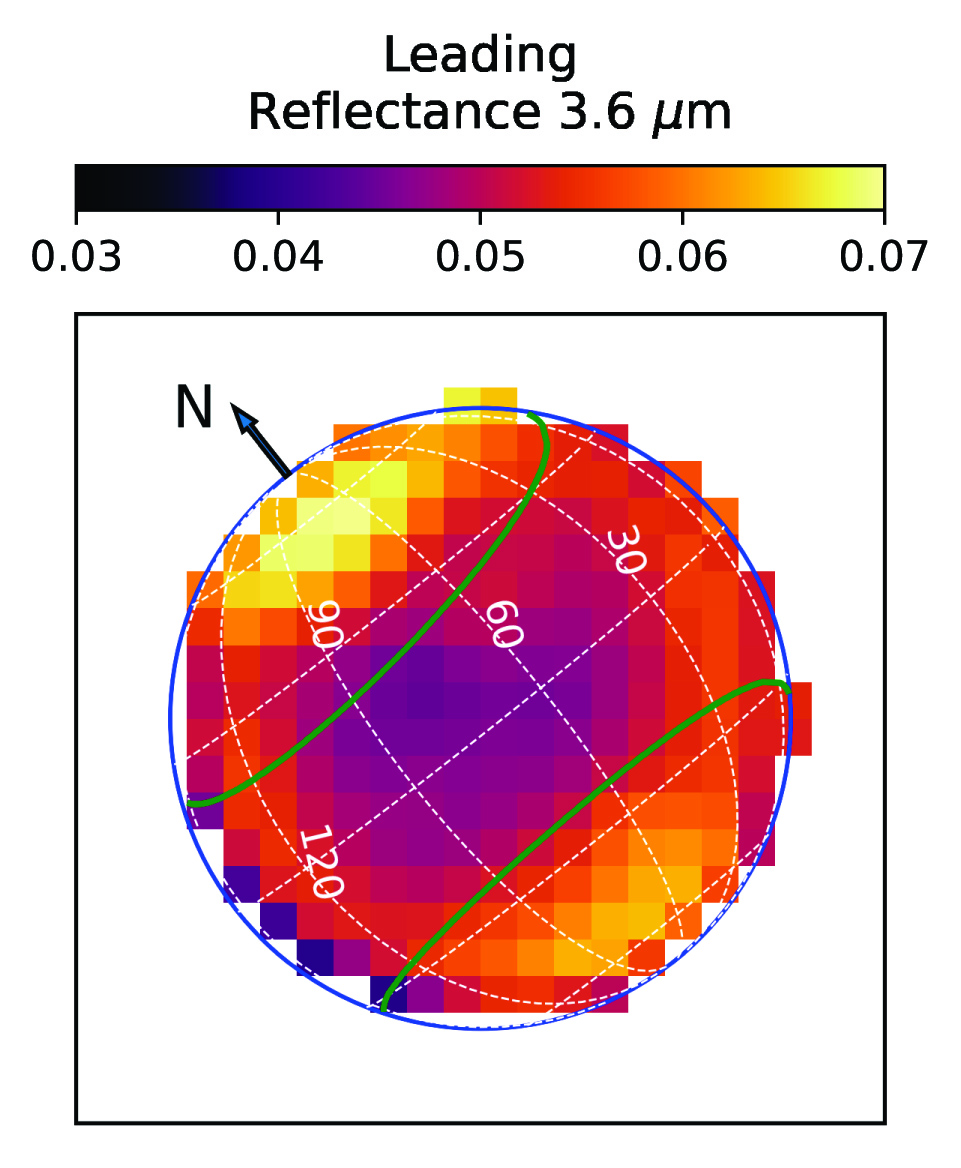}\hfill
\includegraphics[width=4.0cm]{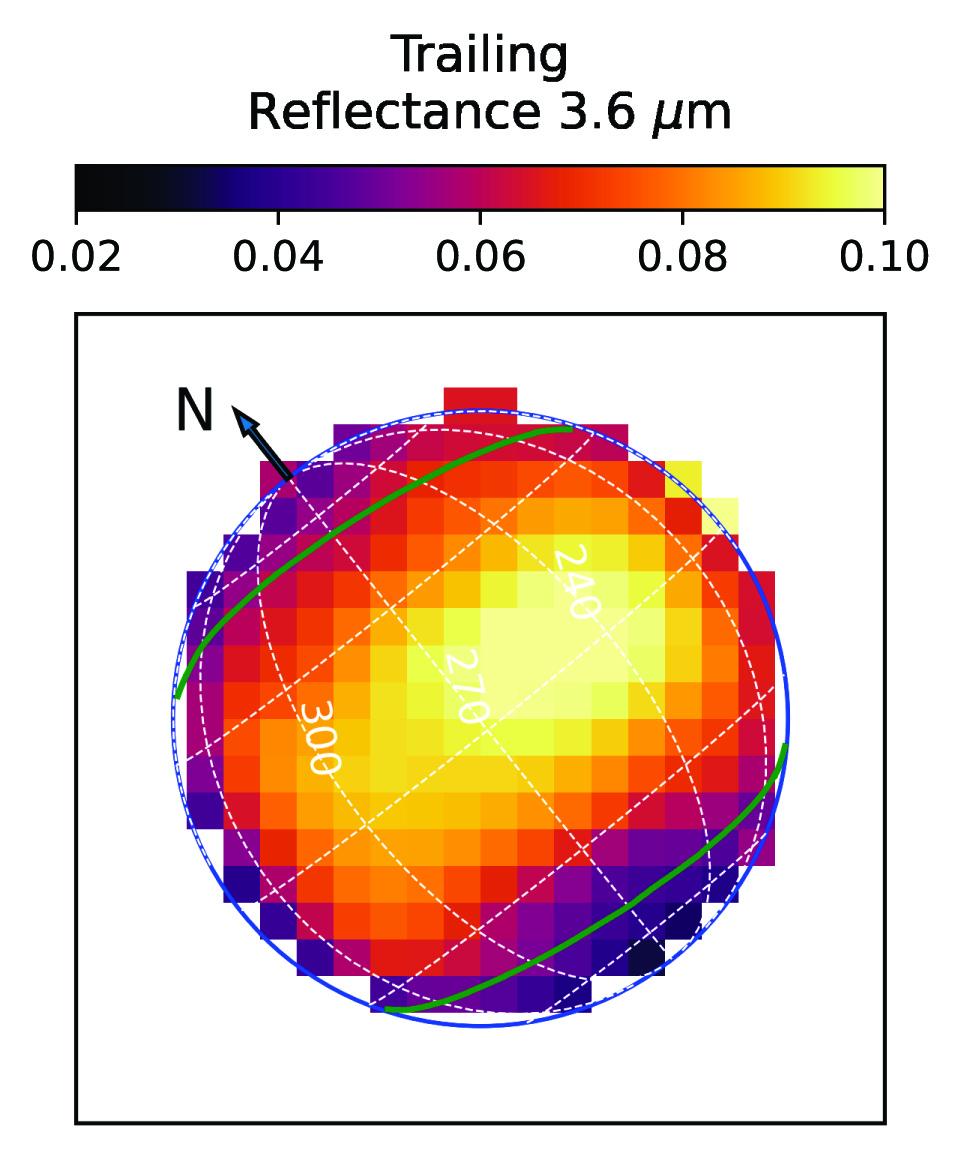}
\end{minipage}
\begin{minipage}{8cm}
\includegraphics[width=4.0cm]{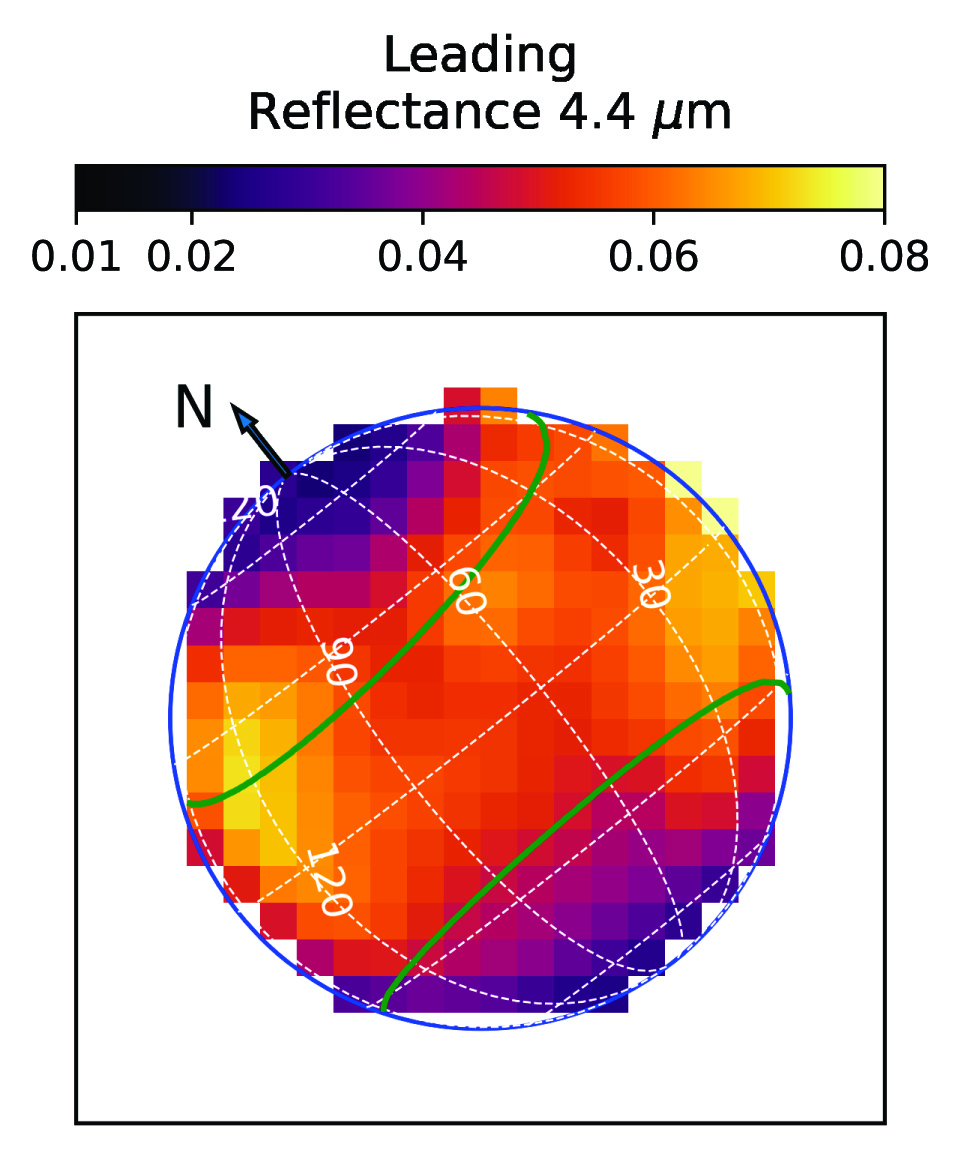}\hfill
\includegraphics[width=4.0cm]{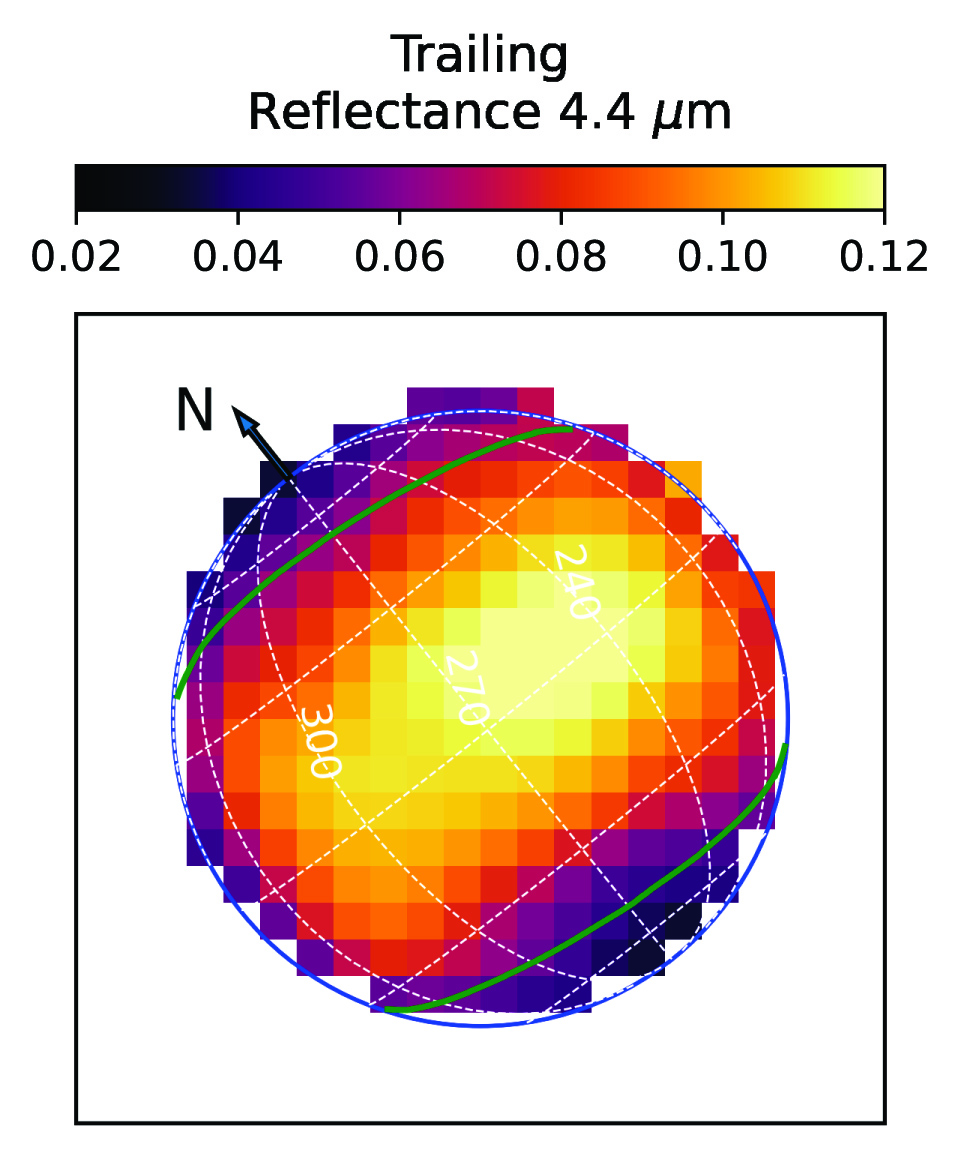}
\end{minipage}
\caption{From top to bottom: Reflectance images at 3.6 (NRS1) and 4.4 $\mu$m (NRS2)  for Leading (left) and Trailing (right) hemispheres. Iso-latitude and longitude lines are drawn by increments of 30$^{\circ}$. The color scale is different for each map. Note that the reflectance at the spaxels near the limb is somewhat inaccurate, as no PSF correction was applied. }
\label{fig:reflectancemaps}
\end{figure}

The possible contribution of straylight from Jupiter was found to be negligible. On the example of the leading side (Obs. 18) data, Fig.~\ref{fig:straylight}
shows the variation of mean radiance in each band (normalized to its maximum value) as a function of distance to Ganymede center. This radial distribution 
indicates a smooth decrease with distance, with no sign of plateauing at large distances that could indicate a ``background" due to Jupiter straylight, except may be in band 2C at the $\sim$ 0.01 level. In Band 2C, band-averaged equivalent brightness temperatures for Obs. 18 (resp. 27) are 145 K (resp. 153 K) at disk center and $\sim$ 132 K (resp. 136 K) at a distance of  0.65" (5 spaxels) from disk center (see Fig.~\ref{fig:globalspectra}). At this mean wavelength (11 $\mu$m), a ``contamination" level corresponding to 1 \% of the maximum flux is equivalent to a 0.2 K offset in the T$_B$ at disk center, and 0.35 K at 0.65", and can be safely ignored in the analysis.

%%%%%%%%%%%%%%%%%% SECTION 3 - NIRSpec DATA ANALYSIS 

\section{NIRSpec data analysis and results}
\label{sec:NIRSpec-an}

\subsection{NIRSpec reflectance spectra and maps}
\label{sec:3.1}
Figure~\ref{fig:albedo} shows maps of the radiance  at selected wavelengths of 3.6 $\mu$m (NRS1) and 4.4 $\mu$m (NRS2) (3$^{rd}$ and 4$^{th}$ row). In the first row of this figure are shown Bond albedo ($A_B$) maps at 1\dg~ resolution derived by \cite{2021PSJ.....2....5D} from the {\it Voyager-Galileo} mosaic\footnote{\url{https://astrogeology.usgs.gov/search/map/Ganymede/Voyager-Galileo/Ganymede_Voyager_GalileoSSI_global_mosaic_1km}}. For each NIRSpec spaxel, we averaged the $A_B$ values of these maps projecting on that spaxel. The resulting “projected maps" of the Bond albedo are shown in the second row of Fig.~\ref{fig:albedo}, and will be used to study correlations of derived parameters with $A_B$.

In Fig.~\ref{fig:albedo} and some following figures, we show the open-closed-field-line boundary (OCFB). This is the boundary between Ganymede's closed field line region, i.e. where both ends of field lines from Ganymede intersect with Ganymede's surface,  and open field lines,  i.e. where one end of the field line intersects with Ganymede's surface and the other end reaches Jupiter. The open-field-line region, which covers a relative large area over the poles, is populated with energetic ions and electrons of Jupiter's magnetosphere and thus the surface is much more exposed to plasma irradiation than on the closed field line regions. 
The location of the OCFB is in the range 20--30\dg and 40-50\dg~latitude N/S for the leading and trailing sides, respectively. It oscillates and additionally shifts latitudinally by up to 10\dg~on the downstream side of Jupiter's magnetospheric plasma flow (i.e., leading side) as a function of Ganymede's position in Jupiter's magnetosphere, i.e. primarily with a period of about 10 hours \citep{Saur2015}. On the upstream (i.e. trailing) side, the variability might be larger, but a dedicated study has not been made. The OCFB shown on the figures is that calculated for the Juno PJ34 flyby of Ganymede, when the moon was near the center of the plasma sheet, i.e., exposed to comparably large plasma ram pressure \citep{2022GeoRL..4901688D}.

Maps of the reflectance are given in Fig.~\ref{fig:reflectancemaps}. The reflectance at each spaxel was determined from the $I/F$ value, applying the photometric model of \citet{Oren1994}. Indeed, it has been shown that the Lambert model is not appropriate for Ganymede at UV wavelengths \citep{Alday2017}, as well as in the near-IR \citep{Ligier2019,King2022-is}, and we found that this turns out to be also the case for the NIRSpec data. The Oren-Nayar model is a widely used reflectivity model for rough diffuse surfaces, where the roughness is defined by a gaussian distribution of facet slopes, with variance $\sigma^2$. The Oren-Nayar model simplifies to the Lambert model (radiance $\propto$ cos($\theta_i$), where $\theta_i$ is the solar incidence angle) for $\sigma$ = 0. Following \cite{Ligier2019}, we adopted a value of $\sigma$ = 20$^{\circ}$. Varying $\sigma$ by $\pm$ 5$^{\circ}$ (\cite{Ligier2019} derived values from 16$ \pm 6^{\circ}$ to 21$ \pm 6^{\circ}$), the reflectance maps look similar. We did not correct for PSF filling effects which are significant for the spaxels near the limb. Therefore reflectance values are inaccurate for those spaxels. 

\begin{figure*}[htp]
\begin{minipage}{18cm}
\includegraphics[width=9.cm]{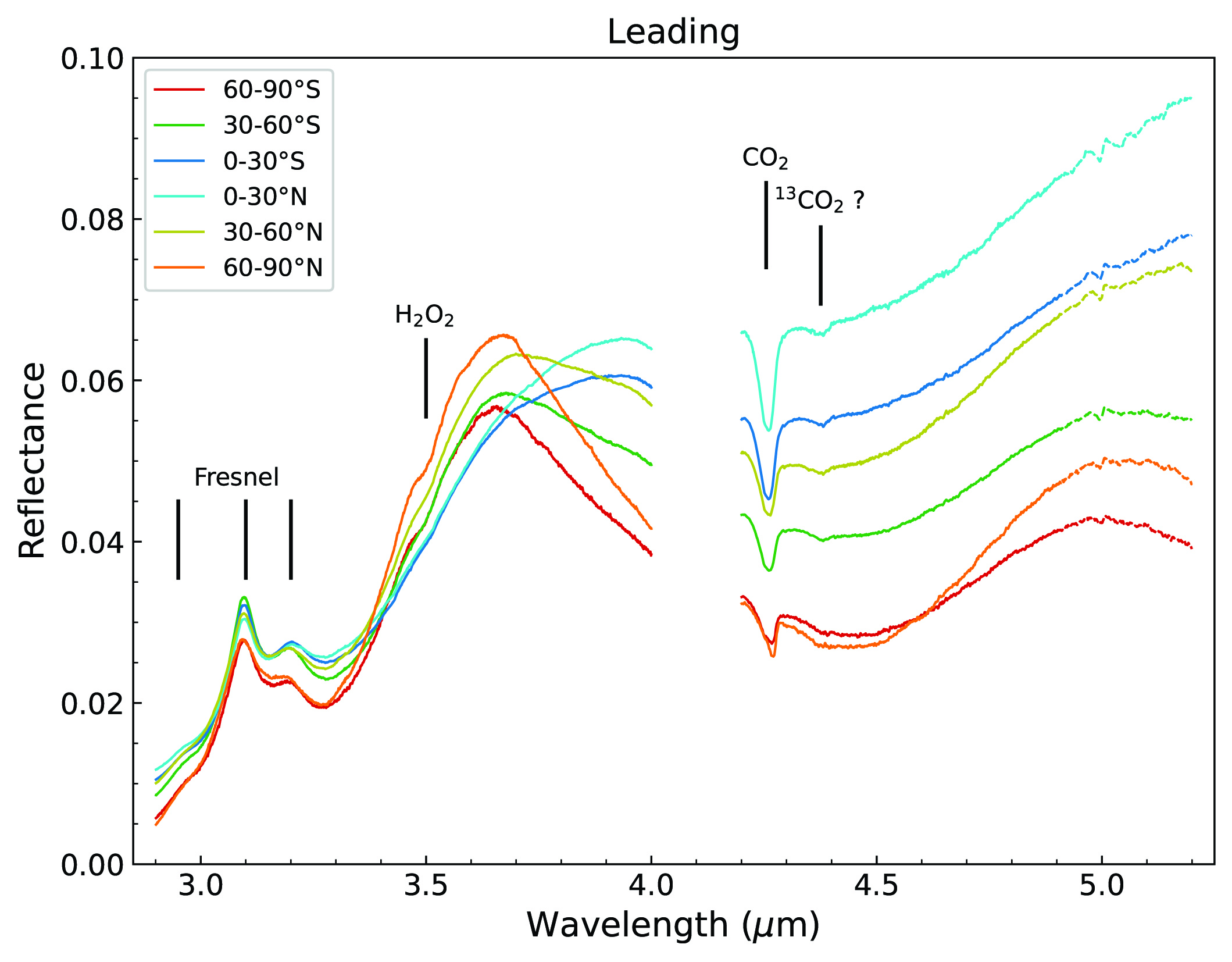}\hfill
\includegraphics[width=9.cm]{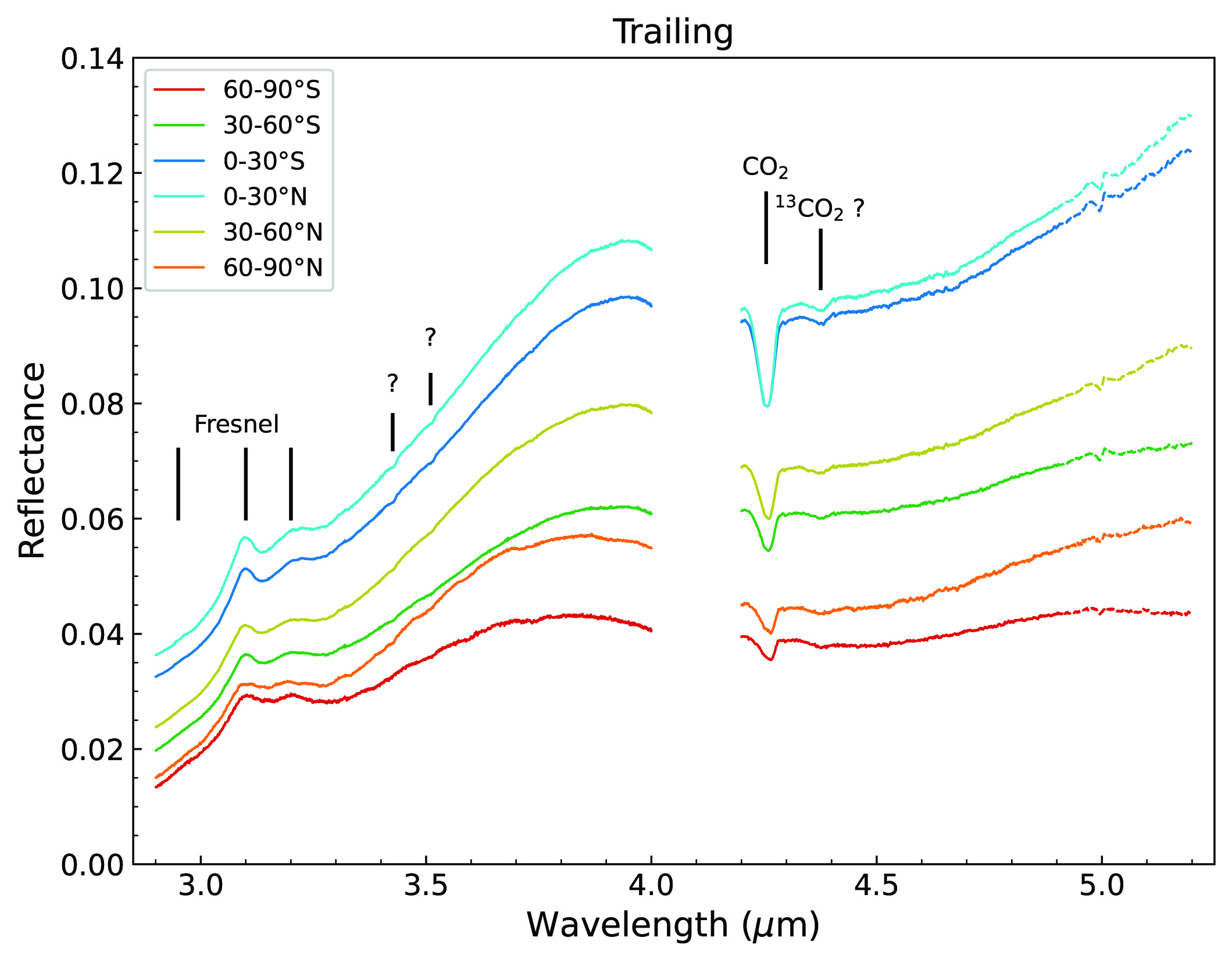}
\end{minipage}
\caption{NIRSpec reflectance spectra of Ganymede surface for the leading (left) and trailing (right) hemispheres. Averages over bins of latitudes are plotted, as labelled in the upper left panels. Thermal emission was not subtracted, so reflectance values are significantly overestimated at $\lambda$ $>$ 4.9 $\mu$m (dashed lines, see text). The absolute scale is also inaccurate for the highest latitudes due to PSF filling effects (see text). Spectral identifications of the three H$_2$O Fresnel peaks (2.95, 3.1 and 3.2 $\mu$m), absorption bands of H$_2$O$_2$ (3.5 $\mu$m) and CO$_2$ (4.26 $\mu$m) are indicated. The absorption signatures at the position of the $^{13}$CO$_2$ band (4.38 $\mu$m), and those at 3.426 and 3.51 $\mu$m indicated by question marks, are most likely spurious (see text). The feature seen in absorption at 4.995 $\mu$m at the 2\% level is not of Ganymede origin as it is not observed in MIRI spectra. The spectral continuum is dominated by water ice for the mid and high latitudes of the leading hemisphere, with local maxima at 3.65 and 5 $\mu$m (H$_2$O inter-bands) and a minimum at 4.45 $\mu$m (H$_2$O absorption band). In other regions, the higher concentration of non-ice materials (minerals etc.) induces the increase of the reflectance with wavelength, the reduction of the amplitude of the water ice spectral features, and the shift of the water inter-band to 3.9 $\mu$m.   \label{fig:sp-fullrange}}
\end{figure*}

Reflectance spectra from 2.86 to 5.28 $\mu$m across Ganymede's leading and trailing hemispheres are shown as average spectra over latitude bins in Fig.~\ref{fig:sp-fullrange}. On average in this wavelength range, the trailing hemisphere is brighter than the leading hemisphere. This is due to a higher abundance of non-ice materials at the surface of the trailing hemisphere, contaminating the water ice. Some of these non-ice materials (such as opaque minerals) are darker than water ice in the visible and near infrared, but brighter than water ice in the mid-infrared \citep{Hibbitts2003, pappalardo2004, calvin1991, sultana2023}. The Bond albedo and 4.4-$\mu$m reflectance maps are indeed anti-correlated on both hemispheres (Figs~\ref{fig:albedo}--\ref{fig:reflectancemaps}). Ganymede's spectral reflectance variations are dominated by the relative surface abundance of water ice and non-ice materials.
 
The spectra averaged over latitude bins shown in Fig.~\ref{fig:sp-fullrange} highlight several spectral features, with the highest signal-to-noise and spectral resolution available to date: 1) Strong spectral features due to water ice (absorption bands around 3.1, 4.5 $\mu$m; inter-bands at 3.65, 5 $\mu$m; Fresnel peaks at 2.95, 3.1, 3.2 $\mu$m); 2) CO$_2$ absorption band at 4.26 $\mu$m together with an absorption band at 4.38 $\mu$m, which coincides in wavelength with $^{13}$CO$_2$, but whose attribution  is uncertain (see below); 3) the H$_2$O$_2$ absorption band at 3.505 $\mu$m analysed by \cite{Trumbo2023}; and 4) weak signatures at 3.426 $\mu$m and 3.51 $\mu$m, best seen at low latitudes on spectra of the trailing hemisphere.

The strongest spectral features are caused by water ice. Water ice absorption bands are centred at 2.94 and 3.1 $\mu$m for the fundamental symmetric ($\nu_1$) and asymmetric stretching modes ($\nu_3$), at 3.18 $\mu$m for the overtone of the bending mode (2$\nu_2$), and at 4.48 $\mu$m for the combination of fundamental bending and libration modes ($\nu_2$ + $\nu_L$) \citep{Ockman1958,Mastrapa2009}. The narrow local maxima at 2.95 (weak), 3.1 and 3.2 $\mu$m are the Fresnel reflection peaks due to the strong increase of water ice refractive index near these maxima of absorption, which are characteristic of the  presence of surficial water ice. The broad local maxima at 3.65 and 5 $\mu$m are due to light scattering by the water ice grains between absorption bands centered at 3.1, 4.48 and 6--6.2 $\mu$m. The intensity and position of the maximum around 3.65 $\mu$m is known to depend on the ice grain size \citep{Hansen2004} and temperature \citep{Filacchione2016}, and is also affected by the presence of non-ice materials. In the following paragraphs, this spectral feature will be called the "inter-band" (see Sect.~\ref{discussion_h2o_nonice}).

Spectra of the dark (i.e. low albedo) terrains -- in the trailing and at low latitudes of the leading hemisphere-- present a reduction of the amplitude of the water ice spectral features whose positions are shifted to higher wavelengths (e.g., the position of the inter-band is shifted from 3.65 to 3.9 $\mu$m, Fig.~\ref{fig:sp-fullrange}). Maps in Fig.~\ref{fig:reflectancemaps} show that the reflectance at 3.6 $\mu$m is the highest on the darkest terrains of the trailing hemisphere, whereas on the leading hemisphere, it is maximum at the poles, due to the higher abundance and possibly smaller grain size of water ice, causing an increased reflectance at the inter-band (see Sect.~\ref{discussion_h2o_nonice}).

The absorption band at 4.26 $\mu$m can be firmly attributed to the asymmetric stretching mode of C=O in CO$_2$ in solid phase. However, pure CO$_2$ ice is not expected to be thermodynamically stable under Ganymede's surface temperature and pressure conditions, so the CO$_2$ molecule is trapped in a host material, or is interacting with other materials (see Sect.~\ref{sec:CO2}). Figure~\ref{fig:sp-fullrange} shows that, the darker the terrains are (trailing versus leading hemispheres, and equatorial versus polar regions), the more the inter-band shifts to longer wavelengths. The CO$_2$ band is also deeper on the equatorial regions  than at the poles. 

The faint (2\% depth, Sect.\ref{sec:13CO2}) band at 4.38 $\mu$m coincides in wavelength with the corresponding stretching mode of $^{13}$CO$_2$. However, the NIRSpec IFU flux calibration is not yet definitive, especially for the G395H grating, so definitive conclusions about the reality of individual features at the level of 1 or 2\% cannot be drawn (Charles R. Proffitt, NIRSpec Team, private communication). Another faint feature is observed at 4.995 $\mu$m, which is not observed in the MIRI spectra of Ganymede. This feature is an artifact seen in several NIRSpec modes that has been confirmed to us by the NIRSpec Team.

Spectra of Ganymede acquired with NIMS/Galileo and JIRAM/Juno instruments suggest the presence of several faint additional absorption features (e.g., at 3.4, 3.88, and 4.57 $\mu$m, \cite{McCord1998}, at 3.01, 3.30, 3.38 and 3.42 $\mu$m,  \cite{Mura2020}). But it must be noted that these features from Juno/JIRAM are very shallow, and not visible in all spectra acquired by Juno. With the exception of the 3.01-$\mu$m water ice signature, which appears as a distortion in the blue wing of the Fresnel peak observed on the leading hemisphere (Fig.~\ref{fig:sp-fullrange}), none of them are firmly identified in JWST spectra. Two faint and narrow absorption bands at 3.426 and 3.51 $\mu$m (i.e. nearby the H$_2$O$_2$ band), that may be attributable to organic C--H stretch signatures, are present in the NIRSpec spectra. Their depths (about 1\% for both features and for both hemispheres) do not show measurable trends with latitude. Their confirmation is pending a more robust calibration. 
 The 3.88-$\mu$m band detected with a depth of 2-3\% in NIMS/Galileo spectra (a possible signature of H$_2$CO$_3$, see discussion in Sect.~\ref{sec:5.9mu}) is not seen. A marginal signal at the 0.3-0.6\% level is not excluded, but again a more robust calibration is needed.

\subsection{Spectral analysis of NIRSpec data}

The relative strength, position and shapes of the above mentioned spectral features are diagnostic of the physical and compositional properties of the surface which vary across the disk \cite[e.g.][]{Ligier2019,King2022-is}. Noticeable in Fig.~\ref{fig:sp-fullrange} is the variation of the CO$_2$ band shape and position with latitude, shifting to higher wavelengths at the poles, thereby confirming Juno/JIRAM results \citep{Mura2020}. In this section, we describe how we characterized each band to study their spectral and spatial variations.

\subsubsection{Fresnel reflection peaks and H$_2$O 4.5-$\mu$m water band} 
\label{sec:fresnel}

We characterized the Fresnel reflection peak by measuring the central wavelengths of the two components at 3.1 and 3.2 $\mu$m (referred to as as peak 1 and peak 2, respectively) and the equivalent width of the whole reflection peak. Formally, the equivalent width is defined as:

\begin{equation}
 EqW =    \int \frac{F_c - F_s}{F_c} d\lambda,
\end{equation}
\noindent
where $F_s$ is the intensity (here $I/F$) of the spectrum, and $F_c$ is the continuum beneath the reflectivity peaks. In other words, the equivalent width is the peak area normalized by the underlying continuum. In the following, the equivalent width (in $\mu$m) will be referred to as the peak area. 

\begin{figure}[ht]
\begin{minipage}{9cm}
\includegraphics[width=4.5cm]{VISmaps_19_v3.jpg}\hfill
\includegraphics[width=4.5cm]{VISmaps_28_v3.jpg}
\includegraphics[width=4.5cm]{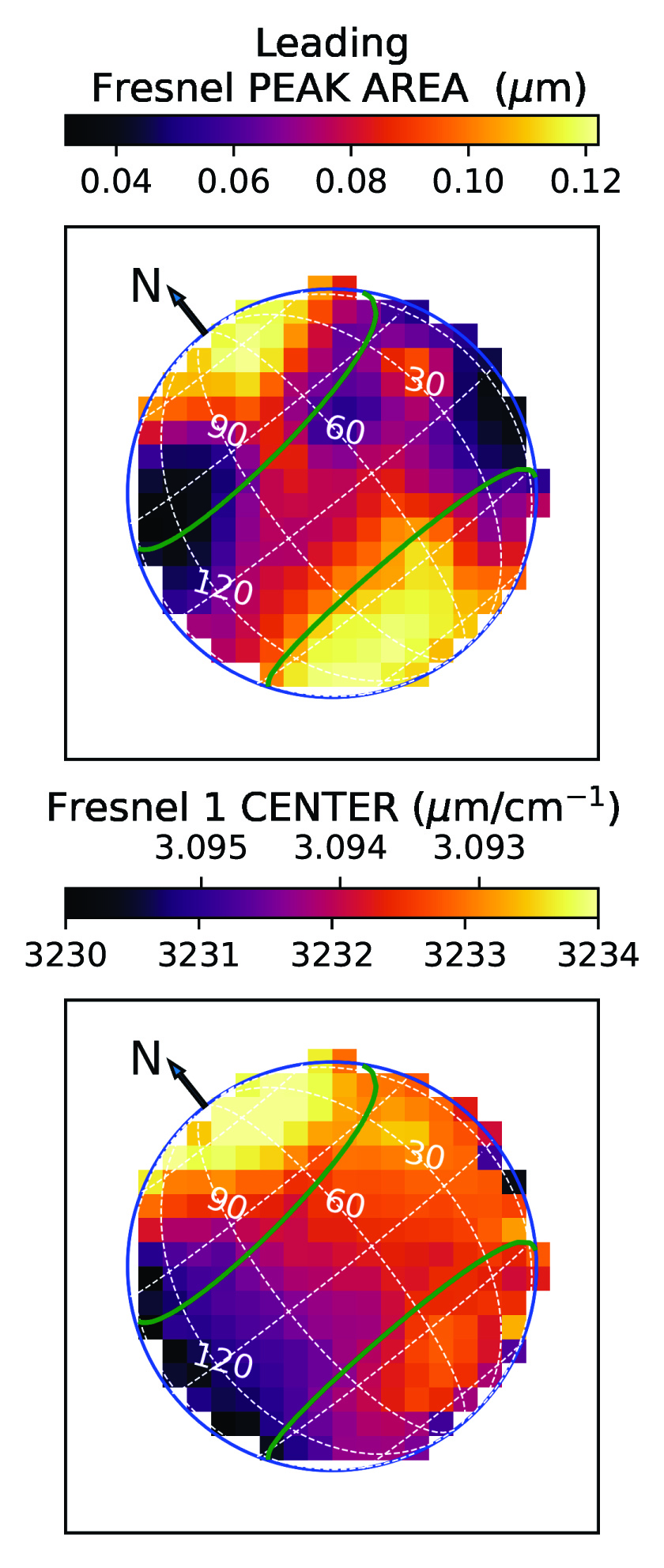}\hfill
\includegraphics[width=4.5cm]{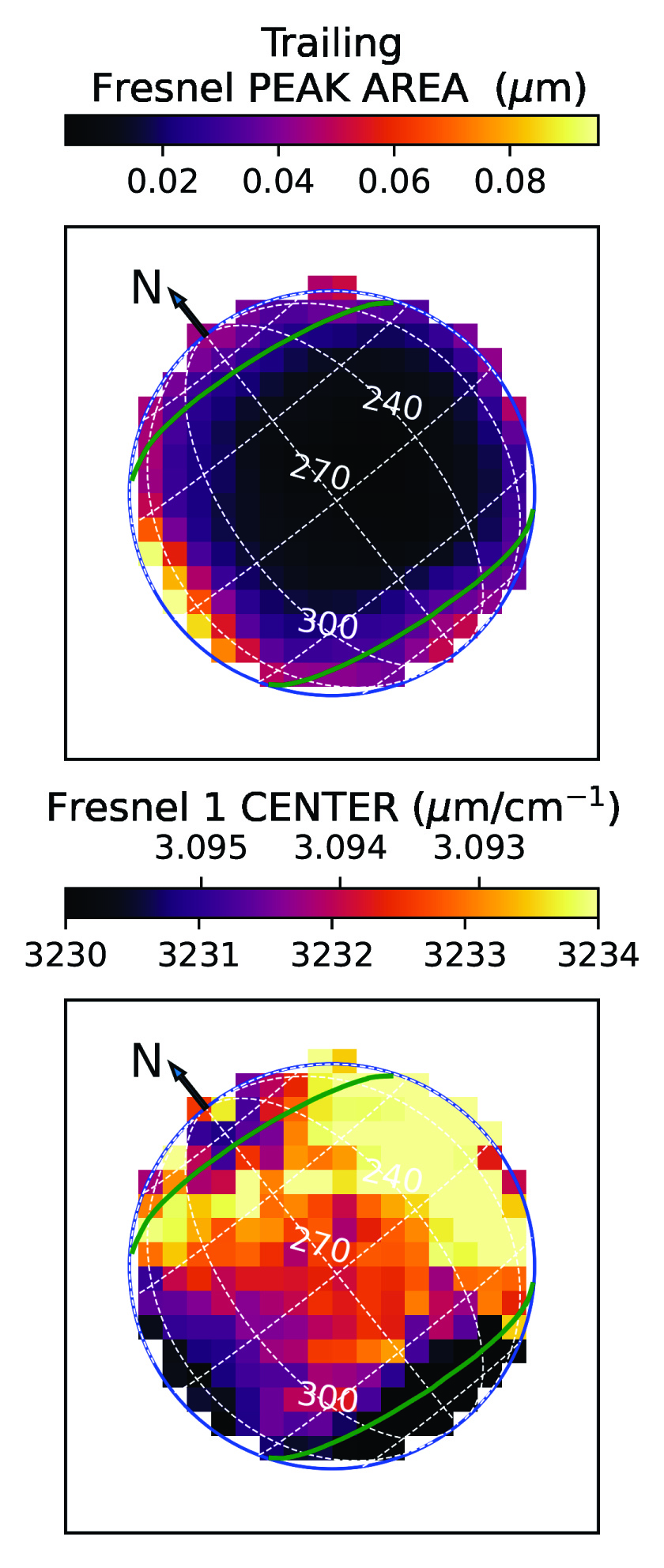}
\end{minipage}
\caption{From top to bottom: context Bond albedo maps, maps of the area of Fresnel reflection peak and central wavelength of Fresnel peak 1 from NIRSpec (NRS1) data. Maps of the leading and trailing hemispheres are shown on the left and right, respectively. 
\label{fig:maps-Fresnel}}
\end{figure}

To determine the central wavelengths of the Fresnel peaks, we modelled the $I/F$ emission in the 3.0--3.4 $\mu$m range by the sum of two Gaussians and a 4th-order polynomial.
The Fresnel-peak area was computed from 3.0 to 3.27 $\mu$m, taking as underlying continuum a straight line passing through the mean of the intensity in the 2.97--3.0 $\mu$m and 3.27--3.30 $\mu$m adjacent ranges. This was done for each spaxel. The resulting distributions are shown in Fig.~\ref{fig:maps-Fresnel}. 

\begin{figure}[ht]
\begin{minipage}{9cm}
\includegraphics[width=4.5cm]{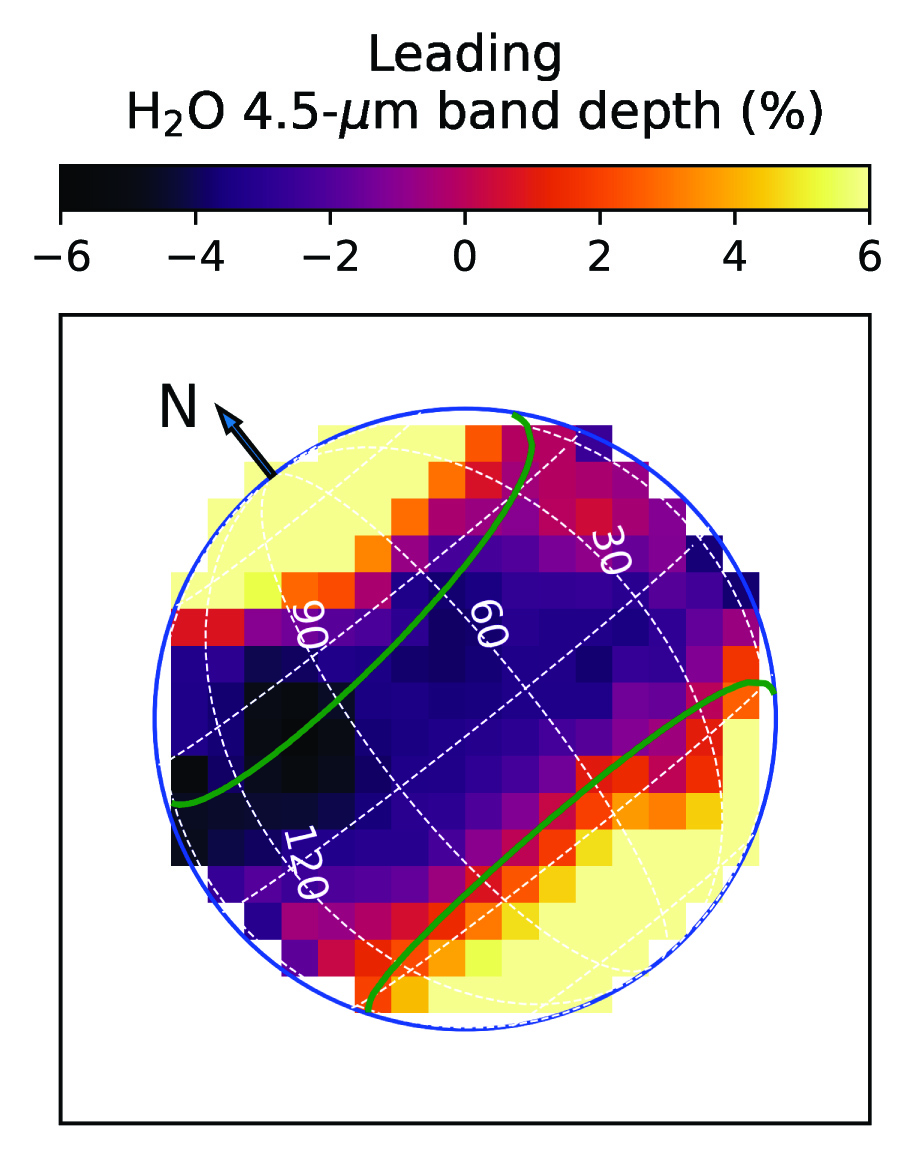}\hfill
\includegraphics[width=4.5cm]{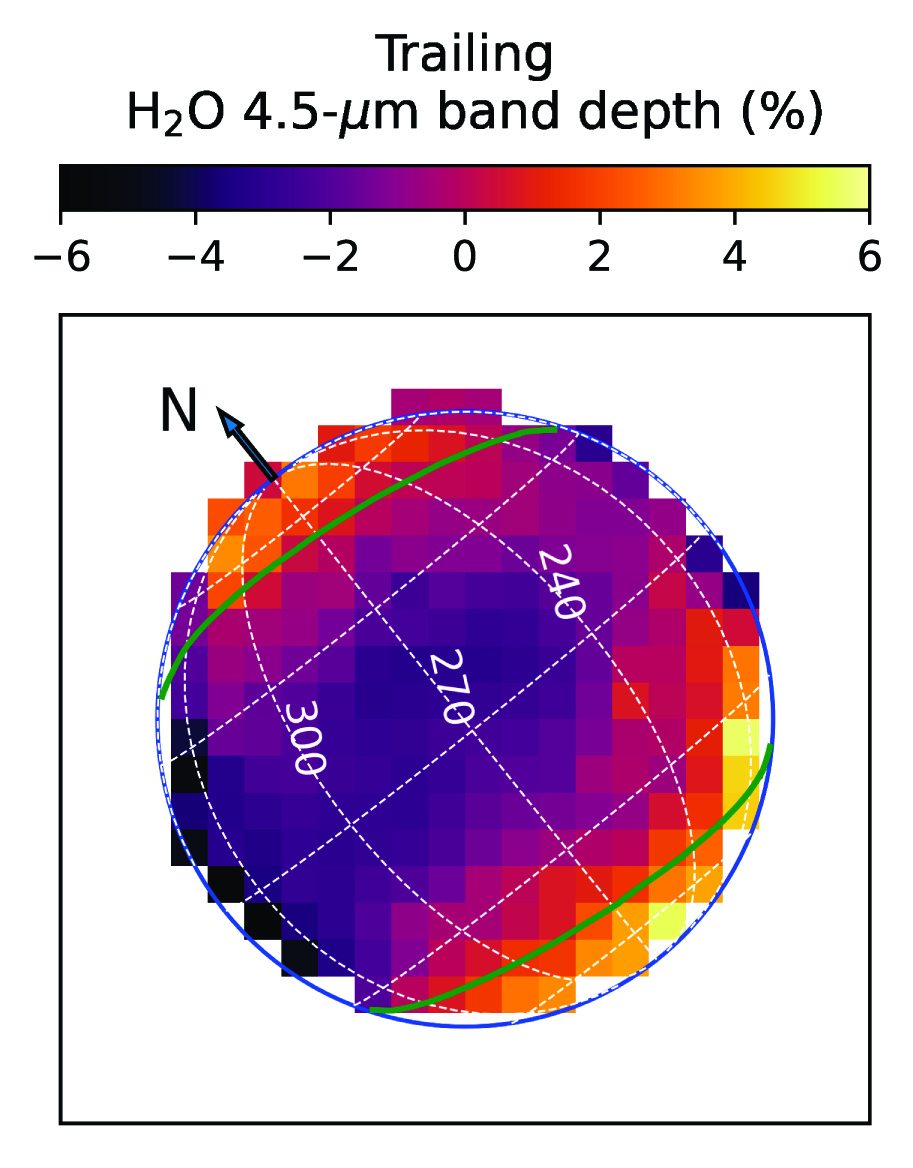}
\end{minipage}
\caption{H$_2$O 4.5-$\mu$m band-depth indicator. This band depth $BD$ is calculated with respect to the reflectance value at 4.205 $\mu$m ($BD$ = (1--$R_{\rm 4.48{\mu}m}$/$R_{\rm 4.205{\mu}m}$)$\times$100, where $R_{\rm 4.48{\mu}m}$ and $R_{\rm 4.205{\mu}m}$ are the reflectance values at 4.48 and 4.205 ${\mu}$m, respectively). Negative values indicate a significant  contribution of non-ice material in the continuum. 
\label{fig:maps-4p5}}
\end{figure}

Similar JWST/NIRSpec maps of the Fresnel-peak area were previously presented in \citet{Trumbo2023}. The Fresnel-peak area is lower on the trailing hemisphere (between 0.0032--0.04 $\mu$m) than on the leading hemisphere (between 0.03 to 0.12 $\mu$m). Both hemispheres present strong latitudinal, as well as longitudinal, variations that correlate with the Bond albedo (Fig.~\ref{fig:CO2-albedo}C). The Fresnel peak area is maximum in the polar regions of the leading hemisphere. 
Noticeable is the local maximum on the Tros crater (27\dg W, 11\dg N). The Fresnel peak area is small on the  Galileo, Perrine, and Nicholson dark regiones, comparing to their surroundings. Interestingly, the distribution of the Fresnel peak area, which is probing ice at the very top surface, closely matches the water ice distribution inferred from the depth of the H$_2$O 1.65 $\mu$m and 2.02 $\mu$m bands \citep{Ligier2019,King2022-is}, which are probing ice below the surface. It is also in good agreement with the distribution of the 4.5-$\mu$m band-depth proxy shown in Fig.~\ref{fig:maps-4p5} (see definition in the caption), characterizing shallower sub-surface layers than the 1.65 $\mu$m and 2.02 $\mu$m bands. Therefore, the maps shown here (Fig.~\ref{fig:maps-Fresnel}) are indicative of the presence of water ice not only at the top surface but also at shallow depth. However, one should keep in mind that the ice band depths and Fresnel peak area strongly depend on the way ice is mixed with other components \citep{Ciarniello2021}, and also on the crystallinity of the ice \citep{Mastrapa2009}. Therefore, their variations over the disk cannot be uniquely interpreted in terms of variations of areal abundance of water ice, and the potential influences of the ice crystallinity and of the mixing mode with non-ice materials should be kept in mind.

The distribution of the Fresnel peak area on the trailing side shows a ''bull's-eye'' pattern (Fig.~\ref{fig:bullseye-trailing}) with the minimum centred near the equator (261\dg W, 7\dg N, 13.21 h local time) and inside the dark Melotte Regio (245\dg W, 12\dg S). The same pattern is observed on near-IR H$_2$O-ice data of Ganymede \citep{Ligier2019} and is also observed for Europa, with the difference that for Europa the pattern is centred on the apex of the trailing hemisphere (270\dg W). For Europa, which does not possess an intrinsic magnetic field, this distribution might be related to the electron precipitation pattern on the surface \citep{Liuzzo2020-ww}, or to the implantation of iogenic sulfur ions \citep{Cassidy2013}.

The Fresnel peak area and the Bond albedo correlate well, but with a linear correlation which is different for the two hemispheres (Fig.~\ref{fig:CO2-albedo}C). One can notice high peak area values near the morning limb of the trailing hemisphere (longitudes 310-330\dg W), which are in the range of those measured in the leading side in regions with similar Bond albedo (Fig.~\ref{fig:CO2-albedo}C). This ice excess is not observed on the H$_2$O 4.5-$\mu$m map (Fig.~\ref{fig:maps-4p5}).

The central wavelength of the Fresnel peak 1 shows striking regional variations (Fig.~\ref{fig:maps-Fresnel}). The mean value is 3232.3 cm$^{-1}$ (3.0938 $\mu$m) for both hemispheres.  It is significantly blue-shifted (by 2 cm$^{-1}$, i.e. 0.0019 $\mu$m, with respect to the mean) at the north polar cap of the leading hemisphere. This trend is not seen at the south polar cap of the leading hemisphere nor for the trailing hemisphere. In addition a trend for red-shifted central wavelengths for earlier local times is observed on both hemispheres. The central wavelength of  Fresnel peak 2 (not shown in Fig.~\ref{fig:maps-Fresnel}) presents similar regional variations on the leading side (no conclusions can be drawn for the trailing side due to the low S/N) and is on average  3124.7/3125.7 cm$^{-1}$ (3.2003/3.1993 $\mu$m) for leading/trailing hemispheres.

\subsubsection{Inter-band}
\label{sec:inter-band}

\begin{figure}[ht]
\begin{minipage}{9cm}
\includegraphics[width=4.5cm]{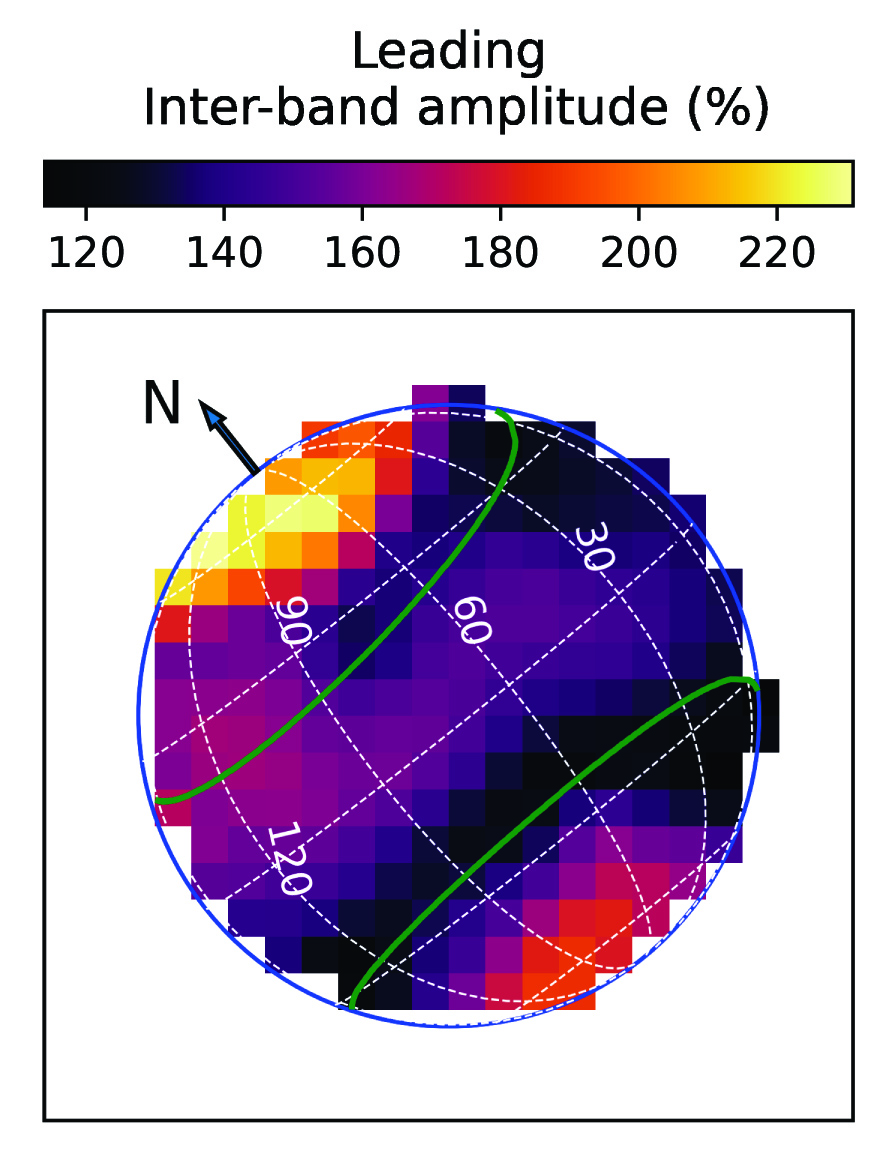}\hfill
\includegraphics[width=4.5cm]{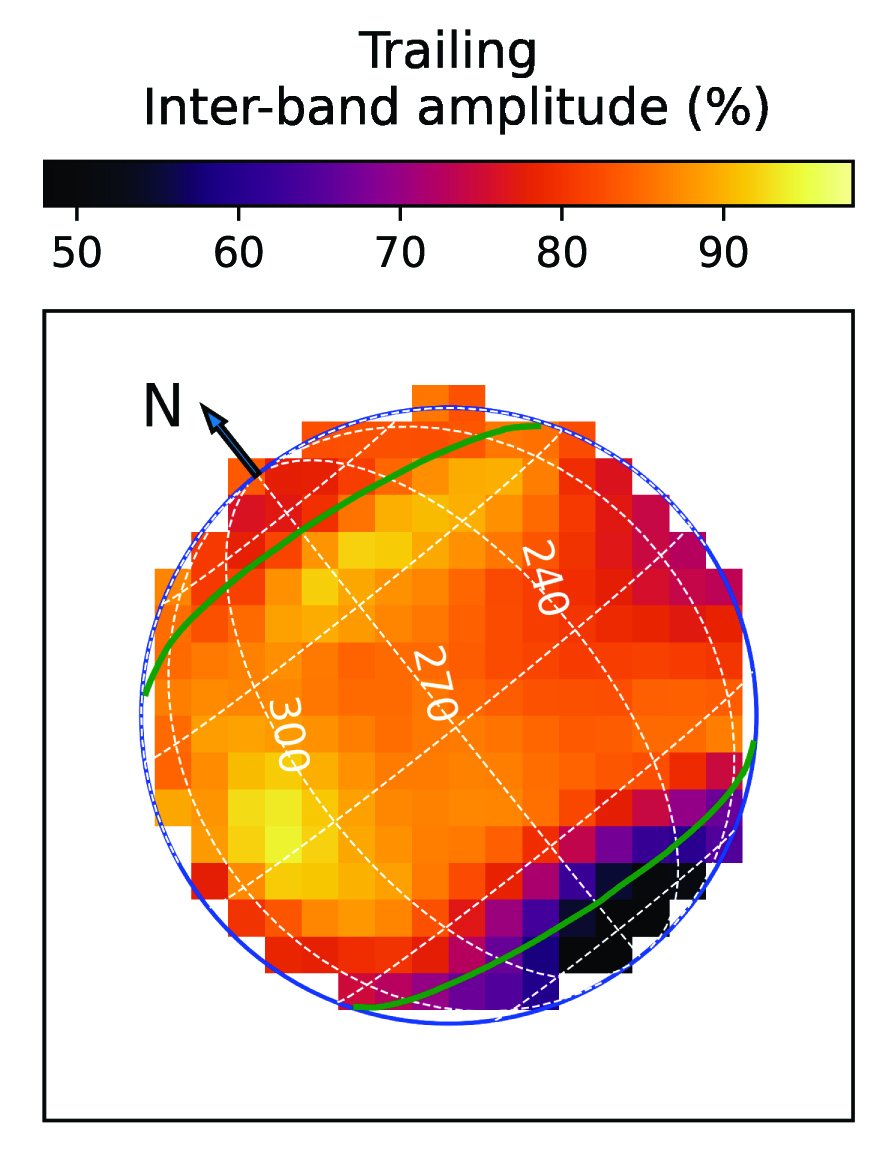}
\end{minipage}
\begin{minipage}{9cm}
\includegraphics[width=4.5cm]{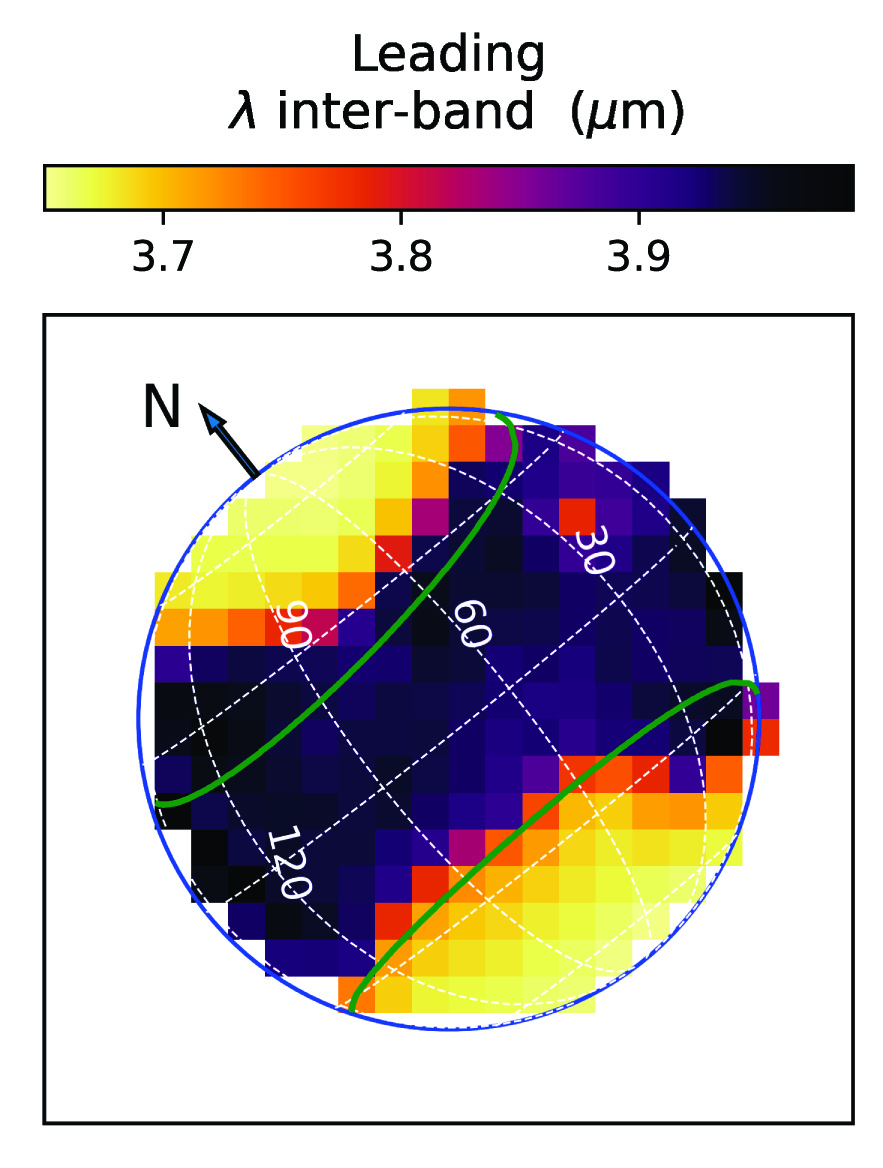}\hfill
\includegraphics[width=4.5cm]{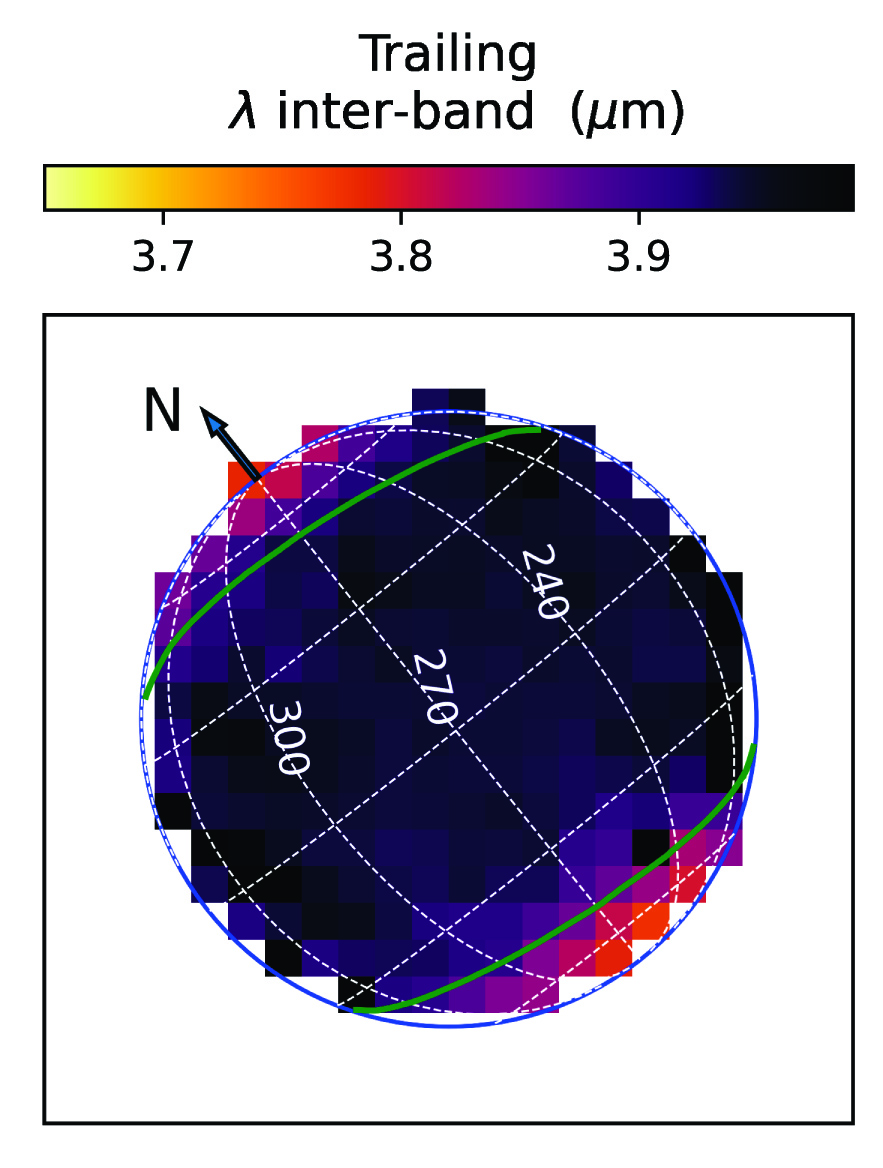}
\end{minipage}
\caption{Inter-band amplitude $A_{\rm IB}$ (top) and wavelength position  (bottom) for the leading (left) and trailing (right) hemispheres. 
\label{fig:maps-interband}}
\end{figure}

The wavelength of the continuum reflectance peak (referred to as the inter-band) longward of 3.6 $\mu$m was determined by fitting the 3.3--4.0 $\mu$m $I/F$ spectra by a 6th-order polynomial, and finding the position of the maximum of the fitted curve. The inferred distributions are plotted in Fig.~\ref{fig:maps-interband} (bottom maps). On the leading hemisphere, the distribution of the inter-band wavelength position shows a strong contrast between high and mid/equatorial latitudes and overall follows the distribution of the Fresnel-peak area (Fig.~\ref{fig:maps-Fresnel}), with a minimum value of 3.65 $\mu$m at the highest latitudes ($>$ 60\dg N/S), and a mean value of 3.93 $\mu$m at equatorial latitudes. The bright/icy Tros crater also stands out.  Latitudinal variations are also observed for the trailing hemisphere, with a mean value of 3.95 $\mu$m in the equatorial regions, and in the range 3.81--3.92 $\mu$m poleward of 40\dg N/S, consistent with the ice excess observed in polar regions. 

Figure~\ref{fig:maps-interband} (top) also shows maps of the amplitude of the reflectance peak at the position of the inter-band, measured in \% with respect to the reflectance at 3.28 $\mu$m:

\begin{equation}
 A_{\rm IB} = (\frac{R_{\rm \lambda_{IB}}}{R_{\rm 3.28{\mu}m}}-1)\times100,
\label{eq:int-IB}
\end{equation}
where $R_{\rm \lambda_{IB}}$ and $R_{\rm 3.28{\mu}m}$ are the reflectance values at the wavelength of the inter-band and at 3.28~$\mu$m, respectively.
For both hemispheres the inter-band is more intense in the north polar caps compared to the south polar caps.  

\subsubsection{CO$_2$ band}
\label{sec:CO2-an}

\begin{figure}
\includegraphics[width=9.cm]{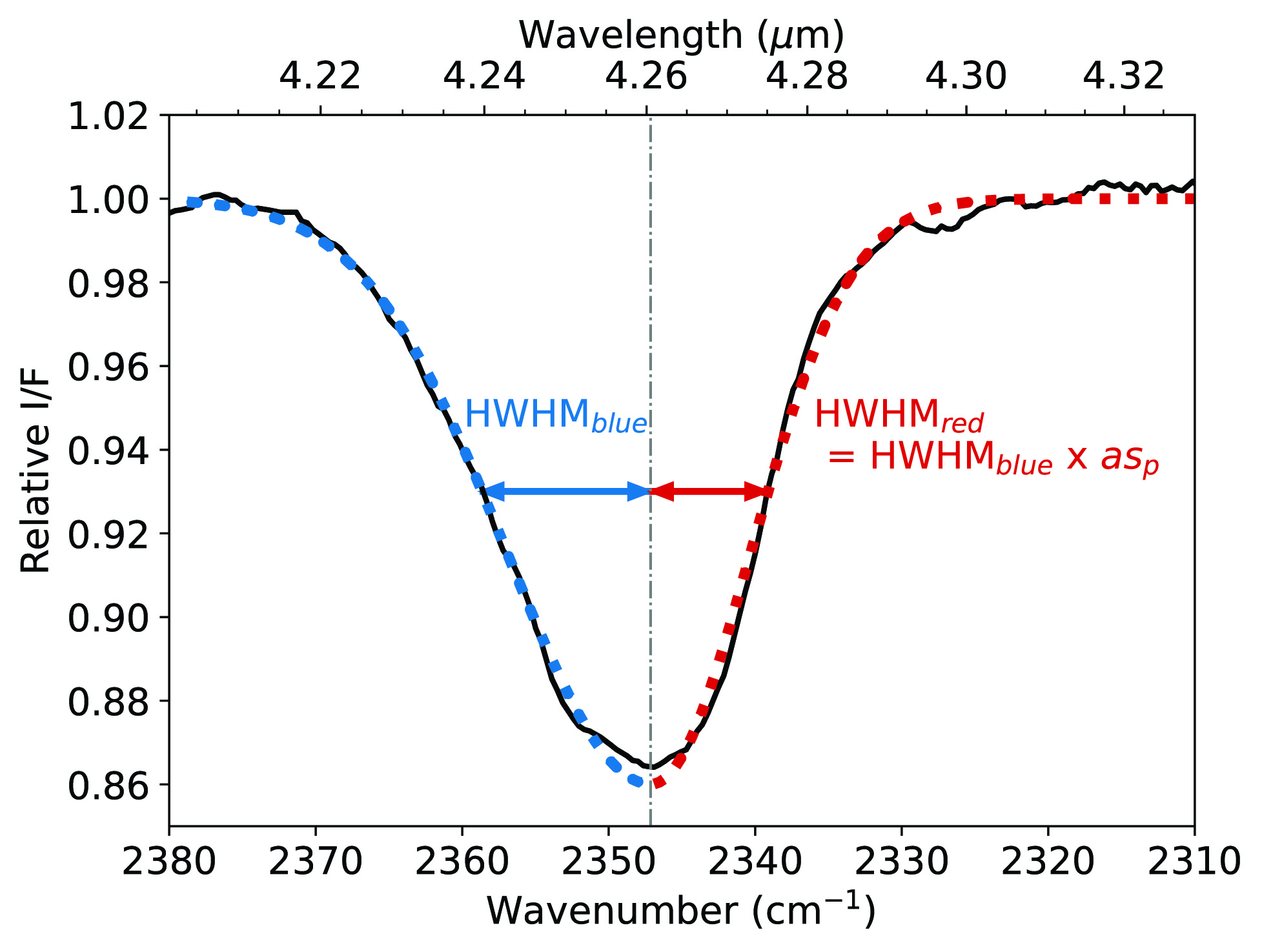}
\caption{Illustration of asymmetric gaussian fitting of the CO$_2$ band. In black, a latitude average NIRSpec spectrum (30-45\dg S/leading in Fig.~\ref{fig:sp-CO2}). In blue and red dotted lines, the two fitted half-gaussians, with their width ($HWHP_{\rm blue}$ and $HWHP_{\rm red}$, respectively) indicated.  The vertical line shows the central position of the fitted asymmetric gaussian.    \label{fig:fit-CO2}}
\end{figure}

\begin{figure}
\includegraphics[width=9.cm]{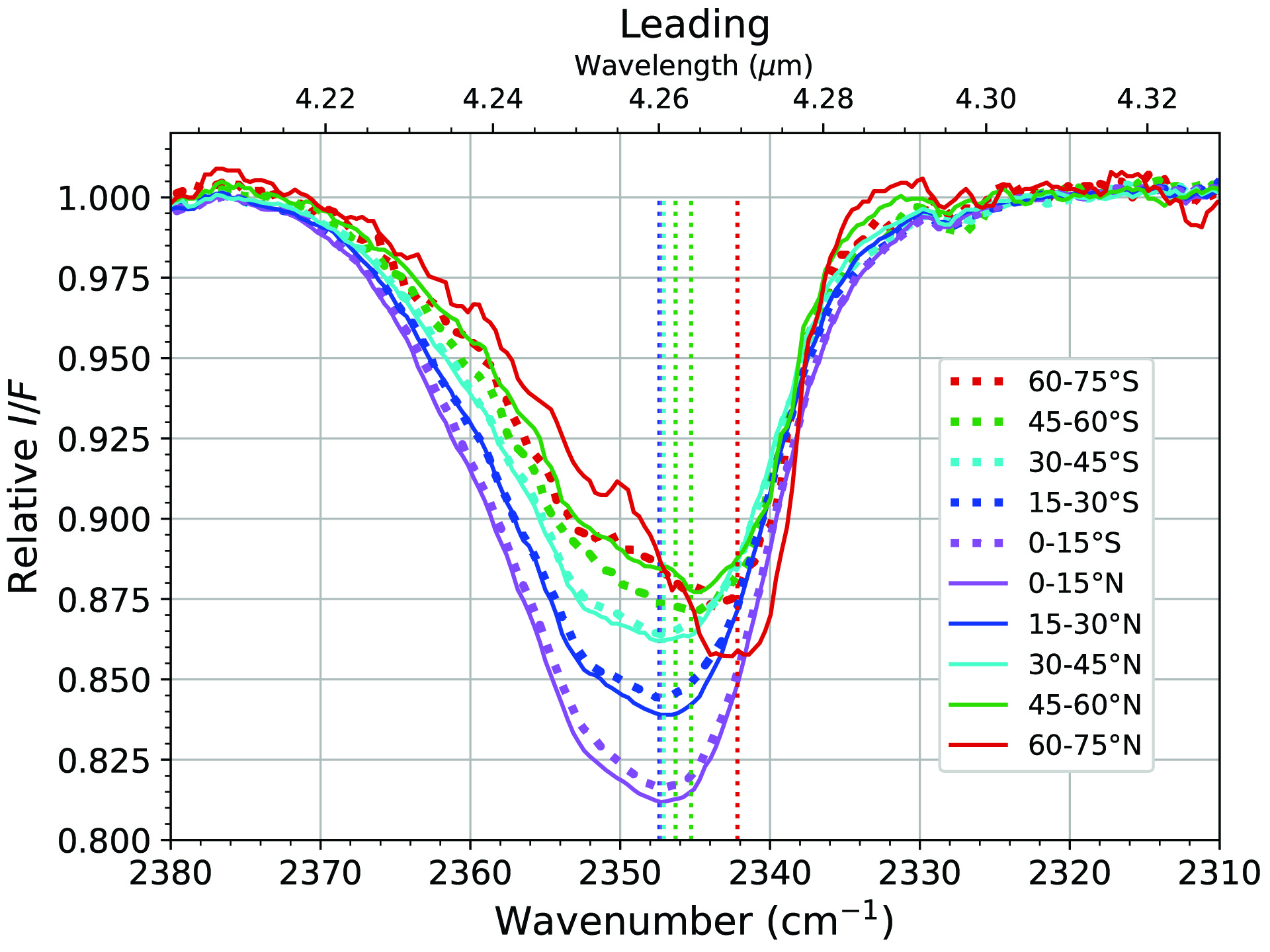}
\includegraphics[width=9.cm]{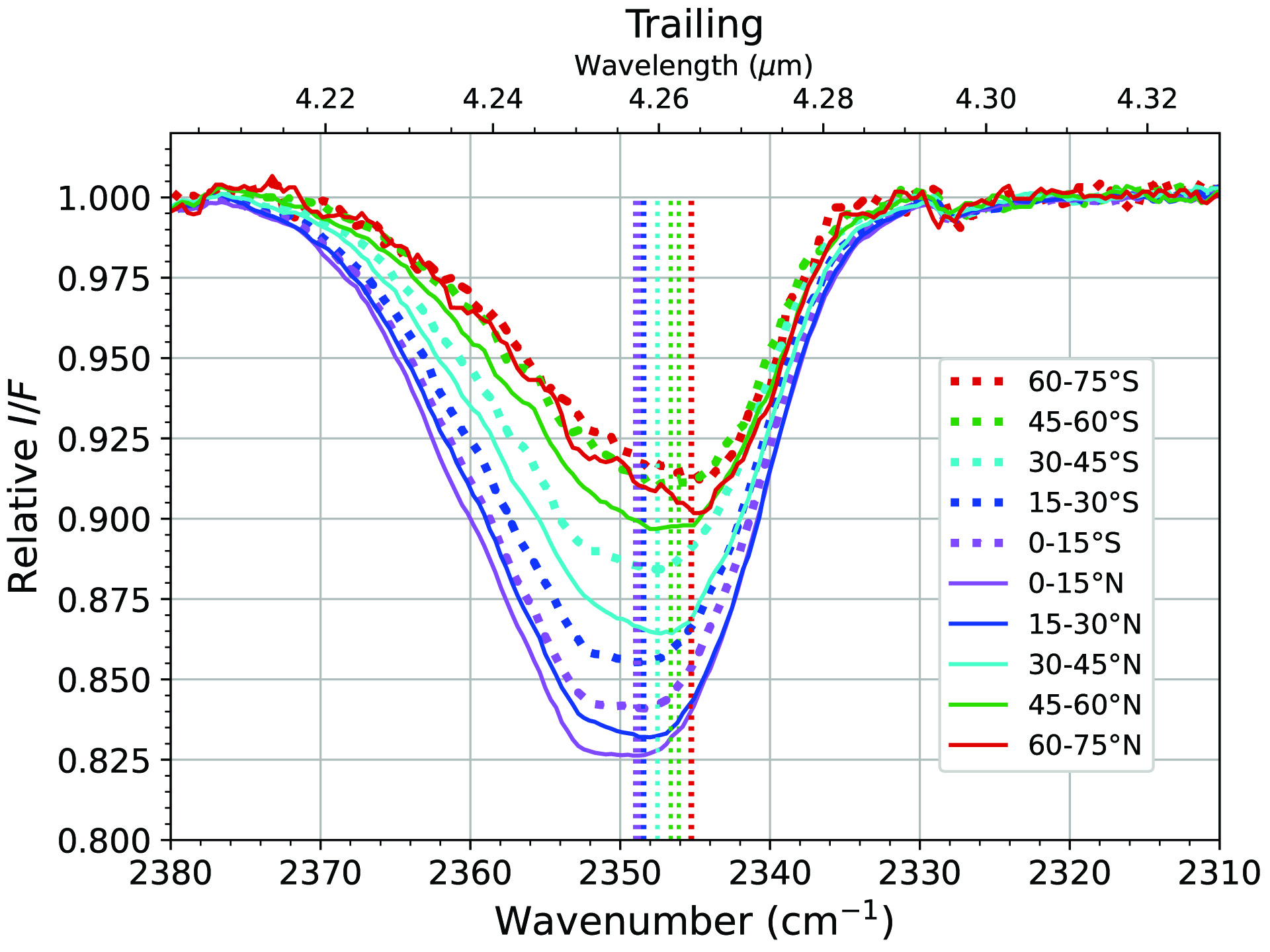}
\caption{CO$_2$ band normalized to the underlying continuum, from averages over latitude bins of $I/F$ spectra. Latitude bins are given in the right panels, with spectra from southern latitudes drawn in dashed lines and northern latitudes in solid lines. The vertical lines show the central positions (given in Table~\ref{tab:CO2_resultfit}) of the fitted asymmetric gaussians with the same colour code as for the spectra.    \label{fig:sp-CO2}}
\end{figure}

The shape of the CO$_2$ band at $\sim$ 4.26 $\mu$m is asymmetric. Therefore, to best measure the position, shape, and width of the band, the band was fitted by an asymmetric gaussian, composed of two adjacent half gaussians of same peak intensity, and peak wavelength, but of different width. This method allows us also to measure the asymmetry of the band through the asymmetry parameter $as_p$, defined as the ratio between the width of the half-gaussian in the red side to that in the blue side (see example in Fig.\ref{fig:fit-CO2}). We considered the spectral region from 4.2 to 4.5 $\mu$m (excluding the 4.335--4.40 $\mu$m range where the 4.38-$\mu$m feature is present) for each spaxel on Ganymede, and fitted the combination of a 4th-order polynomial and an asymmetric gaussian through curve fitting. The CO$_2$ band depth is measured with respect to the local continuum in \%. 

\begin{table}[h]
\caption{CO$_2$ band parameters for average spectra over bins of latitudes.}
    \label{tab:CO2_resultfit}
    \centering
    \begin{tabular}{lccccc}
    \hline
    \hline
    Latitude bin & Center & Center & Width & $as_p$ & Depth \\
    (\dg,\dg) & ($\mu$m) & (cm$^{-1}$) & (cm$^{-1}$) & & (\%) \\
    \hline
{\it Leading} &&&&&\\
(-75,-60) & 4.2639 & 2345.3 & 18.3 & 0.58 & 13.1 \\
(-60,-45) & 4.2620 & 2346.3 & 19.0 & 0.64 & 13.4 \\
(-45,-30) & 4.2605 & 2347.1 & 19.6 & 0.69 & 14.0 \\
(-30,-15) & 4.2601 & 2347.4 & 19.6 & 0.69 & 16.0\\
(-15,0) & 4.2605 & 2347.1 & 19.3 & 0.67 & 18.8\\
(0,15) & 4.2603 & 2347.3 & 19.4 & 0.68 & 19.4\\
(15,30) & 4.2606 & 2347.1 & 19.1 & 0.65 & 16.5\\
(30,45) & 4.2606 & 2347.1 & 19.1 & 0.65 & 14.3\\
(45,60) & 4.2639 & 2345.3 & 17.8 & 0.54 & 12.7\\
(60,75) & 4.2695 & 2342.2 & 15.6 & 0.35 & 14.2\\
\hline
{\it Trailing} &&&&&\\
(-75,-60) &  4.2640 & 2345.2 & 17.1 & 0.48 & 9.0\\
(-60,-45) & 4.2624 & 2346.1 & 17.7 & 0.53 & 9.2\\
(-45,-30) & 4.2598 & 2347.5 & 18.6 & 0.60 & 11.9\\
(-30,-15)& 4.2582 & 2348.4 & 19.3 & 0.67 & 15.0\\
(-15,0) & 4.2576 & 2348.7 & 19.7 & 0.70 & 16.5\\
(0,15) & 4.2571 & 2349.0 & 20.0 & 0.72 & 18.2\\
(15,30) & 4.2581 & 2348.5 & 19.6 & 0.69 & 17.5\\
(30,45) & 4.2598 & 2347.5 & 18.9 & 0.64 & 13.9\\
(45,60) & 4.2614 & 2346.6 & 18.4 & 0.59 & 10.6\\
(60,75) & 4.2638 & 2345.3 & 17.6 & 0.52 & 9.8\\
\hline
    \end{tabular}
\end{table}

Figure~\ref{fig:sp-CO2} shows averages of CO$_2$ spectra over latitude bins, normalized to the underlying continuum. The results of the asymmetric gaussian fit are given in Table~\ref{tab:CO2_resultfit}. For these averages, the CO$_2$ band depth varies from 9 to 18\% on the trailing side, and from 13 to 19\% on the leading side,  increasing  with decreasing latitudes, except for the leading side where a deep CO$_2$ band is observed poleward of 60\dg N. The band center is at smaller wavelengths in the trailing side (varying in the range 4.2571--4.2640 $\mu$m) than in the leading side  (4.2600--4.2695 $\mu$m) especially at low to mid latitudes ($<$ 35\dg S/N), and shifts towards larger wavelengths as the latitude increases. The band shape also changes with latitude, getting narrower and more asymmetric as we reach polar latitudes.

\begin{figure}[ht]
\begin{minipage}{9cm}
\includegraphics[width=4.5cm]{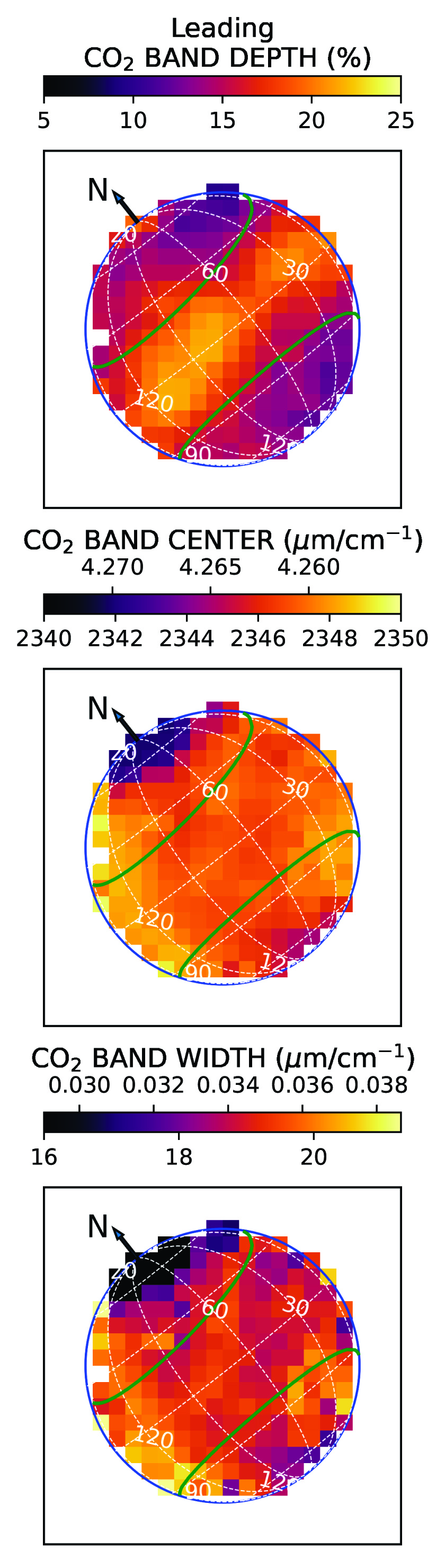}\hfill
\includegraphics[width=4.5cm]{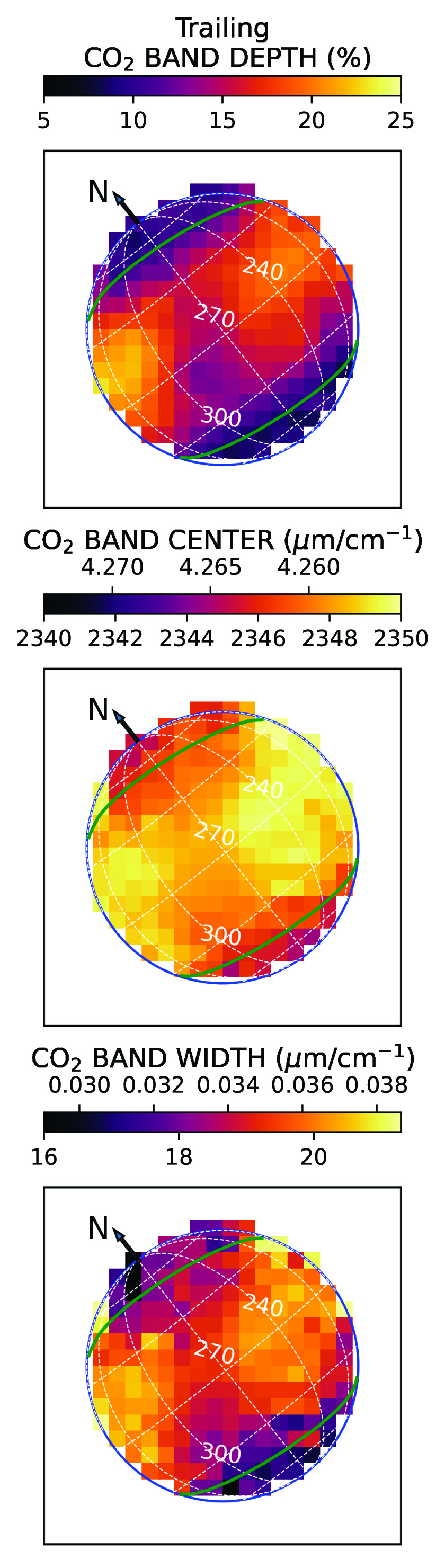}
\end{minipage}
\caption{ Maps of CO$_2$ band depth (top), band center (middle), and band width (bottom) from NIRSpec NRS2 data. Maps of the leading and trailing hemispheres are shown on the left and right, respectively.
\label{fig:maps-CO2}}
\end{figure}

\begin{figure}
\includegraphics[width=9.cm]{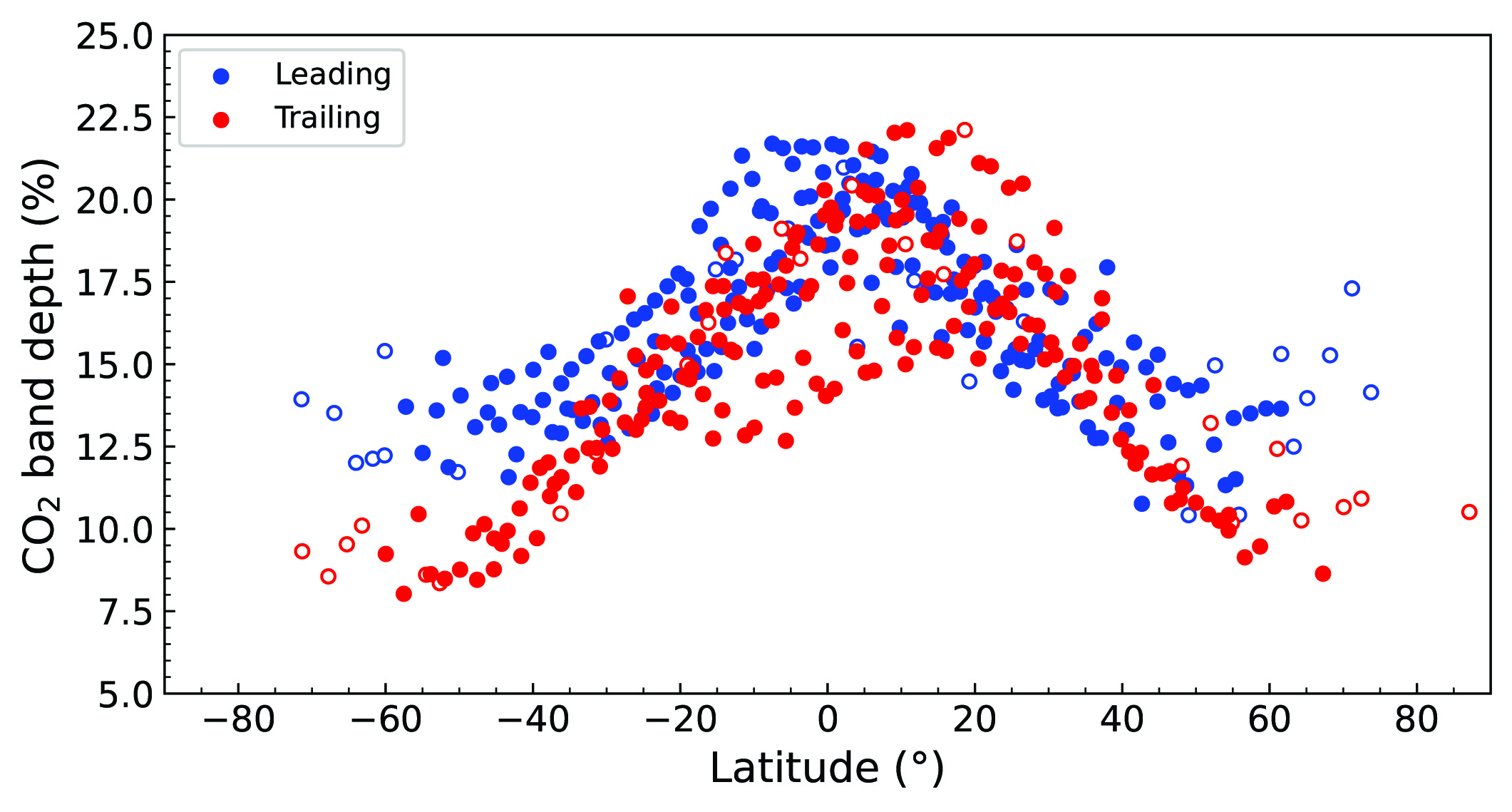}
\caption{The dependence of CO$_2$ band depth with latitude for the leading (blue dots) and trailing (red dots) hemispheres. Each data point corresponds to a spaxel in Fig.~\ref{fig:maps-CO2}. Empty symbols correspond to spaxels which are not entirely on Ganymede disk and can be less reliable.   \label{fig:depth-CO2}}
\end{figure}

Figure~\ref{fig:maps-CO2} shows maps of CO$_2$ band depth, band center position and band width across the two hemispheres. The CO$_2$ band depth ranges from 11 to 21\%, with a median at 16\% for the leading hemisphere, and from 8 to 22\% with a median at 15\% for the trailing hemisphere. These values are overall consistent with the NIMS/Galileo measurements (values of 5 to 20\%) which probed primarily the anti-Jovian hemisphere but acquired a few dataset on the leading and trailing side \citep{Hibbitts2003}. Therefore, JWST data confirm that there is no large leading/trailing side difference in the depth of the CO$_2$ band, nor is there a difference with the anti-Jovian hemisphere. However, the NIRSpec data show that regions poleward of 30\dg N/S on the trailing side present lower CO$_2$ band depths, comparatively to the leading side. This is clearly shown in the scatter plot of Fig.~\ref{fig:depth-CO2}, where the CO$_2$ band depth is plotted as a function of latitude. Poleward of 30\dg N/S, the band depth slowly varies with latitude on the leading side, in contrast to the trailing side which shows a steep drop with increasing latitude. In addition, the CO$_2$ band depth exhibits striking longitudinal variations on the trailing side, causing a large scatter at mid and equatorial latitudes in Fig.~\ref{fig:depth-CO2}. 

\begin{figure}
\includegraphics[width=9.cm]{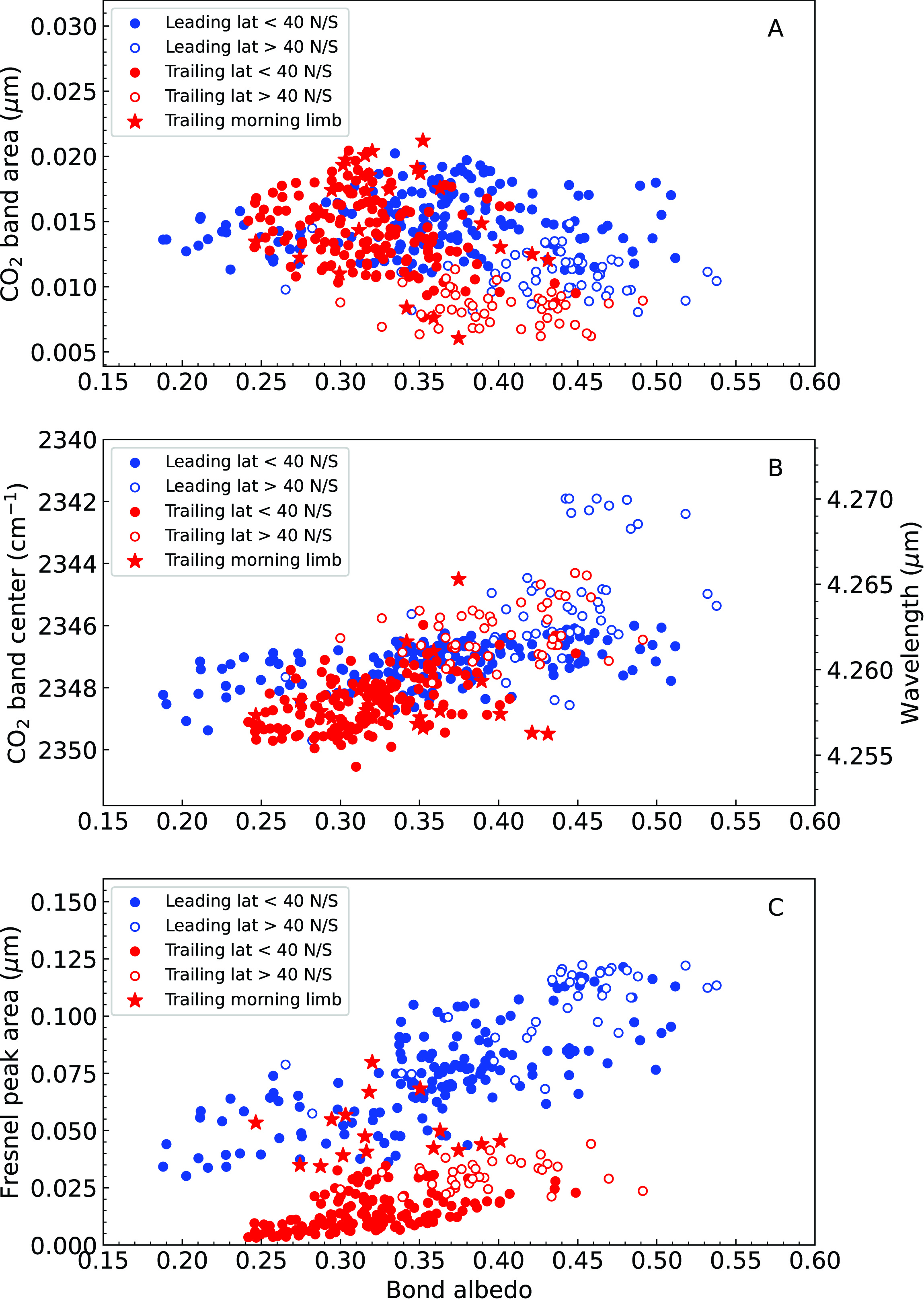}
\caption{CO$_2$ band area (A) and band center (B), and Fresnel-peak area (i.e. $EqW$, C) as a function of Bond albedo. Data for the Leading and Trailing hemispheres are shown with blue and red symbols, respectively. Filled and open dots are for latitudes lower and higher than 40$^{\circ}$ N/S, respectively. Data in the morning limb of the Trailing hemisphere (namely longitudes > 320\dg) are shown with the star symbol. \label{fig:CO2-albedo}}
\end{figure}

In the anti-Jovian hemisphere from NIMS/Galileo data, the largest band depths of CO$_2$ were found to be  generally associated with the less icy regions \citep{Hibbitts2003-rm}.  This trend is observed for the trailing hemisphere (see CO$_2$ maps in Fig.~\ref{fig:maps-CO2} compared to the Bond albedo context map in Fig.~\ref{fig:albedo}). The scatter plot in Fig.~\ref{fig:CO2-albedo}A shows a moderate anticorrelation between CO$_2$ band area (i.e., equivalent width) and Bond albedo for the trailing hemisphere (Spearman rank correlation $r$= --0.56 with 8.2 $\sigma$ significance), whereas this trend is less obvious for the leading hemisphere ($r$= --0.26 with 3.9 $\sigma$ significance). 
However, when considering only equatorial regions ($<$ 40\dg N/S) or within the open-closed-field line-boundary, the relationship between the two quantities is weak to absent for both hemispheres (2--4 $\sigma$ significance). A similar result is obtained when looking how the CO$_2$ band area correlates with surface ice abundance (Fig.~\ref{fig:CO2-H2O}A), using the Fresnel peak area as a proxy (see  Sect.~\ref{sec:fresnel}). In summary, latitudinal and regional trends are observed for the CO$_2$ band area, but the link with ice abundance is not observed.  

\begin{figure*}
\sidecaption
\begin{minipage}{12cm}
\includegraphics[width=12.cm]{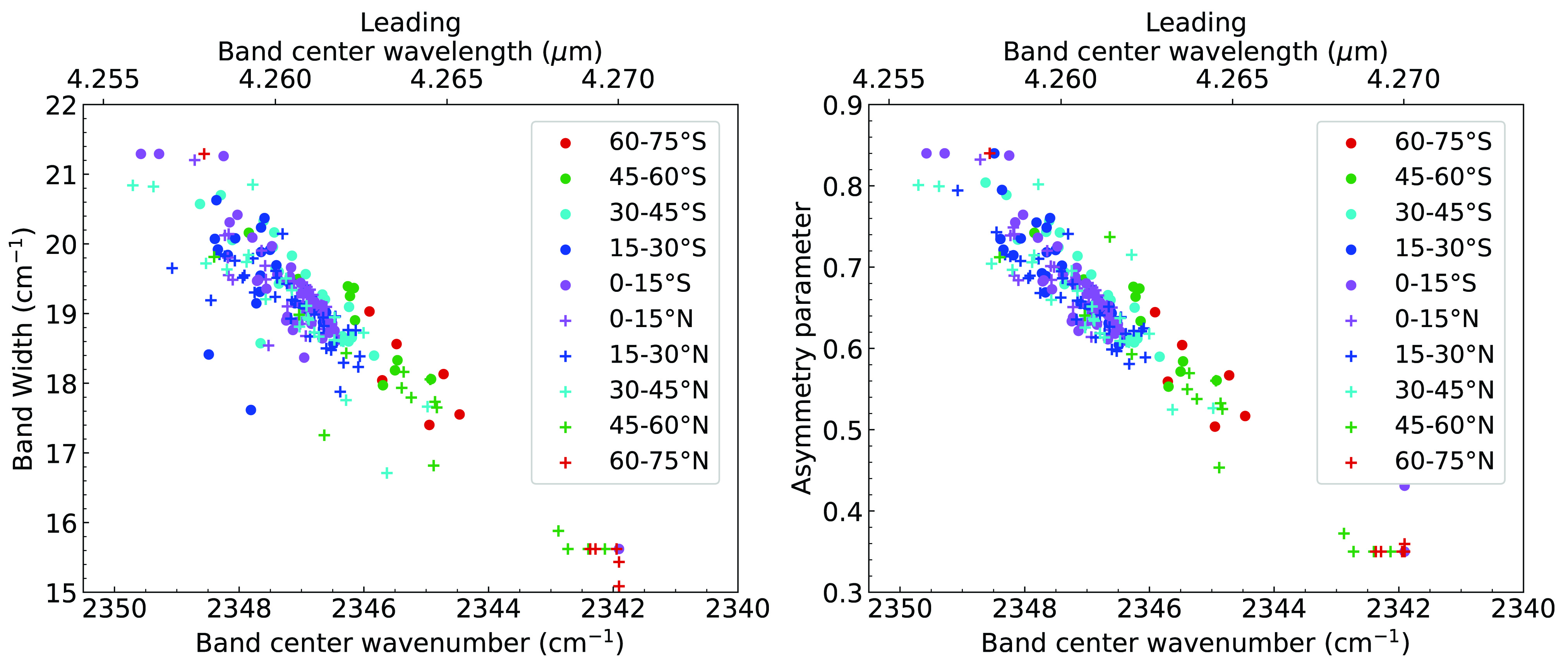}\hfill
\includegraphics[width=12.cm]{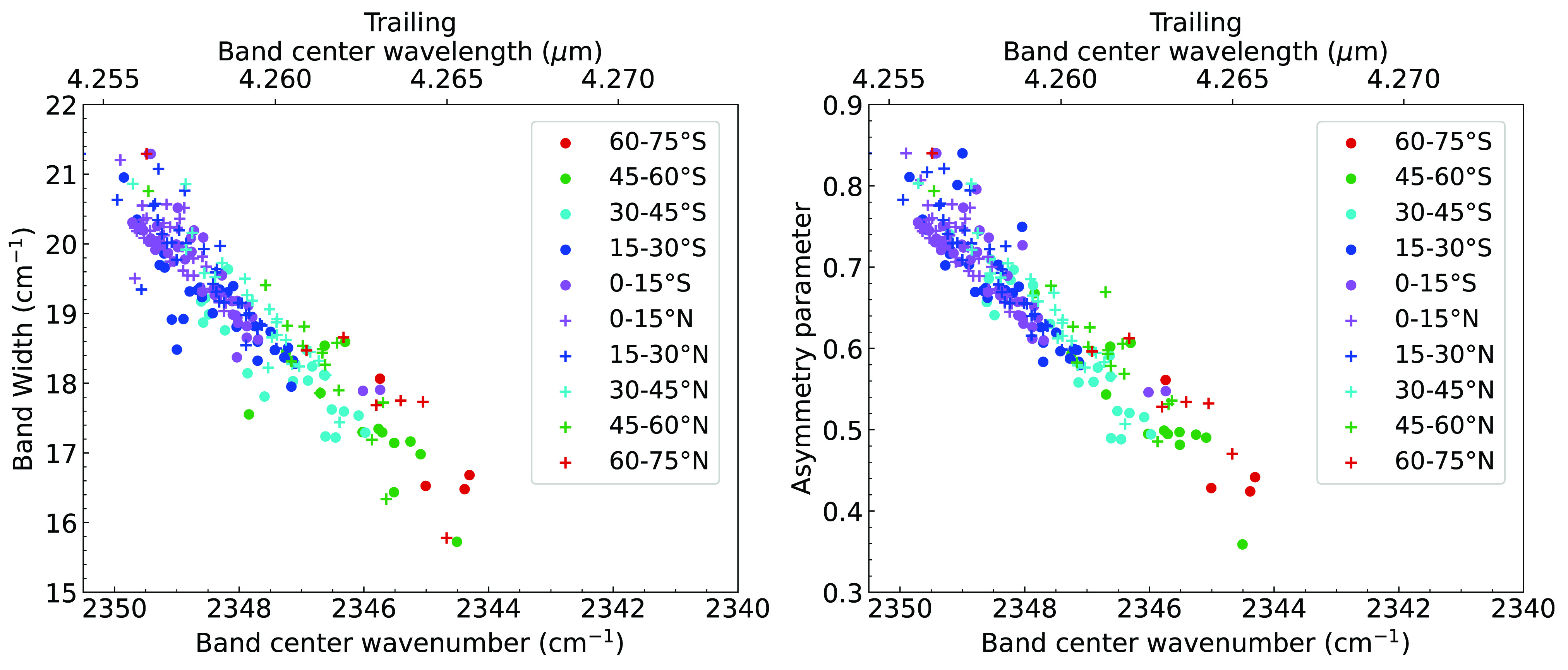}
\end{minipage}
\caption{Correlations between CO$_2$ band width and band center (A, C), and between asymmetry parameter $as_p$ and band center (B, D). Each data point corresponds to one NIRSpec spaxel on the Ganymede disk. The different symbols and colours indicate the latitude range as given in the inset. Upper plots (A, B): leading hemisphere; bottom plots (C, D): trailing hemisphere. \label{fig:correlations-CO2}}
\end{figure*}

The trends between band depth, width and position observed in the latitudinal averages are seen at the pixel scale in the maps. This is conspicuous for the trailing hemisphere which presents large CO$_2$ band depth variations at latitudes $<$ 30\dg~that are associated with variations in band center, band width (Fig.~\ref{fig:maps-CO2}) and band asymmetry parameter $as_p$ (not plotted in Fig.~\ref{fig:maps-CO2}). The anticorrelations between $as_p$ and the band center position (when expressed in wavelength), and between the band width and the band center position  are shown as scatter plots in Fig.~\ref{fig:correlations-CO2}.  It is striking to see how the data points at the poles (especially for the leading north pole) are detached from the rest of the data, indicative of the peculiar physical state of CO$_2$ in this region. At the poles, the band asymmetry parameter $as_p$ reaches values below 0.4, and the band width is up to 40\% lower than in equatorial regions.

Both hemispheres show a striking trend between CO$_2$ band center position and Bond albedo ($r$ = 0.62 and 0.66  with a significance of 9.1 $\sigma$ and 9.3 $\sigma$, for leading and trailing, respectively), with the CO$_2$ band center shifting towards longer wavelengths as the surface brightness increases (Fig.~\ref{fig:CO2-albedo}B). Unlike the CO$_2$ band area, the trend is also present when considering only the equatorial regions ($r$ = 0.47 and 0.43  with a significance of 6.1 $\sigma$ and 5.5  $\sigma$, for leading and trailing, respectively). Since the surface brightness is correlated with water ice abundance (Fig.~\ref{fig:CO2-albedo}C) a similar correlation is expected between the CO$_2$ band center and the ice abundance (i.e. Fresnel peak area). This is indeed observed (at 6-7 $\sigma$) for the leading hemisphere even when considering only equatorial latitudes  (Fig.~\ref{fig:CO2-H2O}B). However, the equatorial latitudes ($<$ 40\dg N/S) of the trailing hemisphere do not show a significant trend between CO$_2$ band center and ice abundance ($r$= 0.20, 2.4 $\sigma$), although the ice-rich polar regions do present peculiar CO$_2$ spectra, as discussed in the previous paragraph. So, in summary, the CO$_2$ band center and shape show a strong relationship with surface brightness for both hemispheres. Their correlation with the ice abundance is only seen on the leading hemisphere, and at the poles of the trailing hemisphere.      

\subsubsection{The 4.38-$\mu$m band} 
\label{sec:13CO2}
A faint signature is observed at the wavelength of the 4.38-$\mu$m $^{13}$CO$_2$ band.
We characterized the depth (in \%) and position of this band fitting the 4.29--4.50 $\mu$m region in the $I/F$ spectra by the combination of a Gaussian and a third-order polynomial.  The mean central wavelength of the band over the disk is 4.3779 $\mu$m (leading) and 4.3766 $\mu$m (trailing)  with a standard deviation of 0.0025 $\mu$m, so in average at 4.377 $\mu$m. The depth does not show significant spaxel-to-spaxel variations over the two hemispheres, and has a mean value of 1.67$\pm$0.39\% and 1.53$\pm$0.35\% on the leading and trailing sides, respectively. In addition to the uncertain reality of features at 1--2\% level in NIRSPec spectra discussed in Sect.~\ref{sec:3.1}, this faint feature cannot be firmly attributed to $^{13}$CO$_2$ as spatial variations following those observed for the CO$_2$ band (Sect.~\ref{sec:CO2-an}) are not observed. In addition, this band is not present when using the data reduction procedure of \citet{Trumbo2023}, where Ganymede spectra are divided (for flux calibration and solar-line removal) by those acquired on a solar-type G0V star also observed with JWST/NIRSpec and calibrated using the same JWST pipeline as for Ganymede. It is also worth noting that this feature at 4.380 was also only
visible in one Juno/JIRAM spectrum out of seven in \citet{Mura2020}, so it was not firmly ascribed to $^{13}$CO$_2 $.

\section{MIRI data analysis and modelling}
\label{sec:MIRI-an}
\subsection{Qualitative analysis of spectra}
\label{qualitative}

\begin{figure}
   \includegraphics[angle=90,width=9cm]{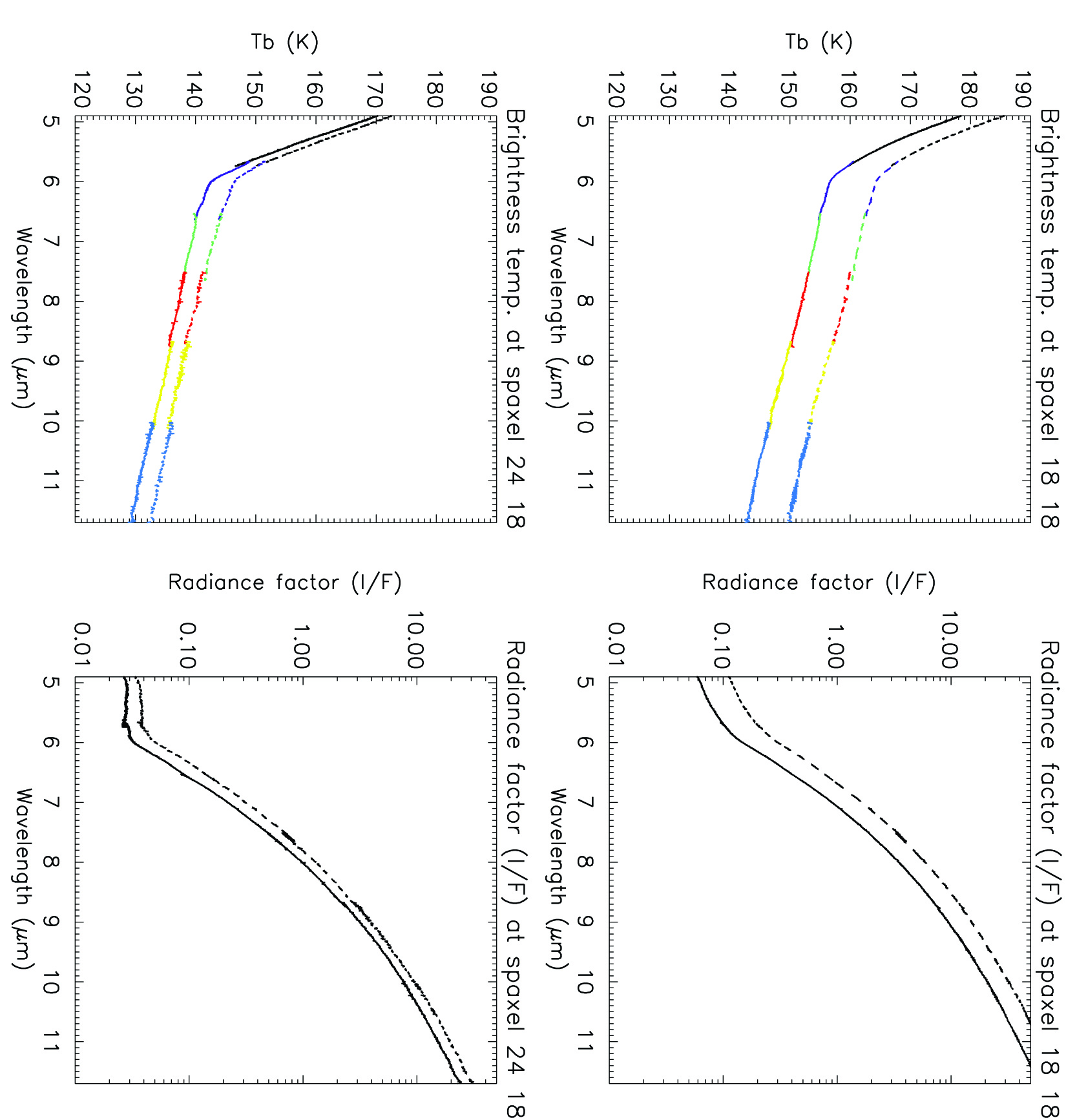}
   \caption{Sample spectra at 4.9-11.7 $\mu$m from Obs. 18 (leading hemisphere, solid line) and 27 (trailing hemisphere, dashed line), at reconstructed disk center (spaxel 18, 18, top) and at offsets RA~=~0.65", DEC~=~0" from this position (spaxel 24, 18, bottom). Spectra are calibrated either in brightness temperature (T$_B$, left panels) with different colors illustrating different bands from 1A to 2C, or radiance factor ($I/F$, right panels). They are not corrected for any account of "filling factor" associated with the fact that for positions close to limb, the PSF partly samples empty space. }
   \label{fig:globalspectra}
    \end{figure}

 The 4.9-11.7 $\mu$m range corresponds to the overlap of the solar reflected and thermal components. We show sample spectra at two spatial positions and for both visits in Fig.~\ref{fig:globalspectra}. Spectra are calibrated either in brightness temperature (left panels) or radiance factor (I/F, right panels). Radiance
factors are typically 5-10 \% in band 1A (and larger on trailing side than leading, unlike the behaviour in the visible), and the fact that they sharply increase above unity beyond $\sim$7 $\mu$m demonstrates the dominance of the thermal component there. Beyond this expected behaviour, a remarkable feature is the monotonic (and quasi-linear) decrease of the brightness temperatures  with increasing wavelength ($\lambda$), by more than 10 K from 7 $\mu$m to 11 $\mu$m at disk center. Qualitatively, this can result from at least three effects, which may well be at play simultaneously, and stem from the non-linear character of the Planck function \footnote{Note that combining SOFIA data over 8-37 $\mu$m with radio data, \citet{depater2021} find that such a trend persists up to the mm-cm range, where the very low T$_B$ ($\sim$ 60-70 K) results from a dramatic collapse of the spectral emissivity with wavelength.}.

\begin{enumerate}
\item A spectrally constant but less than unity spectral emissivity.
\item A decrease of the spectral emissivity with increasing wavelength.
\item Planck-weighted mixing of a variety of surface temperatures within the PSF. This effect may be amplified by the fact that the PSF width increases with $\lambda$, causing long wavelengths to ``see" more near-limb (i.e. colder) terrains than short-wavelengths for points within the disk.
\end{enumerate}
   
\begin{figure}
   \includegraphics[angle=90,width=9cm]{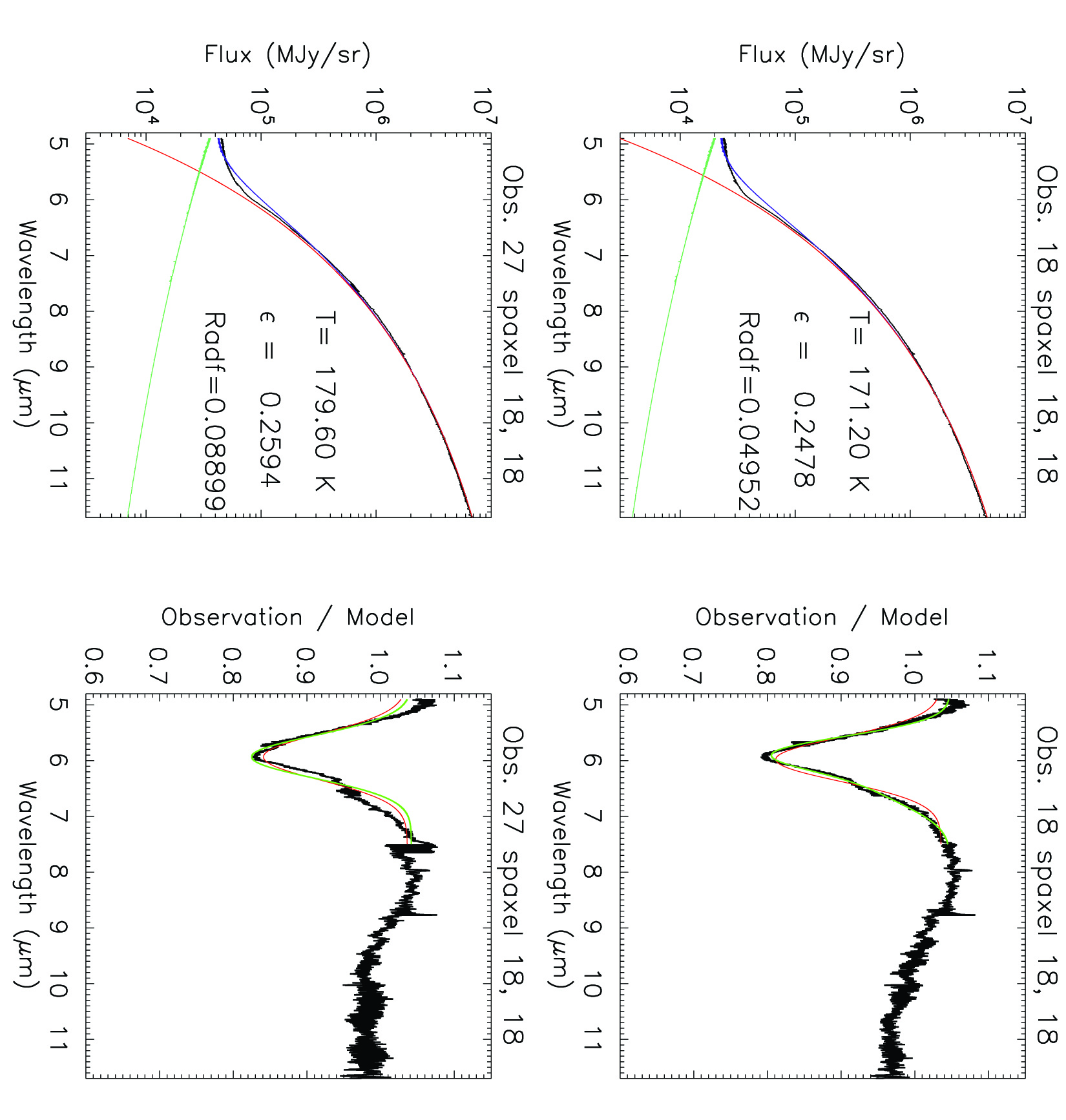}
   \caption{Left panels: Fit of disk-center spectra (spaxel 18, 18) in Obs. 18 (leading, top) and 27 (trailing, bottom) in radiance units with the sum of thermal and solar reflected components. Black: data. Red: thermal component. Green: solar reflected component. Blue: total model. Right panels: observation/model ratios. }
   \label{fig:showband}
    \end{figure}

The other feature obvious in Fig.~\ref{fig:globalspectra} is the presence of an absorption band near 5.9 $\mu$m, which shows up as an inflexion in the brightness temperatures or radiance factors there. A first assessment of the band parameters is shown in Fig.~\ref{fig:showband}. For this, for the two disk center spectra, we fit the observed radiances by the sum of the two components, where the thermal one is modelled phenomenologically by a single surface temperature T$_{surf}$ and spectral emissivity $\varepsilon$ (case 1 above), and the solar component is characterized by a single free parameter, the $I/F$ radiance factor ($Radf$). T$_{surf}$, $\varepsilon$, and $Radf$ are found from Levenberg-Marquardt minimization.
This approach, whose physical meaning is admittedly limited -- given the unrealistically large (resp. low) values of T$_{surf}$ (resp. $\varepsilon$) -- confirms a large band depth and width ($\sim$ 20 \% and $\sim$1 $\mu$m in these examples) and permits the assessment of the relative contribution of the thermal and solar reflected components, which show crossover near 5.5 $\mu$m. 
 
       \begin{figure}
   \includegraphics[angle=0,width=9cm]{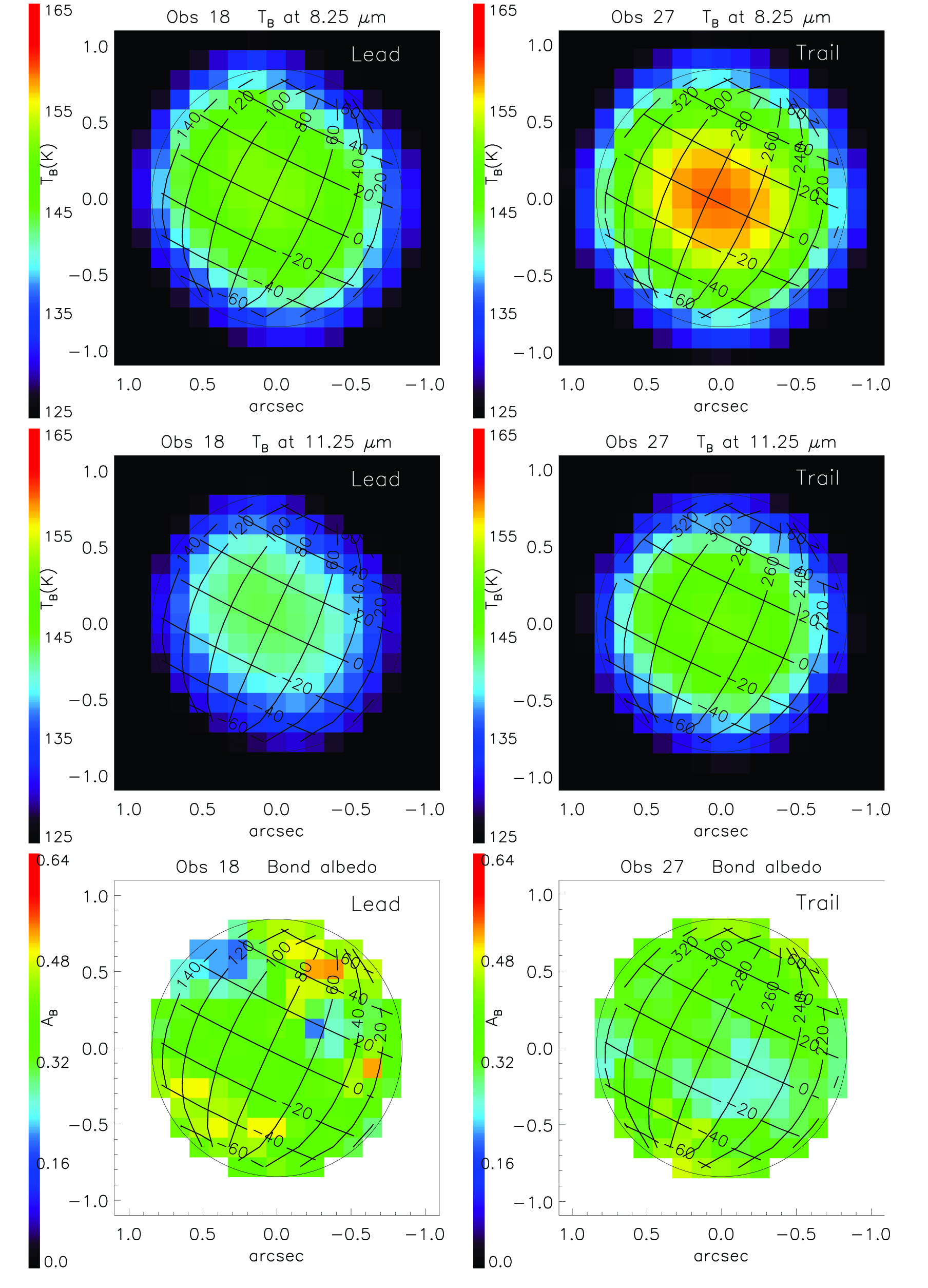}
   \caption{Maps of the 8.25-$\mu$m (top) and 11.25-$\mu$m (middle)  brightness temperature, and spaxel-averaged Bond albedo maps (bottom) from \citet{2021PSJ.....2....5D} (see text). Left: leading side (Obs. 18). Right: trailing side (Obs. 27). }
   \label{fig:thermalmaps}
    \end{figure}
    
\subsection{Thermal maps and models}
\label{sec:thermal}
We now analyse the thermal emission continuum, focussing on the long-wavelength part ($>$ 8 $\mu$m) unaffected by the 5.9-$\mu$m band. To describe the T$_B$ decrease 
with $\lambda$ we consider two representative intervals, 8.0-8.5 $\mu$m and 11.0-11.5 $\mu$m and averaged the T$_B$ in those, yielding the so-called T$_{B, 8.25}$ and T$_{B, 11.25}$
brightness temperatures. Fig.~\ref{fig:thermalmaps} shows maps of T$_{B, 8.25}$ and T$_{B, 11.25}$ for the two datasets, along with ``projected maps" of the Bond albedos ($A_B$).
To obtain the latter we used the 1\dg\ resolution Bond albedo map from \citet{2021PSJ.....2....5D}; for each observed spaxel, we averaged (in a (1-$A_B$)$^{0.25}$ sense) the $A_B$ values of the map elements projecting into that spaxel. As indicated above, the four dithers were averaged at this step and in future analyses. Rms differences in T$_{B, 8.25}$ and T$_{B, 11.25}$
between data from individual dithers and their average are typically 0.4 -- 0.7 K, i.e. comparable and somewhat smaller than the best fits we achieve (see below).

Brightness temperatures are consistently higher on the trailing vs leading side, as may be expected from the somewhat darker albedos there. However, correlations between 
T$_{B}$'s and Bond's albedos, while present, are not very pronounced (Spearman correlation coefficient $r$~=~--0.50 and 5.7 $\sigma$ significance on the trailing side, $r$~=~--0.23 and 2.7 $\sigma$ on the leading), being presumably erased by PSF smoothing and dwarfed by other effects (latitude and local time variability).

Focussing on the T$_{B, 8.25}$ maps, we performed standard thermophysical modelling, separately for the two observations. The main model parameters are (i) the thermal inertia ($\Gamma$, assumed constant over one side, parameterized by the thermophysical parameter $\Theta$ \citep{1989Icar...78..337S}\footnote{$\Theta$ is defined by $\Theta$ = $\frac{\Gamma \sqrt{\omega}}{\varepsilon\sigma T_{SS}^3}$, where $\omega$ is Ganymede's rotation rate, and $T_{SS}$ is the subsolar equilibrium temperature }; for Ganymede at 4.96 au, $\Theta$ = 1 typically corresponds to $\Gamma$ = 75 SI units (J m$^{-2}$ K$^{-1}$ s$^{-0.5}$) (ii) the Bond albedo; we either let it be a free parameter (constant for a given side) or use the projected Bond albedo maps from Fig.~\ref{fig:thermalmaps}, as such or with an adjustable scaling factor. A Bond emissivity (and a spectrally constant emissivity) $\varepsilon$ = 0.90 as adopted, i.e. we purposedly do not include spectral emissivity effects. As detailed below, we also introduced surface roughness in the model using a description of slopes. For each model, local fluxes were calculated on the 0.13'' spaxel grid and then convolved by the beam, for which we used FWHM (arcsec) = 0.0328$\times$$\lambda$($\mu$m) and a simplified description of the secondary Airy pattern, and finally converted in a 8.25 $\mu$m brightness temperature. Model parameters were determined by Levenberg-Marquardt fit.

       \begin{figure*}
%\vspace*{-5cm}
   \includegraphics[angle=0,width=18cm]{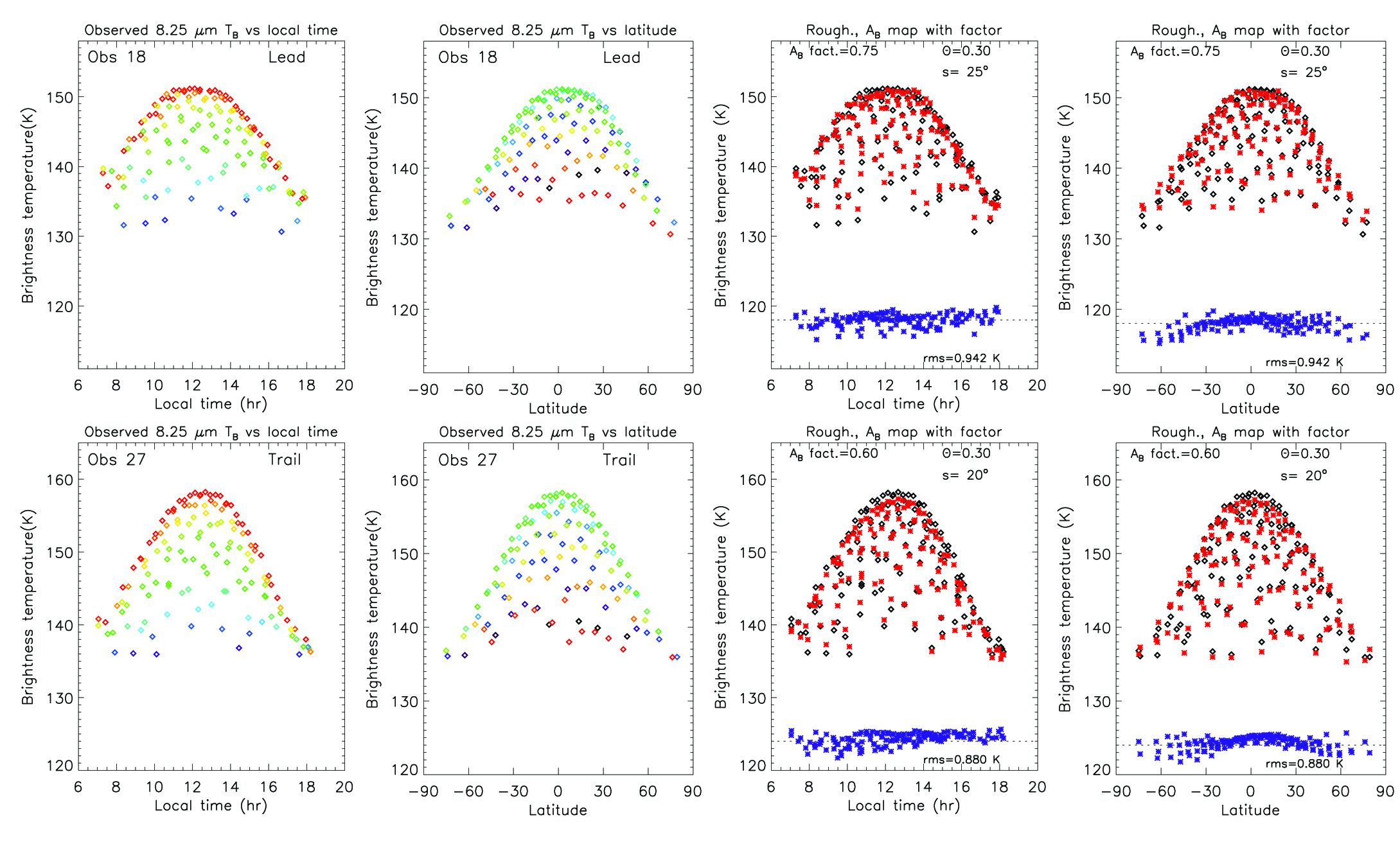}
%\vspace*{-5.25cm}
   \caption{8.25-$\mu$m brightness temperature (T$_B$) for Obs. 18 (leading side, first row) and Obs. 27 (trailing side, second row), plotted as a function of local time (first column) or latitude (second column). The data are color-coded according to latitude (i.e. to equal values of the latitude difference from the sub-observer point) in the T$_B$ vs local time plots, and according to local time in the T$_B$ vs latitude plots. In the third and fourth columns, the data are compared to one of the best models, that includes both a rescaling of the Bond albedo map and surface roughess (see text). In these panels, data are in black, models in red, and the data minus model difference in blue, with the 0 value indicated by the dashed line. Parameter values and rms of fit are indicated in each panel. }
   \label{fig:thermalfits_only_best_fits}
    \end{figure*}

Data (in the form of T$_{B,8.25}$ vs local time and T$_B$ vs latitude) are displayed in the first column of Fig.~\ref{fig:thermalfits} and in the first and second columns of Fig.\ref{fig:thermalfits_only_best_fits}. They clearly show that peak T$_{B,8.25}$  occur near the equator and near 12:30 pm local time, suggestive of low thermal inertia effects. Results for the best-fit models without roughness are shown in columns 2-4 of 
Fig.~\ref{fig:thermalfits}, for the three descriptions of the Bond albedo respectively. A common feature of these models is that they overestimate the diurnal variation of 
T$_{B,8.25}$, overpredicting the temperatures near noon and underpredicting the dawn and dusk temperatures, leading to relatively large model-observation residuals (2 - 3.5 K rms).
Temperatures at high latitudes ($\sim$50\dg ) are also underpredicted. Increasing the thermal inertia could improve the diurnal contrast, but would unacceptably shift the maximum temperatures towards afternoon hours and further lower the model high-latitude temperatures. We also note that models using the fixed Bond albedo map from \citet{2021PSJ.....2....5D} 
result in a zero thermal inertia and provide a relatively bad fit. They can be improved by scaling the $A_B$ by factors of 0.60 (trailing) - 0.80 (leading) but the problem of the too large diurnal contrast remains.

Including surface roughness is a possible solution to improve fits. The effect of a distribution of local surface slopes at any scale within each spaxel is to decrease temperatures and emission near disk center / subsolar point and increase them near the limb (due to facets preferentially oriented towards the Sun). We followed the description of \citet{1984Icar...59...41H} for macroscopic roughness, in which facets (with dimensions much larger than particle size) are randomly tilted from the local smooth surface normal, with a uniform distribution in azimuth and a gaussian distribution in zenith angles. The distribution of slopes is characterized by a mean slope angle $s$ \citep[noted $\bar{\theta}$ and defined by Eqs. 5 and 44 of][]{1984Icar...59...41H}. At a given position on Ganymede, a facet with a given orientation owns an ``effective" latitude and longitude, which determines its diurnal insolation pattern and local time. In practice, we generated 10000 facets with random orientation on a sphere and calculated their temperature according to their effective coordinates. Then at each spaxel position, Planck emissions from these 10000 facets were combined, with a weighting defined by the angle between the facet normal and the local smooth surface normal \citep[i.e. Eq. 44 from][]{1984Icar...59...41H}. This approach thus adds one free parameter, $s$. As shown in col. 5-7 of Fig.~\ref{fig:thermalfits} and in col. 3-4 of Fig.~\ref{fig:thermalfits_only_best_fits}, substantially improved fits are achieved, with rms residuals below 1 K in the best cases, and subdued systematic trends in the residuals. Mean slopes $s$= 15-20\dg\ on the trailing side and $s$= 20-25\dg\ on the leading are found, somewhat lower but reasonably consistent with the surface roughness estimates from optical photometry using the same Hapke definition of slopes \citep[$s$= 35\dg\ and 28\dg, respectively,][]{1997Icar..128...49D}. Although in the model, slopes may occur at any scale smaller than the spaxel (450 km across), the relatively large values of $s$ probably result from roughness at the scale of tens of meters at most, as for example stereo and photoclinometric analysis of Galileo and Voyager images of Ganymede indicate mean slopes of 3.5--8\dg\ only at 630 m scale \citep{2013P&SS...77...40B}. Presumably, the roughness measured here pertains to scales of 0.1 mm-10 cm, as indicated by roughness statistics of the lunar surface \citep{1999Icar..141..107H}.

Once again, we find that the fixed Bond albedo map, being "too bright", gives worse results and drives zero thermal inertia, and recovering good fits requires multiplying the $A_B$ map by factors of 0.60-0.75. The best fits in this case are shown in Fig.\ref{fig:thermalfits_only_best_fits}. We do not have a simple interpretation for this, since the \citet{2021PSJ.....2....5D} map is based on well-documented maps of normal albedos and phase integrals from Voyager measurements. A possible explanation is that our thermal models describe roughness purely as a slope effect, and do not account for other more complex effects associated with topography, such as shadowing and self-heating due to scattering and re-absorption of solar and thermal radiation within craters \citep[see e.g][]{2011MNRAS.415.2042R}. Applying such more advanced thermophysical models is left to future investigations. 

For now, restricting ourselves to the two best classes of models (col. 5 and 7 in 
Fig.~\ref{fig:thermalfits}) and col. 3 in Fig.~\ref{fig:thermalfits_only_best_fits} , the suite of solutions indicates thermal inertia parameters $\Theta$  = 0.3-0.5, i.e. thermal inertias in the range $\Gamma$ = 20--40 SI, with no obvious difference between leading and trailing side. These values are slightly below the ``canonical" value $\Gamma$ = 70 derived from Voyager/IRIS spectra over 7-50 $\mu$m \citep{1987PhDT........81S,1989Icar...78..337S}, but further analyses including Galileo/PPR data have indicated that two-component models provided better ﬁts, with end member thermal inertias values between 16 (associated with dust) and 1000 (ice), with dark and grooved terrains covering the $\Gamma$ = 70-150 range (\citet{2004jpsm.book..363P}; see also \citet{2021PSJ.....2....5D}). Given that PPR covered the 17-110 $\mu$m range and that radiation at the shorter wavelengths is progressively dominated by warmer (i.e. lower thermal inertia, for dayside measurements) areas, it is probably not surprising that the $\Gamma$ values we derive at 8.25 $\mu$m are in the lower range of the PPR results. Similarly, \citet{2021PSJ.....2....5D} determined $\Gamma$ = 400-800 from ALMA data at 0.9-3 mm; such wavelengths sample all temperatures equally and (unlike the mid-IR emission which originates from the topmost surface layers) probe the subsurface, where material compaction may account for the higher thermal inertias.  Low thermal inertias at the surface are probably indicative of material porosity.

\begin{figure}
   \includegraphics[angle=0,width=9cm]{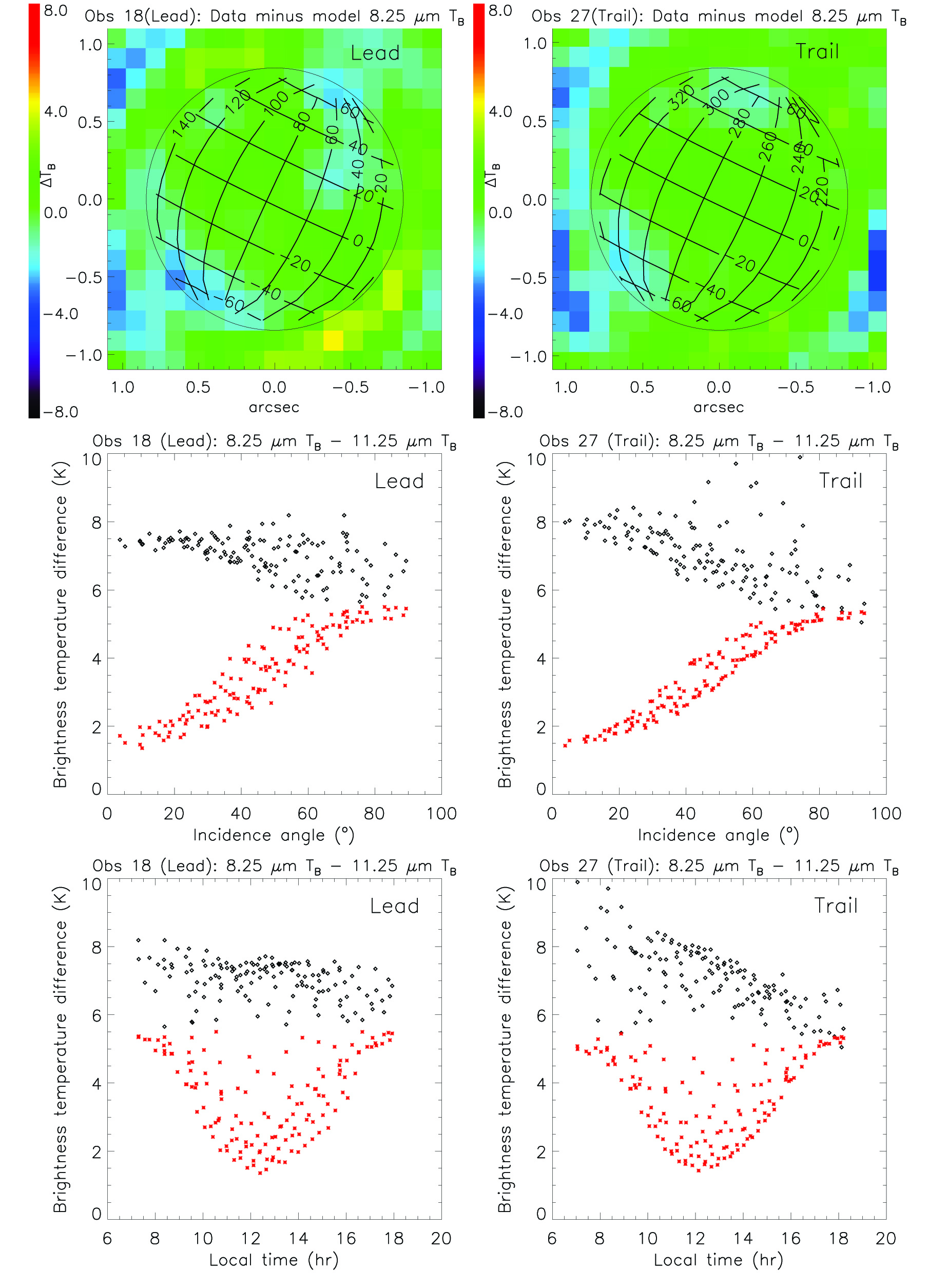}
   \caption{Difference plots: Top: Difference between observed and modelled T$_{B,8.25}$, for the rough, Bond albedo-rescaled, model (last column of Fig.~\ref{fig:thermalfits}).
   Middle: T$_{B,8.25}$ - T$_{B,11.25}$ brightness temperature difference as a function of incidence angle.
    Bottom: T$_{B,8.25}$ - T$_{B,11.25}$ brightness temperature difference as a function of local time. Black: data. Red: model. Left: leading side (Obs. 18). Right: trailing side (Obs. 27). }
   \label{fig:mapsdiffs}
    \end{figure}

We searched for thermal anomalies, by examining the difference between observed and modelled T$_{B,8.25}$ for one of the best models. The top two panels of 
Fig.~\ref{fig:mapsdiffs} do not indicate clear outliers in the difference, although there is a suggestion that the model overestimates the observed T$_{B, 8.25}$ in the high northern and southern latitudes, by 2-3 K. If real, these residuals may be due to locally higher than assumed albedos, larger thermal inertia, or higher Bond emissivity, all of which leading to lower T$_{B}$.

However, the main drawback of the models tailored to the 8.25-$\mu$m emission is that as such they do not match the 11.25 $\mu$m fluxes. This is visualized in  
the bottom two rows of Fig.~\ref{fig:mapsdiffs} where the observed and modelled T$_{B,8.25}$ - T$_{B,11.25}$ are compared for the rough model using the rescaled Bond albedos. As anticipated qualitatively in Sec.~\ref{qualitative}, the mixing of temperatures does produce a decrease of T$_B$ with increasing $\lambda$, but our calculated temperatures differences are too small (2-5 K vs 6-8 K observed) and strikingly do not show the observed dependence with incidence angles. In the model, maximum T$_{B,8.25}$ - T$_{B,11.25}$ occur near the terminators (where temperature gradients are larger), but this behaviour (also obtained for the non-rough models) is not seen in the data, which rather show a mild decrease of T$_{B,8.25}$ - T$_{B,11.25}$ from the subsolar region to the terminators. We note that based on Voyager-IRIS data, \citet{1987PhDT........81S} also found that ``the Ganymede spectrum slopes\footnote{defined in his case as the T$_B$ difference between 20 and 40 $\mu$m. \citet{1987PhDT........81S} is available on-line at https://www.boulder.swri.edu/~spencer/dissn/} cannot be due primarily to topographic temperature contrasts. The main effect of topography on the thermal emission spectrum is to increase the spectrum slope with increasing solar incidence angle, a trend not observed on Ganymede''. The similarity of the second row  of Fig.~\ref{fig:mapsdiffs} with Fig. 19 of \citet{1987PhDT........81S} is noteworthy. As shown in the third row of Fig.~\ref{fig:mapsdiffs}, the observed T$_{B,8.25}$ - T$_{B,11.25}$ decreases from dawn to dusk, again a behaviour
not reproduced in our model. We speculate that the T$_B$ negative gradient with $\lambda$ is dominated by (angle-dependent?) spectral emissivity effects, which, for a reason that remains unclear at this point, mask the expected behaviour from thermophysical model. Based on Fig.~\ref{fig:mapsdiffs}, these emissivity effects deplete the T$_{B,11.25}$ by $\sim$5 K, which for T$_{B,11.25}$ $\sim$ 145 K, corresponds to a spectral emissivity of  $\sim$0.75. More elaborate thermophysical models than presented here are probably needed to address this question with more realism.

\subsection{5-$\mu$m reflectance and 5.9-$\mu$m band maps}
\label{5umband}

The blue part of the MIRI spectra shows the onset of solar reflected radiation and evidence for a 5.9-$\mu$m absorption band (Fig.~\ref{fig:showband}). Given the difficulties, outlined above, to fit simultaneously the spatial distribution of T$_B$'s and their spectral dependence, we now restrict the modelling to the 4.9-8.5 $\mu$m
and fit the spectra on a spaxel-by-spaxel basis, using the three types of continuum models envisaged above, with the following free parameters:

\begin{itemize}
\item Model 1 (lower than unit emissivity): surface temperature (T$_1$), spectrally constant emissivity ($\varepsilon_1$), spectrally constant $I/F$ reflectivity (Radf$_1$). 
\item Model 2 (wavelength-dependent emissivity): surface temperature (T$_2$), spectral emissivity gradient ($\varepsilon$$^\prime_2$ = d$\varepsilon$/d$\lambda$, in $\mu$m$^{-1}$), spectrally constant $I/F$ reflectivity (Radf$_2$). In this case, the emissivity is taken as 1.0 at 4.9 $\mu$m and $\varepsilon$($\lambda$) = 1 + $\varepsilon$$^\prime_2$ ($\lambda$($\mu$m) - 4.9).
\item Model 3 (distribution of temperatures): central surface temperature (T$_3$) and temperature range ($\pm$$\Delta$T), and spectrally constant $I/F$ reflectivity (Radf$_3$). In this case, we consider a distribution of surface temperatures $T$ over T$_3$$\pm$3$\Delta$T, assigning a gaussian weight to each contribution according to 2$^{- (\frac{T - T_3}{\Delta T})^2}$.
\end{itemize}

       \begin{figure*}
%\vspace*{-3cm}
   \includegraphics[angle=0,width=18cm]{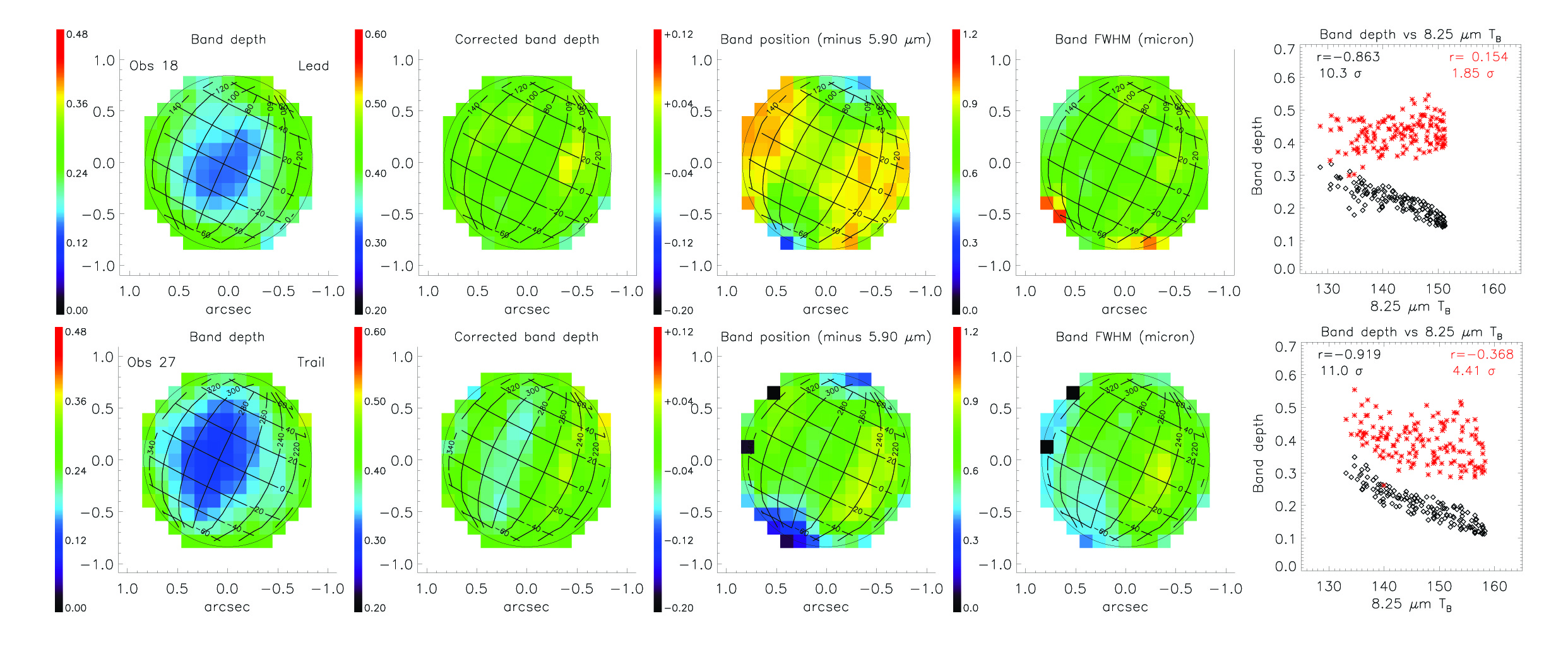}
%\vspace*{-1.25cm}
   \caption{Fitted characteristics of the 5.9-$\mu$m band in Obs. 18 (leading side, first row) and Obs. 27 (trailing side, second row), for model 2 described in the text: observed band depth relative to total continuum, band depth corrected for thermal contribution (i.e. relative to model solar reflected component), band position (offset from 5.90 $\mu$m), and band FWHM width. The last column shows the correlation between the observed (black) and corrected (red) band depths with the observed 8.25 $\mu$m T$_B$. Spearman correlation coefficients and number of standard deviations significance are indicated as insets. Similar figures for models 1 and 3 can be found in Fig.\ref{fig:5mu9_band}.}
   \label{fig:5mu9_band_only_model2}
    \end{figure*}

The three models are meant to represent end-member solutions for the spectral slope over 4.9-8.5 $\mu$m, providing robustness on the 5.9-$\mu$m band characterization. Unlike the $I/F$ reflectivities which have a clear physical meaning, the other parameters are more phenomenological; in particular $\varepsilon_1$, $\varepsilon$$^\prime_2$, and $\Delta$T are proxies for the amount of deviation of the thermal part of the spectrum from a blackbody; e.g. $\varepsilon_1$ $\sim$1 (resp. $<<1$), $\varepsilon$$^\prime_2$ $\sim$0 (resp. $<<$ 0), and small (resp. large) $\Delta$T are associated with a weak (resp. large) departure from Planck emission.  For each of the three models, Fig.~\ref{fig:fitted_param} shows maps of three model fit parameters, and of the rms residuals between observation and models, expressed in brightness temperature units. Such residuals are of the order of 0.3-0.7 K at most spaxels. As visible in the first column of Fig.~\ref{fig:fitted_param}, T$_1$, T$_2$, and T$_3$ are correlated to the surface temperature, with T$_2$ being the closest representation of T$_{B,8.25}$ (see Fig.~\ref{fig:thermalmaps}). Given also that model 2 yields the best rms fit (see Fig.~\ref{fig:fitted_param}) it is adopted as the nominal one. Median values of $\varepsilon$$^\prime_2$ are -0.061 and -0.058 $\mu$m$^{-1}$ for Obs. 18 and 27, respectively, indicative of a decrease of the spectral emissivity from 1 at 4.9 $\mu$m
to $\sim$0.81 at 8 $\mu$m. The sign of $\varepsilon$$^\prime_2$ is negative as expected, and its magnitude is large but reasonable given the end-member character of the model.
The last column of Fig.~\ref{fig:fitted_param} shows the reflectance. To convert $I/F$ to reflectance, 
we followed \citet{Ligier2019} and \citet{King2022-is} 
in adopting the \citet{Oren1994} model which generalizes the Lambert model for rough surfaces. Specifically, we used the small phase angle limit of the model and adopted a surface roughness $s$ = 20\dg, which (perhaps coincidentally given that the definitions of roughness are not exactly the same) matches both our above estimate of the slopes based on the thermophysical model and the near-IR results of \citet{Ligier2019}. The resulting reflectances appear anti-correlated with the Bond albedo (see Fig.~\ref{fig:thermalmaps}), with a Spearman correlation coefficient of $r$ = -0.62 (trailing) to -0.56 (leading) and a significance level of 6--7 $\sigma$.
The trailing side also appears to be $\sim$1.5 times more 5-$\mu$m reflective than the leading (median reflectances of 0.092 and 0.061, respectively), while being $\sim$7 K warmer. Both aspects argue for 5-$\mu$m bright material being optically dark, which is consistent with NIRSpec observations (see Sect.~\ref{sec:dust-rocks}).  

The 5.9-$\mu$m band properties were then determined from gaussian fitting of the Obs./Model ratios. Results (see Fig. \ref{fig:5mu9_band_only_model2} for the preferred Model 2 and the comparison to  models 1 and 3 in Fig.~\ref{fig:5mu9_band}) are reassuringly consistent for each of the three continuum models. The inferred band depths vary over 12-30 \%, with the shallowest bands occurring near disk center, and minimum absorptions on the trailing side. Band depths are correlated with local time and latitude, but the clearest aspect is a strongly negative ($r$ $\sim$ --0.9, 10~$\sigma$ significance) correlation between band depth and surface temperature (T$_{B,8.25}$ proxy).  Since the 5.9-$\mu$m band occurs near the cross-over of the thermal ($TH$) and solar reflected ($SR$) components, the most likely explanation is that apparent band depth variations reflect relative variations of the two components. The second column of Fig.~\ref{fig:5mu9_band_only_model2} and Fig.~\ref{fig:5mu9_band} shows band depths corrected for the thermal contribution, i.e. rescaled by 1. + $TH$/$SR$ at band center. This correction, meant to yield the band contrast in the solar reflected component, removes most of the correlation between band depth and surface temperature (red points in fifth column in  Fig.~\ref{fig:5mu9_band_only_model2} and Fig.~\ref{fig:5mu9_band}). Spatial variations of this corrected band contrast are still present, with values ranging typically over 35-54 \% (resp. 30-52 \%) on the leading (trailing) side, but with no clear geographical trend, so we regard them with caution. We also note that this correction assumes that the band does not show up as an emissivity feature in the thermal component; therefore, the presented corrected band depths may in fact be lower limits to the actual depths in the reflected component.

\begin{figure}
   \includegraphics[angle=0,width=9cm]{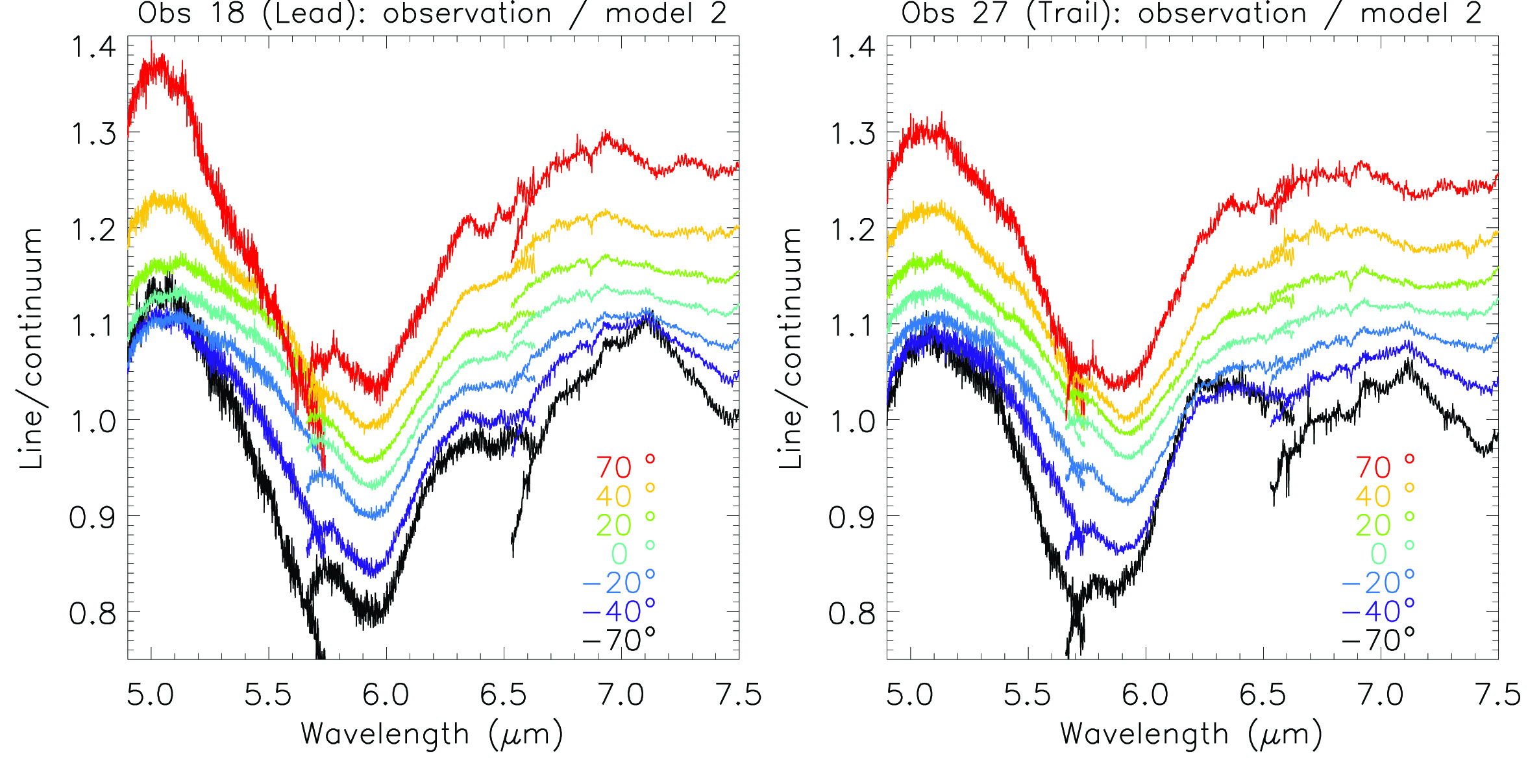}
%\vspace*{-6.5cm}
   \caption{5.9 $\mu$m band (Observation / Model 2) in 7 latitude bins ([-90\dg,-50\dg], [-50\dg,-30\dg], [-30\dg,-10\dg], [-10\dg,10\dg], [10\dg,30\dg], [30\dg,50\dg] and [-50\dg,-90\dg]. Left: leading side (Obs. 18). Right: trailing side (Obs. 27). The band appears blueshifted in the high-latitude bins, but spectra are of more modest quality there.}
   \label{fig:band_bylat}
    \end{figure}

Band central wavelengths determined from gaussian fits have mean values of 5.92$\pm$0.04 $\mu$m (leading) and 5.87$\pm$0.05 $\mu$m (trailing), where the error bar reflects the 1-$\sigma$ dispersion over the disk. On both hemispheres, the band seems blueshifted in high-latitude regions (third column of Fig.~\ref{fig:5mu9_band}). However, this conclusion must be taken with caution. Indeed, Fig.~\ref{fig:band_bylat}, in which the band aspect (with respect to Model 2) is shown in seven latitude bins, illustrates that the "--70\dg" and "70\dg" ratio spectra (which correspond to  the [-90\dg,-50\dg] and  [+50\dg,90\dg] bins) are of relatively low quality, being especially affected by significant flux discontinuities at MIRI channel edges near 5.7 $\mu$m (1A/1B) and 6.6 $\mu$m (1B/1C); the severity of the problem is presumably related to the low number of spectra in these latitude bins and their near-limb position, making fluxes more sensitive to residual pointing reconstruction errors. Another interesting aspect is that the variation in band position on the leading side (third column in Fig.~\ref{fig:5mu9_band_only_model2} and Fig.~\ref{fig:5mu9_band}) seems correlated to that seen for the H$_2$O 3.1-$\mu$m Fresnel peak area and inter-band position (see Sects~\ref{sec:fresnel}, ~\ref{sec:inter-band} and \citet{Trumbo2023}). 

\begin{figure}
   \includegraphics[angle=0,width=9cm]{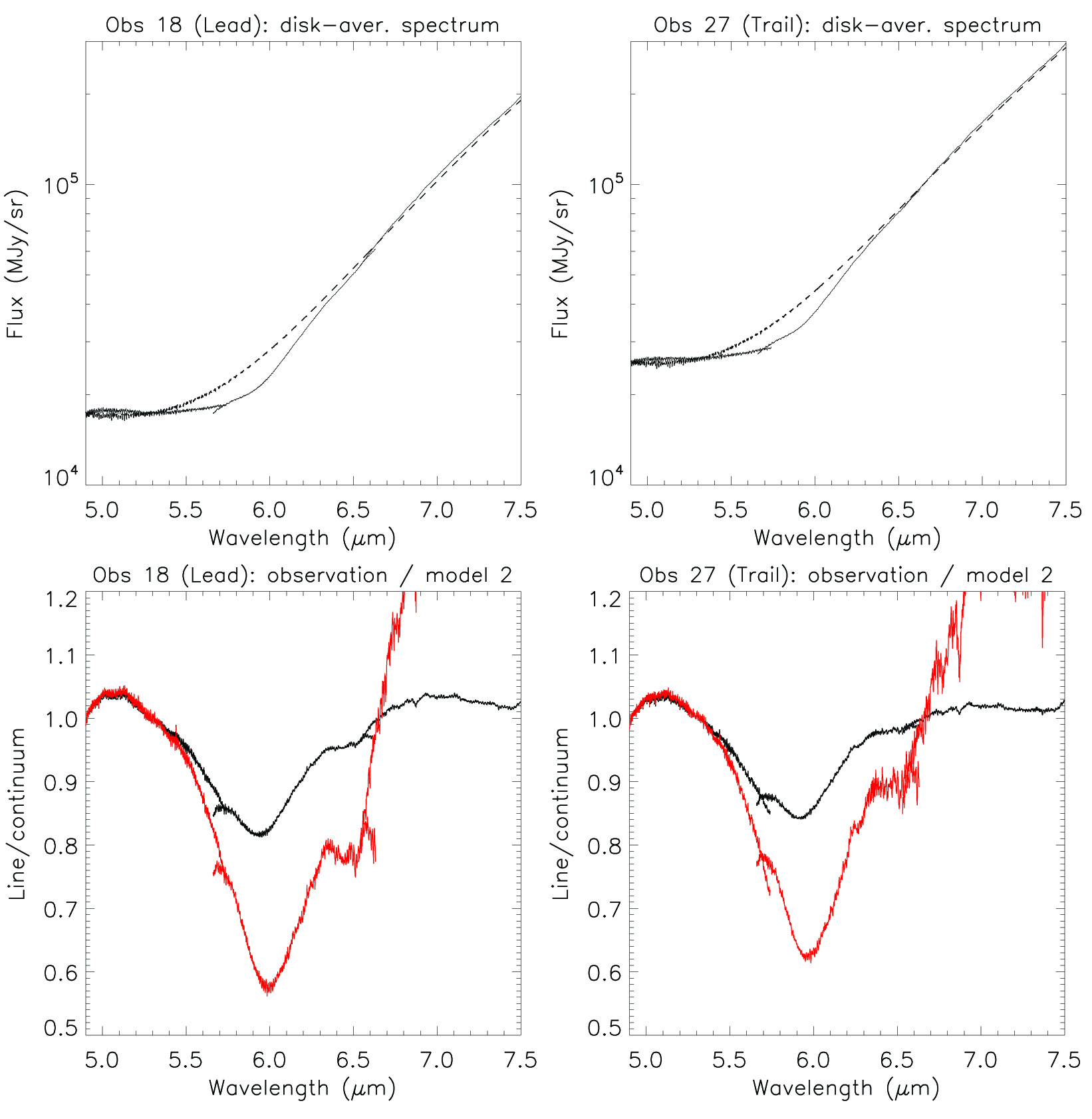}
   \caption{Disk-average spectra at 4.9-7.5 $\mu$m. Top panels: data (solid line) and continuum fit with model 2 (dashed line). Bottom panel. Observation / Model ratios. Black: ratio in the total spectrum. Red: ratio in the solar component (see text for details). Left: leading side (Obs. 18). Right: trailing side (Obs. 27). }
   
   \label{fig:band_bigavg}
    \end{figure}

Finally, to further explore the band shape, we created large spectral averages, separately for leading and trailing sides, by coadding all spectra within the disk. Fig.~\ref{fig:band_bigavg} shows such spectra and their fits with model 2. In addition to the 5.9-$\mu$m minimum, the resulting ratio spectrum shows hints for a secondary absorption near 6.5 $\mu$m in the wing of the main feature, especially on the leading side. We also attempted determining the mean band profile in the solar reflected component, by subtracting the (model) thermal component from both the observed and modelled radiances before ratioing them (red lines in Fig.~\ref{fig:band_bigavg}).  This approach yields reasonable results up to $\sim$6.7 $\mu$m but as expected fails beyond that due to the huge dominance of the subtracted term. The 6.5 $\mu$m secondary absorption and its prevalence on the leading side are confirmed. The relative depth of the 6.5 $\mu$m band with respect to the main 5.9-$\mu$m band is typically $\sim$ 15-25 \% (trailing/leading) in the total spectrum, and $\sim$ 30-50 \% after thermal correction. The possible carriers of these bands are discussed in Sect.~\ref{sec:5.9mu}.

\section{Discussion: interpretation of the results}

\begin{figure}[ht]
\includegraphics[width=9cm]{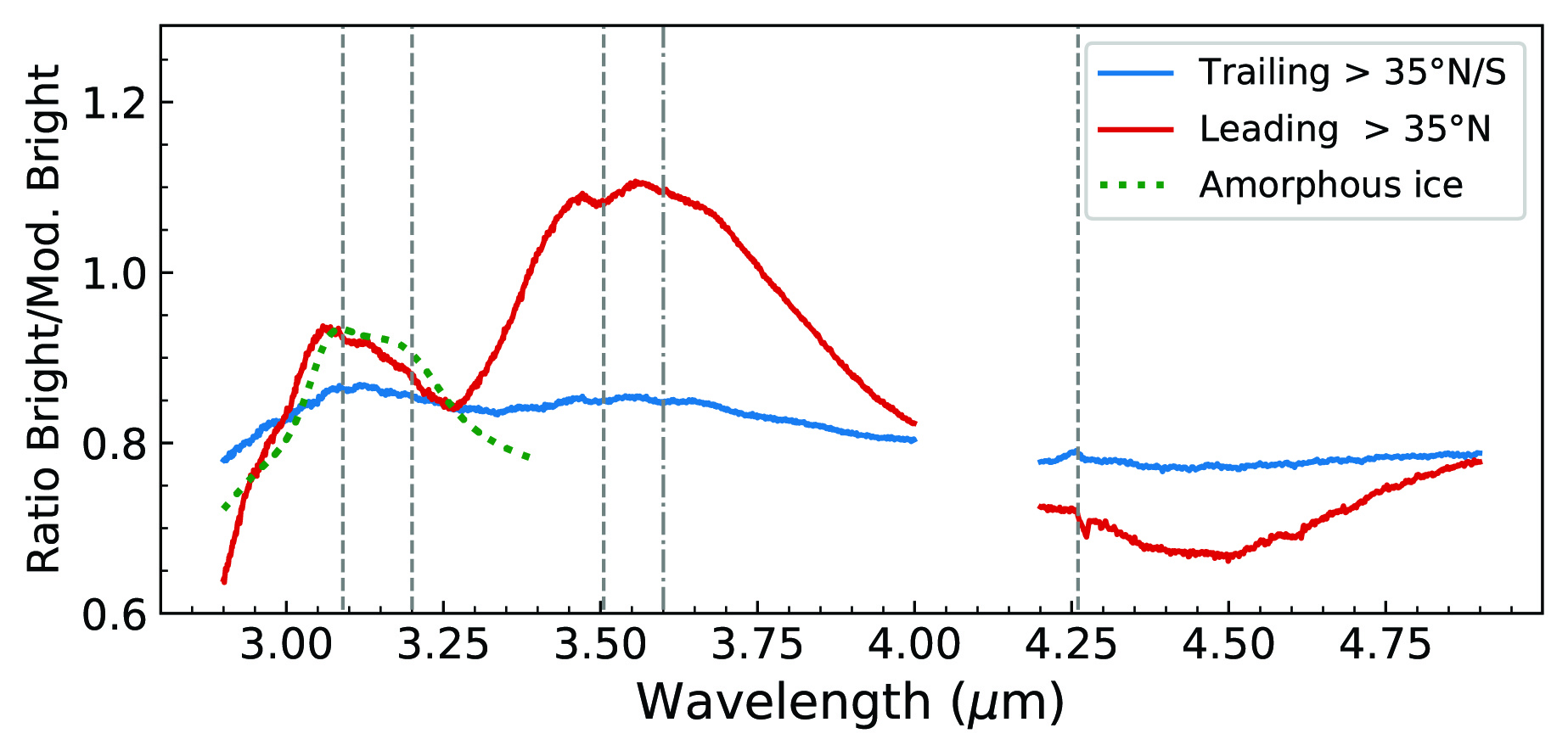}
\caption{Comparison of  Bright and Moderately Bright terrains in Ganymede's leading and trailing hemispheres, through spectral ratios. Moderately Bright and Bright terrains are defined by a Bond albedo $A_{B}$ = 0.35--0.4, $A_{B}$ = 0.4--0.6, respectively. For the leading hemisphere, only the northern regions at latitudes $>$ 35\dg N are considered, whereas for the trailing hemisphere, we consider both northern and southern high-latitude regions ($>$ 35\dg N/S). The spectrum shown in green dashed line is a spectrum of amorphous water ice obtained using Fresnel's formulae (applicable for flat surfaces) \citep{Born1999} and the optical constants at 120 K of \citet{Mastrapa2009}. The vertical dashed lines are at the wavelengths of the two main H$_2$O Fresnel peaks (3.09 and 3.20 $\mu$m), H$_2$O$_2$ (3.505 $\mu$m) and CO$_2$ (4.26 $\mu$m) absorption bands. The vertical dashed-dotted line at 3.6 $\mu$m is roughly at the H$_2$O inter-band position. \label{fig:spectral-ratios}}
\end{figure}

\begin{figure}[ht]
\includegraphics[width=9.00cm]{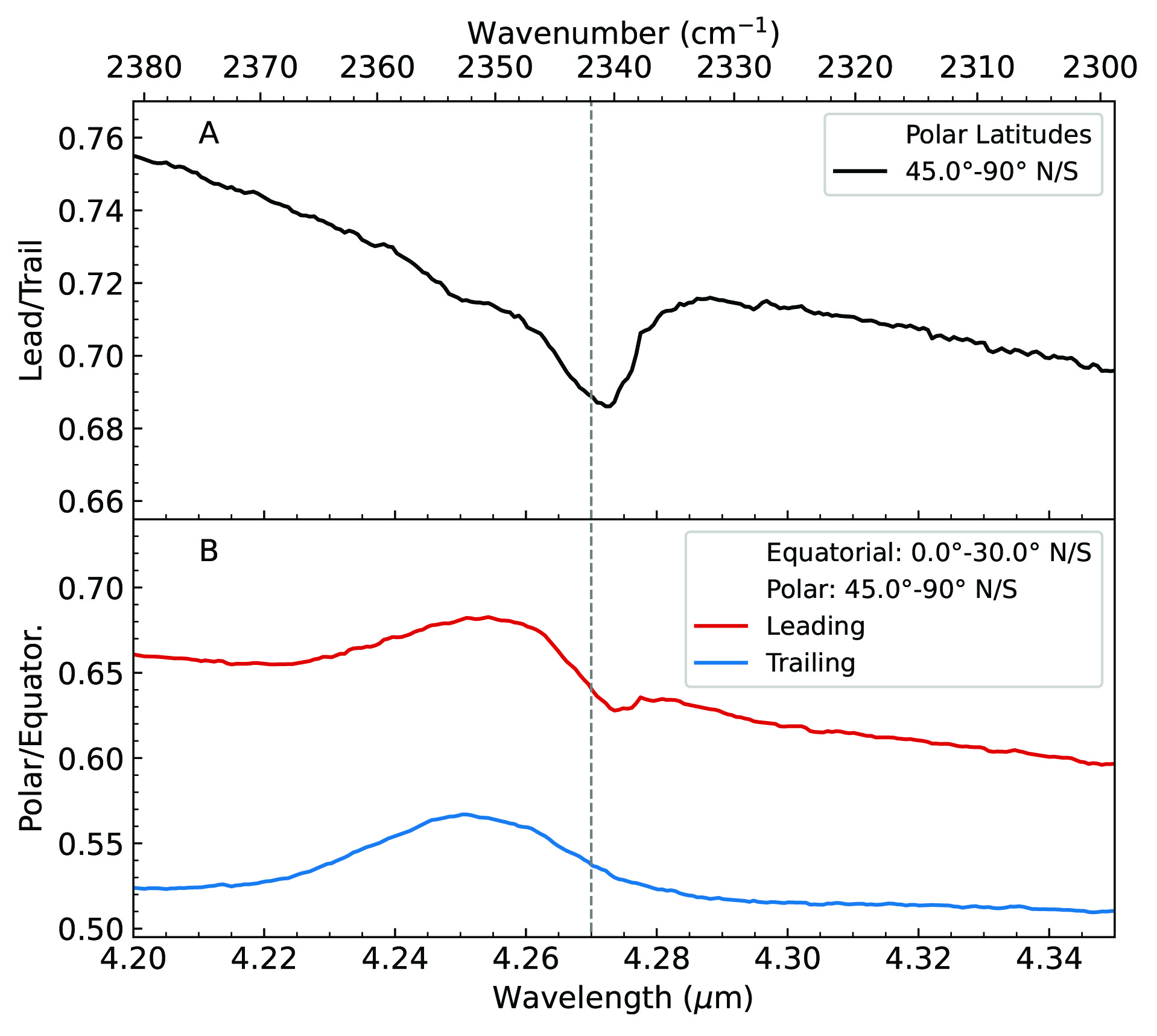}
\caption{Comparison of the CO$_2$ band in polar ($>$ 45 \dg N/S) and equatorial ($<$ 30 \dg N/S) regions through spectral ratios. (A): Spectral ratios Leading/Trailing for polar regions; (B): Spectral ratios Polar/Equatorial for Leading (red) and Trailing (blue) hemispheres. The vertical dashed line is at 4.27 $\mu$m (2341.9 cm$^{-1}$).\label{fig:zoom-CO2}}
\end{figure}

In this section we discuss the properties and distribution of H$_2$O, CO$_2$ and non-ice components as derived from the NIRSpec data, and the possible carriers of the 5.9-$\mu$m band detected with MIRI. Part of the discussion is based on ratios of spectra of different terrains:  moderately bright vs bright on high-latitude regions  (Fig.~\ref{fig:spectral-ratios}), at polar vs equatorial regions (Fig.~\ref{fig:zoom-CO2}). These spectral ratios reveal differences of abundance and/or mixture modality of species (geographical or intimate mixtures, deposits etc.) and/or differences of the chemical and/or physical properties (nature, grain size, solid phase, temperature, etc.) between terrain types and/or hemispheres. In turn, these differences can be due to processes affecting each hemisphere differently (energetic particle bombardment, meteoroid gardening). The spectral ratios Leading/Trailing and between terrains of different Bond albedo show variations of the H$_2$O, H$_2$O$_2$ and CO$_2$ spectral features (other minor spectral features at 3.43, 3.51, 3.88, and 4.38 $\mu$m, do not show up in these ratios, indicating that they do not vary over the surface or may be spurious, as discussed in Sect.~\ref{sec:NIRSpec-an}).

For this discussion, we also produced merged NIRSPec and MIRI  spectra corrected from the contribution of thermal emission (Figs ~\ref{fig:salts}, \ref{fig:phyllo-zarcas}, \ref{fig:Hapkemodel_CO2}), as described in Appendix~\ref{appendix:merged}.

\subsection{H$_2$O ice and non-ice materials}
\label{discussion_h2o_nonice}

The variations of the H$_2$O Fresnel peaks and inter-bands obtained by NIRSpec over the disk of Ganymede provide insights on the properties of the water ice (grain size, temperature, crystallinity, etc.) and on the presence of non-ice materials, complementing previous observations of water ice spectral features at shorter wavelengths \citep{Ligier2019,King2022-is} or at lower spectral resolution \citep{Hansen2004}.

\subsubsection{Known properties of H$_2$O ice spectral features}

The Fresnel peaks are produced after the first reflection on a facet of water ice, so they are indicative of the properties of the very first interface of Ganymede's surface. Crystalline water ice is characterized by three well-defined peaks (at 2.95, 3.1, 3.2 $\mu$m) whereas amorphous water ice has a broad and featureless 3-$\mu$m reflection peak shifted to shorter wavelengths (\citealt{Hansen2004,Mastrapa2009}). The intensity of the Fresnel peaks tends to increase with increasing ice grain size (when larger than several micrometers), but decreases with increasing ice temperature (\citealt{Stephan2021}). Any non-H$_2$O contaminant covering the ice surface decreases the amplitude of the peaks (\citealt{Ciarniello2021}). Finally, the peak position has been observed to slightly shift to longer wavelengths with decreasing temperature (\citealt{Stephan2021,Mastrapa2009}). Concerning the H$_2$O inter-band around 3.6 $\mu$m, its amplitude $A_{\rm IB}$ with respect to the local minimum at $\sim$ 3.3 $\mu$m (Eq.~\ref{eq:int-IB}) increases with decreasing grain size and decreasing temperature (\citealt{Poch2016,Stephan2019}).  The inter-band wavelength position depends on the temperature (\citealt{Clark2012,Filacchione2016}) and also on the amount of amorphous vs. crystalline water ice (\citealt{Mastrapa2009}). The increase of inter-band amplitude with decreasing grain size is due to the increased density of facets with which photons interact, reducing their mean free path and inducing multiple scattering, increasing the reflectance at wavelengths in between absorption bands. Therefore, any reduction of mean free path in H$_2$O ice (due to e.g. grain size, defects, porosity etc.) would increase the inter-band amplitude. Contaminants can increase or decrease the amplitude of the inter-band and may shift its position depending on their spectra. Given the high number of parameters controlling these spectral features, the variations observed on Ganymede are equivocal. Therefore, here we propose a preliminary interpretation of the observed NIRSpec spectra of Ganymede, based on this current knowledge. Additional radiative transfer modelling and experimental measurements would be needed to provide a more detailed interpretation.

\subsubsection{Global distribution of H$_2$O ice}

Maps of the H$_2$O Fresnel peak area (Fig.~\ref{fig:maps-Fresnel}) are consistent with the distribution of water ice inferred from H$_2$O bands at 4.5-$\mu$m (Fig.~\ref{fig:maps-4p5}) and in the near-infrared \citep{Ligier2019,Stephan2020-zk,Mura2020,King2022-is}, and from the visible albedo maps (Fig.~\ref{fig:albedo}), as discussed in Sect.~\ref{sec:fresnel}. Ganymede is globally covered by H$_2$O ice (Fig.~\ref{fig:sp-fullrange}), and all spectral ratios of Fig.~\ref{fig:spectral-ratios} have an inter-band centered at 3.55-3.65 $\mu$m, indicating that variations of H$_2$O ice surface abundance and/or properties are the main sources of differences between terrains/hemispheres. In particular, Ganymede's H$_2$O surface ice is more or less contaminated by non-ice materials in proportions that vary regionally. The strong contrast of the H$_2$O Fresnel peak area, 10 to 40 times higher on the leading hemisphere than on trailing hemisphere (Fig.~\ref{fig:CO2-albedo}C), can be attributed to a higher abundance of non-ice materials covering the ice on the trailing hemisphere, whereas more ice is directly exposed to the surface of the leading hemisphere. As discussed in \citet{Ligier2019} and \citet{King2022-is}, compared to the trailing hemisphere, the leading hemisphere may receive (1) a higher flux of (and/or more energetic) ions and electrons causing surface sputtering and deposition of fresh ice (\citealt{plainaki2022,Liuzzo2020-ww,Poppe2018-wo}), and (2) a higher flux of micro-meteoroids (\citealt{bottke2013}) excavating the ice and also exposing fresh ice at the surface, whereas most of the ice on the trailing side remains below a low-albedo lag deposit of non-ice materials.

\subsubsection{Properties of H$_2$O ice in polar regions \label{h2o_polar}}

The maximum of H$_2$O Fresnel peak area and intensity is reached in the polar regions of the leading side, which appear to have more water ice directly exposed at the surface than other terrains on Ganymede. This is confirmed by the inter-band position, transitioning abruptly from 3.9 to 3.66 $\mu$m poleward of 30\dg~latitude (Fig.~\ref{fig:maps-interband}). These polar regions are where the highest flux of energetic ions is expected to irradiate the surface \citep{Fatemi2016-uy,Poppe2018-wo, Liuzzo2020-ww, plainaki2022}. It is striking to see how close the inter-band transition follows Ganymede's open-closed field line boundary (Fig.~\ref{fig:maps-interband}). The purity of the ice exposed at the surface of the leading polar regions may thus be explained by the combination of micro-meteoroid gardening, excavating the ice, and ion irradiation, sputtering the excavated ice followed by local re-accretion of water vapor forming purer water ice on top of the non-ice materials \citep{johnson1997,Khurana2007-og}. Moreover, the leading north polar region spectrum has the highest inter-band amplitude (Figs~\ref{fig:sp-fullrange}, \ref{fig:maps-interband}), indicating that this region is covered by water ice particles offering the smallest mean free path to photons on Ganymede, which should be smaller than 70 $\mu$m \citep{Stephan2021}, in agreement with the conclusions of previous studies \citep[$\le$50 $\mu$m according to][] {Hibbitts2003,Ligier2019,King2022-is,Stephan2020-zk}. Physically, this smallest mean free path may be due to the presence of (1) the smallest ice grains and/or (2) more internal defects (crystal defects, voids, bubbles, etc.) in large ice grains (\citealt{johnson1997}) and/or (3) a higher micro-roughness/porosity of the ice grains, increasing the scattering of the light like smaller grains. All could result from the processes triggered by energetic particles irradiations \citep{johnson1997,johnsonandq1997,khurana2007}. Dedicated numerical and/or experimental models of NIRSpec spectra could be used to discriminate these possibilities, and improve our knowledge of the ice texture.

On the trailing side, no such abrupt transition at 30$^{\circ}$ latitude is observed, but a progressive shift of the inter-band position toward lower values at the poles (Fig.~\ref{fig:maps-interband}), indicative of a higher proportion of ice and/or smaller optical mean free path in ice as the latitude increases. The polar regions of the trailing side have much less ice and/or larger optical mean free path in ice (so larger grains and/or less internal defects and/or lower micro-roughness/porosity) than the polar regions of the leading side (Fig.~\ref{fig:sp-fullrange}).

Figure~\ref{fig:spectral-ratios} shows the ratio between the spectra of the brightest northern polar regions of the leading hemisphere, with spectra of the surrounding darker terrains. The ratios between these spectra indicate the variations of H$_2$O ice surface abundance and/or properties. This ratio has a Fresnel peak shifted to 3.0583 $\mu$m, indicative of the predominance of amorphous water ice in the north polar cap of the leading hemisphere (see Fig.~\ref{fig:spectral-ratios}, and \citet{Mastrapa2009}), possibly as a deposit on top of the surface. The predominance of amorphous ice in the polar regions is consistent with previous analyses of the shape of the Fresnel reflectance peak from NIMS data \citep{Hansen2004} and with analyses of the H$_2$O near-infrared bands observed with the VLT \citep{Ligier2019}.

Amorphous ice can form (1) via the direct irradiation of crystalline ice by magnetospheric energetic ions and electrons \citep{fama2010}, and/or (2) via the re-condensation of water vapour produced after a micro-meteoroid impact, and/or (3) via the re-condensation after sputtering by energetic particles. In addition, for amorphous ice to be observed, its re-crystallisation timescale must be longer than its formation timescale \citep{faure2015,mitchell2017}. This seems only possible in Ganymede's polar regions because they experience the lowest temperatures (i.e., longest re-crystallisation timescales) and receive the highest flux of energetic ions forming amorphous ice. Amorphous water ice is expected to be present as a layer at the surface of icy crystalline grains. The thickness of this amorphous layer is however unknown, due to the extent of resurfacing processes that controls the actual sputtering rate, and therefore both the kinetics of amorphization and recrystallization (see \cite{Johnson2004} for extensive details).

The wavelength shift of the Fresnel peak is slightly more pronounced on the leading north polar cap than the south polar cap (Fig.~\ref{fig:maps-Fresnel}). This is possibly due to a slightly higher abundance of amorphous ice in the north polar cap. According to the thermal maps and models presented in Sect.~\ref{sec:MIRI-an} (Fig.~\ref{fig:fitted_param}) the poles have similar temperatures, so a temperature difference cannot explain this shift. This asymmetry is intriguing, as it suggests there are differences in the processes transforming the ice in the north and south polar caps.

Contrary to the leading hemisphere, on the trailing hemisphere the Fresnel peak is red-shifted in the polar caps compared to the surrounding regions (Fig.~\ref{fig:maps-Fresnel}). As seen in Fig.~\ref{fig:spectral-ratios}, the spectral ratio between the polar and surrounding regions is extremely different between hemispheres, suggesting that the polar ices have different origins and/or undergo different transformations depending on the hemisphere. However, both hemispheres have their north polar caps with a Fresnel peak blue-shifted and a 3.6-$\mu$m inter-band more intense compared to their south polar caps (Fig.~\ref{fig:maps-Fresnel}, Fig.~\ref{fig:maps-interband}). A possible explanation is that the processes forming amorphous ice also decrease the optical mean free path in ice (forming smaller grains and/or more internal defects and/or higher micro-roughness/porosity). These processes appear to be enhanced on the north polar cap.

The temperature of the surfaces of Saturn's icy satellites has been derived from the spectral position of the 3.6 $\mu$m inter-band, using laboratory reflectance spectra of crystalline water ice at different temperatures \citep{Filacchione2016, Clark2012}. Applying the same approach, and assuming fine-grained crystalline water ice \citep[measurements by][]{Clark2012}, we derive a temperature of 155 K for the polar caps of Ganymede's leading hemisphere (inter-band at 3.66 $\mu$m; see Fig.~\ref{fig:h2o_ice_temp}). However, as previously discussed, water ice is partially in amorphous form at the poles of the leading hemisphere. For amorphous ice, the position of the inter-band exhibits no trend with temperature, and lies at lower wavelengths than for crystalline ice (see Appendix~\ref{appendix:h2o_ice_temp}). This implies that the temperature derived by comparing the position of the inter-band to spectra of crystalline ice only, is an upper limit. Moreover, the presence of larger ice grains and/or of non-ice materials would also decrease the temperature derived via this method. This upper limit of 155 K is consistent with a surface temperature of 135 K poleward of 60\dg~latitude obtained  with the thermal model with surface roughness that most closely reproduces the MIRI data.

NIRSpec observations discussed above, and especially the presence of amorphous ice, tend to favor ion sputtering as the mechanism forming Ganymede's polar caps. Other past studies have also explained their formation by the migration of water ice, sublimating in the equatorial regions and re-condensing at higher/colder latitudes \citep{Purves1980}, a mechanism which may not be excluded.

\subsubsection{Longitudinal/Diurnal variations of H$_2$O ice at low latitudes}

Some spectral variations of the Fresnel peaks over Ganymede's longitudes, shown in Fig.~\ref{fig:maps-Fresnel}, are uncorrelated with Bond albedo variations. They may indicate diurnal processes and/or exogenous processes specific to some longitudes, affecting the surficial water ice.

First, on the trailing hemisphere, the Fresnel peak area is maximum at low latitudes of the morning limb (Fig.~\ref{fig:maps-Fresnel}). The terrains of this morning limb have a Bond albedo (Fig.~\ref{fig:albedo}) and a 3.6-$\mu$m band position (Fig.~\ref{fig:maps-interband}) similar to the surroundings, and the lowest 4.5-$\mu$m H$_2$O band depth on the trailing disk (Fig.~\ref{fig:maps-4p5}), suggesting that the ice responsible for this peculiar morning limb behavior is thin and may be transient, specific to the local hour. These ice crystals could have formed by the condensation of water vapour present in the exosphere, but this cannot explain an excess of ice only seen on the morning limb. In addition, even a H$_2$O column density of 10$^{16}$ cm$^{-2}$ -- which is probably an upper limit given observational \citep{Hartogh2013,2021NatAs...5.1043R} and model \citep[e.g.][]{2017Icar..293..185L,Leblanc2023} results -- corresponds to a mere 0.003 $\mu$m ice layer. Another mechanism is the direct condensation of water vapour sublimating from the subsurface during the night. Transient (e.g. early morning) frost was observed at the surface of comet 67P/Churyumov-Gerasimenko, and explained by the low thermal inertia of the highly porous 67P's nucleus causing a temperature inversion at the beginning of the night with surface layers getting colder than the interior layers \citep{DeSanctis2018}. This mechanism might be at work on Ganymede since the upper layer of the surface has the lowest thermal inertia, as derived from MIRI (Sect.~\ref{sec:thermal}) and Galileo data \citep{2021PSJ.....2....5D}, whereas layers few centimeters deeper have higher thermal inertia \citep{2021PSJ.....2....5D}. Such process should occur globally on all terrains having a low-enough thermal inertia at the top surface, so frost would also be observed on the morning limb of the leading hemisphere. In reality, Fig.~\ref{fig:maps-Fresnel} does not show a maximum of Fresnel peak area on the leading morning limb. However, the albedo and Fresnel peak area of the morning limb of the trailing side are similar to those of terrains of the leading side (Fig. \ref{fig:CO2-albedo}C). Therefore, the formation of such morning frost might be more difficult to detect on the surface of the leading side, because it is covered by more permanent water ice than the trailing side, whose Fresnel peak could mask that of the morning frost. Alternatively, this scenario of diurnal frost condensation/sublimation may be wrong or incomplete, and these terrains may be peculiar because of endogenous processes similar to those affecting the leading side. In particular, the "morning" limb of the trailing side observed here probes Ganymede's sub-jovian (towards Jupiter) hemisphere, which may be affected by different energetic particles bombardments than the trailing apex, altering the surficial ice differently \citep{Liuzzo2020-ww,plainaki2022}.

Another unexpected result is the variation of the central wavelength of the Fresnel peak with longitude (Fig.~\ref{fig:maps-Fresnel}), observed at low latitudes (typically $\leq$ 40\dg) for both hemispheres. The central wavelength has a similar value at the west limb of the two hemispheres (corresponding to the earliest local times). But as the longitude changes from the west to the east, the Fresnel peak position becomes more and more blue-shifted, reaching the maximum blue-shift at the east limb (latest local times), with a bluer wavelength on the trailing hemisphere. Remarkably, neither the H$_2$O Fresnel peak area nor the H$_2$O inter-band position exhibit such longitudinal variation, and one should note that this shift is extremely small (0.004 $\mu$m, Fig.~\ref{fig:maps-Fresnel}). Therefore, the modification of the H$_2$O ice inducing this shift may be limited to the first nano/micrometres of ice grains and/or to very minor geographical extents (much smaller than NIRSpec spatial resolution of about 300 km per spaxel on Ganymede). This longitudinal variation may be geographic and possibly linked to regional differences of magnetospheric energetic particles bombardment, or it may be diurnal and linked to solar radiations. Such as small shift of the H$_2$O Fresnel peak position towards smaller wavelengths can have various physical origins. It could be due to a modification of the surface ice molecular structure, changing its optical index: either an increased fraction of amorphous ice, or a higher ice temperature, or another change of the ice molecular structure (other phase change, chemical change etc.). But it could also be due to a modification of the nano/micrometre-scale texture of the ice surface, changing the relative contribution of absorption versus reflection of the light within the first nano/micrometres of the ice surface: for example an increased surface roughness induced by sublimation and/or re-condensation and/or photodesorption of H$_2$O ice. If this shift was caused by solar heating of the ice, we would expect the Fresnel peak position to be the bluest at the center of the disk (around noon, where the temperature is maximum as seen in Fig.~\ref{fig:thermalmaps}), and not at the longitude corresponding to the evening. More likely, this pattern might be explained by a progressive and cumulative modification of the H$_2$O ice molecular structure and/or nano/micrometre-scale texture, induced by the accumulation of solar radiation from the morning to the evening. Solar radiation not only regulates Ganymede's surface heat, inducing sublimation or re-condensation of H$_2$O ice, but it also induces photochemical processes (including photodissociation of H$_2$O followed by further chemistry, H$_2$O photodesorption, and H$_2$O ice amorphisation), and radiolytic processes due to solar energetic particles. The spectral ratio between more blue-shifted and less blue-shifted low latitudes (not shown) is inconsistent with the spectrum of pure amorphous ice. Moreover, as discussed in Sect.~\ref{h2o_polar}, amorphous ice should not be thermodynamically stable at these latitudes, although a detailed investigation of the dynamic of re-crystallisation and amorphisation under Ganymede's surface conditions may still be needed to conclude. Moreover, this modification of the surface ice is reversible so that during the night the Fresnel peak position is shifted back to its original position on the morning, possibly via a mechanism of relaxation of the ice (re-condensation, metamorphisation, phase change, chemical change, etc. ?). Constraining the mechanism at the origin of this shift requires dedicating modelling and/or laboratory investigations of these various processes, which are out of the scope of this paper. The leading-trailing differences in the Fresnel peak shift indicates that the mechanism is enhanced during the day on the trailing side, possibly because of its higher surface temperature. Maybe this phenomenon could have a link with the maximum sublimation rate of water at the sub-solar point, previously modelled \citep{Leblanc2023} and observed \citep{2021NatAs...5.1043R}.

\begin{figure}
\includegraphics[width=9.cm]{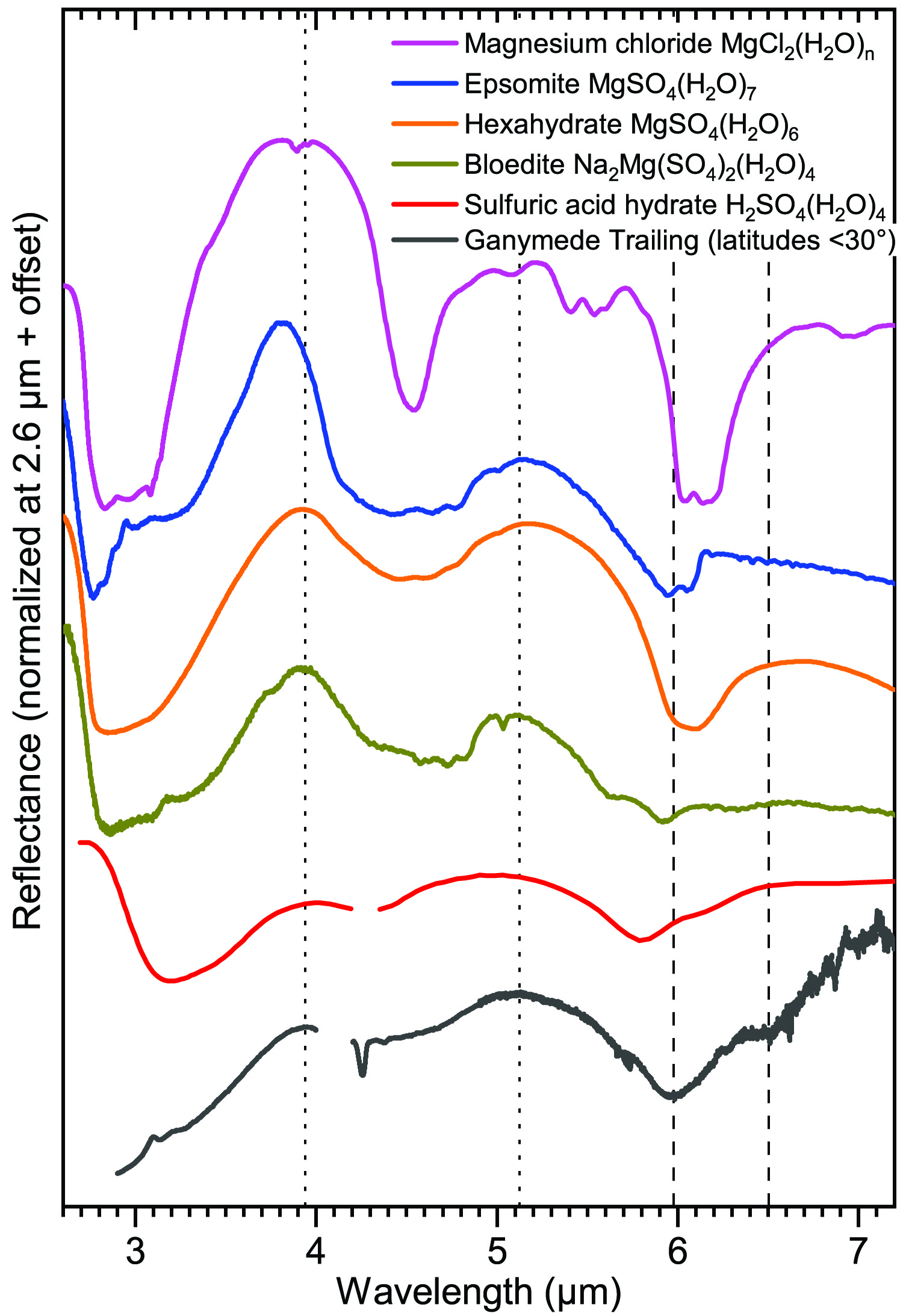}
\caption{Laboratory reflectance spectra of several hydrated salts and sulfuric acic hydrate, possibly present on Ganymede according to previous near-infrared observations \citep{Ligier2019,King2022-is}. The spectrum of Magnesium chloride (<250 $\mu$m, JB-JLB-G32-A) is from the NASA Reflectance Experiment LABoratory (RELAB) spectral database. Spectrum of Epsomite (GDS149) is from the USGS Spectral Library (\citealt{kokaly2017}). Spectra of Bloedite ($\sim$250 $\mu$m, Jim Crowley, USGS) and Hexahydrate (<45 $\mu$m, SPT143) are from the database of the Centre For Terrestrial and Planetary Exploration (C-TAPE) of University Winnipeg. The spectrum of sulfuric acid tetrahydrate at 160 K is from \citet{loeffler2011}. Note that this is a transmittance spectrum. A reflectance spectrum of Ganymede's trailing side equatorial region (<30$^{\circ}$ latitude N/S, see Fig.~\ref{fig:Hapkemodel_CO2}) is scaled for comparison (black line). The vertical dotted lines indicate H$_2$O inter-band positions in NIRSpec spectrum, and the vertical dashed lines indicate the band positions in MIRI spectrum.
\label{fig:salts}}
\end{figure}

\begin{figure}
\includegraphics[width=9.cm]{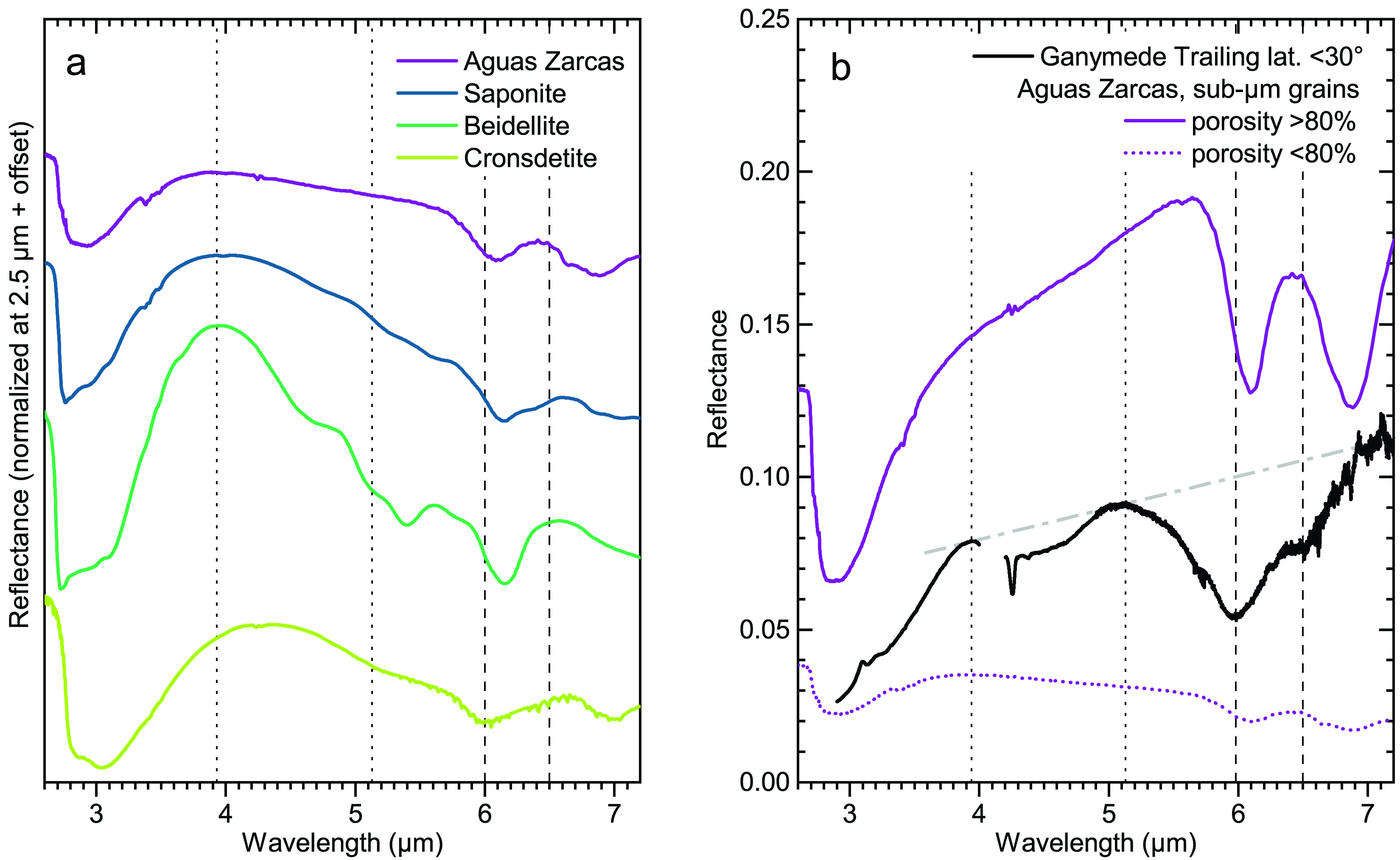}
\caption{Laboratory reflectance spectra of Aguas Zarcas CM2 meteorite powder and several phyllosilicate powders (a) and  influence of the porosity on the reflectance spectrum of Aguas Zarcas sub-$\mu$m powder (b). Spectra of Aguas Zarcas were measured at IPAG (Institut de Plan\'etologie et d'Astrophysique de Grenoble, see Appendix \ref{appendix:aguas_zarcas}). Ganymede's trailing side equatorial region (<30$^{\circ}$ latitude N/S) reflectance spectrum (see caption of Fig.~\ref{fig:Hapkemodel_CO2}) is plotted at the same reflectance scale in plot (b) (black line), with the gray dashed-dotted line in plot indicating the potential continuum obtained by drawing a line between the points at 3.9 and 5.1 $\mu$m (inter-bands). This continuum has a positive (red) spectral slope, possibly due to a relatively high surface porosity of dark terrains. Spectra of Cronsdetite (M3542) is from the USGS Spectral Library (\citealt{kokaly2017}). Spectra of Saponite (<45 $\mu$m, SA-EAC-059) and Beidellite (<12 $\mu$m, JB-JLB-919-A) are from the NASA Reflectance Experiment LABoratory (RELAB) spectral database. The vertical dotted lines indicate H$_2$O inter-band positions in NIRSpec spectrum, and the vertical dashed lines indicate the band positions in MIRI spectrum.
\label{fig:phyllo-zarcas}}
\end{figure}

\subsubsection{Non-ice materials composition and texture} \label{sec:dust-rocks}

Spectra of the dark terrains (see those at mid latitudes 0--30\dg~in Fig.~\ref{fig:sp-fullrange}) all have a maximum H$_2$O inter-band at 3.9 $\mu$m (3.935 $\mu$m for leading, and 3.950 $\mu$m for trailing) and a continuum with a positive (red) spectral slope from 3.9 to 7 $\mu$m (i.e., the reflectance at the inter-bands -- 3.9, 5.1, 7 $\mu$m -- increases with the wavelength, see Fig.~\ref{fig:salts}). Both of these spectral characteristics are incompatible with water ice \citep[water ice inter-band maximum cannot be at more than 3.7 $\mu$m,][]{Clark2012,Filacchione2016} so they are due to the presence of non-ice materials.
Several H$_2$O-bearing minerals, such as phyllosilicates, sulfates and chlorides, exhibit a H$_2$O inter-band around 3.9 $\mu$m, as shown in Figure~\ref{fig:salts} and Figure~\ref{fig:phyllo-zarcas}a. Figure~\ref{fig:salts} shows infrared reflectance spectra of some sulfate and chloride salts and a sulfuric acid hydrate expected to be present on Ganymede's surface based on the shape of the water absorption bands in the near-infrared (\citealt{Ligier2019, King2022-is}). Among these species, bloedite (Na$_2$Mg(SO$_4$)$_2$(H$_2$O)$_4$) and hexahydrate (MgSO$_4$(H$_2$O)$_6$) spectra best match Ganymede's dark terrain spectrum from 3.3 to 4 $\mu$m (general shape and position of the H$_2$O inter-band maximum). However, none of the salt spectra match Ganymede's dark terrain spectrum from 4 to 7 $\mu$m. Ganymede's dark terrain spectrum exhibits a 4.5-$\mu$m H$_2$O absorption band attributed mainly to water ice. Several salts spectra also exhibit a H$_2$O band centered around 4.5 $\mu$m, but bloedite and hexahydrate spectra show an abrupt increase of reflectance from 4.8 to 4.9 $\mu$m which is absent on Ganymede's dark terrain spectrum. Sulfuric acid tetrahydrate (H$_2$SO$_4$(H$_2$O)$_4$) spectrum exhibits smoother spectral variations than the salts, although not identical to Ganymede's dark terrain spectrum. Notably, the shape of the absorption band of sulfuric acid tetrahydrate centered at 5.8-$\mu$m is similar to the one observed by MIRI on Ganymede, although its position is different (Fig.~\ref{fig:salts}). The attribution of the 5.9-$\mu$m band observed on Ganymede will be discussed in Sect. \ref{sec:5.9mu}. Finally, Ganymede's dark terrain continuum has a positive (red) spectral slope from 3.9 to 7 $\mu$m, unlike reflectance spectra of salts. Therefore this spectral feature is probably caused by other components of the dark terrains.

Previous studies have noted similarities between the 3-$\mu$m absorption band of Ganymede, Callisto and Amalthea's dark terrains, with that of phyllosilicates present in carbonaceous chondrites \citep{calvin1991,schenk1991,takato2004}. Phyllosilicates are hydrated minerals known to be formed by aqueous alteration of silicates. Serpentine and smectite phyllosilicates are among the main constituents of the matrix of carbonaceous chondrites (especially CI and CM groups), which are dark meteorites from undifferentiated bodies (C-type asteroids). If present on Ganymede, such phyllosilicates may have endogenous and/or exogenous origins, formed by aqueous alteration in its subsurface, or brought by impacting bodies. Figure~\ref{fig:phyllo-zarcas}a shows reflectance spectra of some of these phyllosilicates, together with the spectrum of a CM chondrite (Aguas Zarcas) in powder form. Cronsdetite is a Fe-rich serpentine, beidellite and saponite are smectites. Beidellite and saponite, as well as Aguas Zarcas, exhibit a 3.9-$\mu$m H$_2$O inter-band having a maximum at similar wavelength than Ganymede's dark terrains (Fig.~\ref{fig:sp-fullrange}), but with a much larger width (Fig.~\ref{fig:phyllo-zarcas}a). Moreover, these phyllosilicates have a negative spectral slope from 4 to 6 $\mu$m, unlike Ganymede's dark terrain continuum with a positive spectral slope from 3.9 to 7 $\mu$m.

In summary, Ganymede's dark terrain 3.9-$\mu$m and 5.1-$\mu$m H$_2$O inter-bands correspond better to Na-/Mg-sulfates or sulfuric acid hydrates than phyllosilicates, but none of these compounds explain Ganymede's dark terrain red spectral continuum between 3.9 to 7 $\mu$m (Fig.~\ref{fig:salts}, Fig.~\ref{fig:phyllo-zarcas}a). In fact, this latter part of the spectrum may be dominated not by absorption but by scattering effects induced by the peculiar surface texture of Ganymede's dark terrains \citep[85\% porosity,][]{2021PSJ.....2....5D}. Fig.~\ref{fig:phyllo-zarcas}b shows that sub-micrometer-sized carbonaceous chondrite powder with porosity >80\% exhibits a strong red (positive) spectral slope reminiscent of the one observed on Ganymede's dark terrains continuum from 4.6 to 7 $\mu$m (see Appendix \ref{appendix:aguas_zarcas} for details on the preparation of this sample). The radiative transfer mechanism explaining how an increase of porosity induces a spectral reddening is not yet established (it will be the subject of an upcoming study), but it appears to require the presence of sub-$\mu$m opaque grains such as iron sulfides or magnetite found in carbonaceous chondrites. We note that opaque minerals have already been suggested as constituents of the dark terrains of Callisto \citep{calvin1991} and they could also be the spectrally flat darkening agent necessary to fit Ganymede's near-infrared spectral properties \citep{Ligier2019,King2022-is}. When mixed with other components, even at few percent levels, fine-grained opaque minerals dominate the light scattering and control the reflectance level, masking absorption bands of the other components \citep{sultana2023}. At visible and near-infrared wavelengths, the salt and hydrate candidates discussed above have relatively high reflectances \citep{carlson1999_sulfuric}. Therefore, the low reflectance of Ganymede's dark terrains could be explained by a mixture of salts and/or hydrates with opaque minerals and/or by radiation-induced color changes of salts \citep{hibbitts2019}.

In conclusion, JWST spectra of Ganymede's dark terrains seem compatible with the presence of H$_2$O-bearing non-ice compounds such as Na-/Mg-sulfate salts, sulfuric acid hydrates, and possibly phyllosilicates, mixed with fine-grained opaque minerals, having an highly porous surficial texture. However, neither previous NIR spectra (NIMS/Galileo and VLT) nor these NIRSpec spectra show diagnostic absorption bands that would enable firm identifications of Ganymede's dark terrain constituents. The 5.9-$\mu$m band observed by MIRI is present all over Ganymede's surface, in dark and bright terrains, so it is due to constituent(s) present on both terrain types, possibly sulfuric acid hydrates as discussed in Sect.~\ref{sec:5.9mu}.

%%%%%%%%%%%%%%%%%%%%%%%%%%%%%%%%%%%%%%%%%%%%%%%%%%%%%%%%%
%%%%%%%%%%%%%%%%%%%%    CO2 band    %%%%%%%%%%%%%%%%%%%% %%%%%%%%%%%%%%%%%%%%%%%%%%%%%%%%%%%%%%%%%%%%%%%%%%%%%%%%%

\subsection{CO$_2$}
\label{sec:CO2}
The band at $\sim$ 4.26 $\mu$m in Ganymede observations is assigned to the asymmetric stretching mode $\nu_3$ of CO$_2$. The analysis of NIRSpec spectra presented before has led to the following conclusions:
\begin{itemize}
\item The band is asymmetric (Fig. \ref{fig:sp-CO2}).
\item The band centre shifts towards higher wavelengths with increasing latitude (Figs \ref{fig:sp-CO2}, \ref{fig:maps-CO2}, Table~\ref{tab:CO2_resultfit}), and with increasing Bond albedo (Fig. \ref{fig:CO2-albedo}) and Fresnel peak area (Fig.~\ref{fig:CO2-H2O}). The CO$_2$ band centre ranges from 4.257 to 4.260 $\mu$m  (2349--2347 cm$^{-1}$) on equatorial regions, is at 4.264 $\mu$m (2345.3 cm$^{-1}$) on trailing north/south poles and leading south pole, and at 4.270 $\mu$m (2342 cm$^{-1}$) on the north polar cap of the leading hemisphere. 
\item The band depth decreases with increasing latitude, for both the leading and trailing hemispheres (Fig. \ref{fig:depth-CO2}, Table~\ref{tab:CO2_resultfit}).
\end{itemize}

A rough estimate of the CO$_2$ concentration can be estimated through Hapke modelling \citep{1993tres.book.....H}. We considered two terrains geographically separated: (1) a granular intimate mixture with grains made of pure crystalline water ice and pure carbon dioxide (optical constants from \citet{Mastrapa2009}, and \citet{trotta96}); (2) a dark terrain, simulated by a mature coal, which intends to mimic opaque ultramafic assemblages as observed in primitive carbonaceous chondrites and cometary grains \citep{Quirico2016}. The optical constants $n$ and $k$ of this dark material are equal to 1.9 and 2.9 over the whole spectral range, respectively, and are averages of values taken from \citet{FOSTER68}.

\begin{figure*}[h!]
\begin{minipage}{17cm}
\includegraphics[angle=0, width=9.cm]{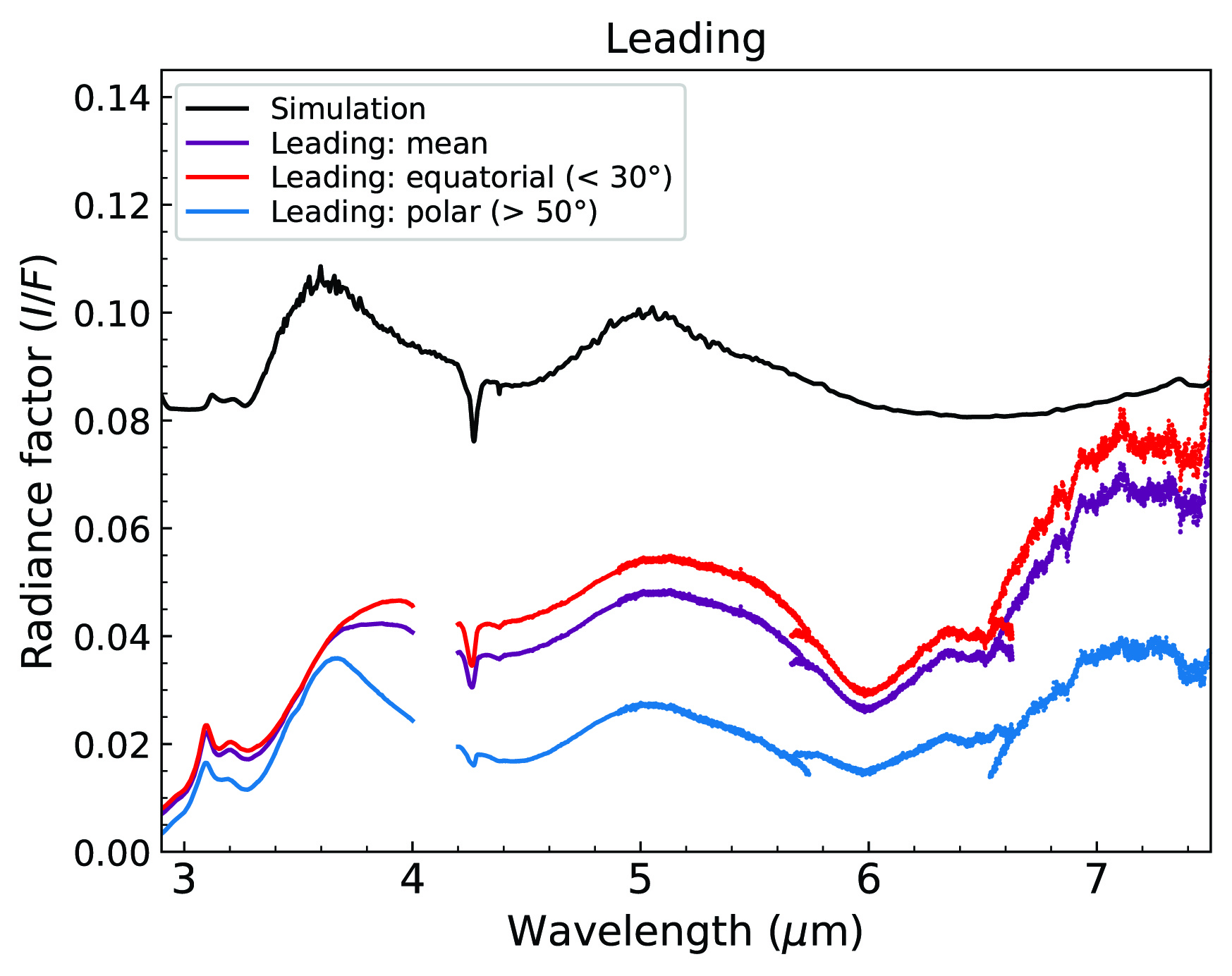}
\includegraphics[angle=0, width=9.cm]{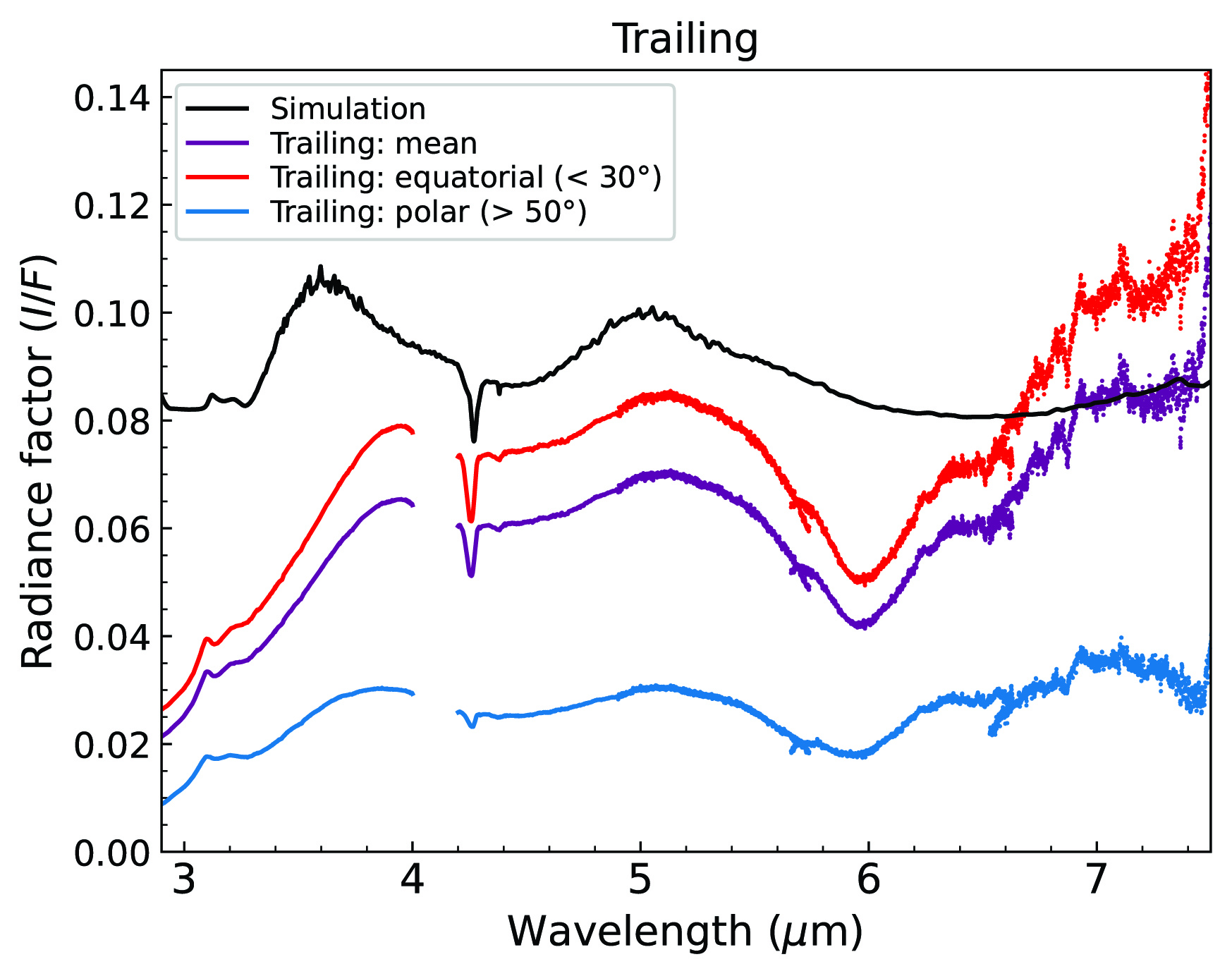}\hfill
\end{minipage}
\caption{Simulated spectrum compared with merged NIRSpec/MIRI $I/F$ spectra of the leading (left) and trailing (right) hemispheres. The model parameters are: grain sizes D$_{CO2}$ = 0.05 $\mu$m, D$_{H2O}$ = 10 $\mu$m, D$_{dark}$ = 1 $\mu$m; mass ratio relative to water m$_{CO2}$ = 1\%; relative areas of the ice:dark terrains = 10:90. Averages NIRSpec/MIRI merged spectra covering the full disk, low and high latitudes regions are shown, as indicated in the inset. NIRSpec and MIRI spectra are corrected from the contribution of thermal emission (see Appendix \ref{appendix:merged}). The simulated spectrum cannot explain the 5.9-$\mu$m feature. \label{fig:Hapkemodel_CO2}}
\end{figure*}

As mentioned above, pure CO$_2$ ice is not expected at Ganymede's surface due to its high volatility in the temperature range 90--160 K \citep{fray2009}. Therefore, optical constants of pure CO$_2$ ice are used here to estimate the CO$_2$ abundance in water ice, i.e. to reproduce the CO$_2$ band area but not its shape, width and band centre (Fig. \ref{fig:Hapkemodel_CO2}). Our simulations show that no set of parameters is able to accurately match the reflectance level, nor the shape of the continuum at low latitudes (in particular the position of the inter-band region at 3.9 $\mu$m). The misfit of the inter-band region could be due to the absence of salts in the surface model, which are likely present (see Sect. \ref{discussion_h2o_nonice}), but unfortunately, no optical constants are available \citep{Ligier2019}. The global reflectance misfit cannot be fully resolved by tuning the relative area of the dark terrain, because at some point the continuum amplitude gets too weak compared to that in observations. In addition, the reflectance of the sole dark terrain does not reach values below 0.06. A more sophisticated model is definitely needed, but despite its limitations, the abundance of CO$_2$ trapped in water ice can be estimated to $\sim$ 1 \% (in mass). This crude estimate is used below to discuss the relevance of laboratory experiments to interpret the physical state of CO$_2$.

\subsubsection{CO$_2$ in water ice \label{CO2_water}}
The latitudinal variations of the CO$_2$ band and its asymmetric band shape support the view that different CO$_2$ physical states are involved. At high latitudes, the band is dominated by a long wavelength (low wavenumber) component. At lower latitudes (and especially on dark terrains) the band broadens and shifts towards the blue, thus showing that other components of CO$_2$ dominate (Figs \ref{fig:sp-CO2}, \ref{fig:CO2-albedo}B, \ref{fig:correlations-CO2}). As pointed out above, pure CO$_2$ is not stable at the surface of Ganymede, but CO$_2$ trapped in water ice can be stabilized in the range of temperature 90--160 K.

The position of the $\nu_3$ CO$_2$ band diluted at ppm levels in crystalline water ice, i.e. isolated in the lattice, lies at 2340.7 cm$^{-1}$ at 20 K (4.2722 $\mu$m) (unpublished data from IPAG). Similarly, we made a compilation of the position of the $\nu_3$ CO$_2$ band trapped as an impurity in several ices: Ar, N$_2$, N$_2$:O$_2$, CH$_4$, C$_2$H$_4$, C$_2$H$_6$, NO and SO$_2$ (\citet{1998ASSL..227..199S}; \citet{1997Icar..127..354Q}; Unpublished data from Institut de Plan\'etologie et d'Astrophysique de Grenoble (IPAG)). The peak position varies between 2333.5 and 2348.5 cm$^{-1}$ (4.2854 and 4.2580 $\mu$m), which does not encompass the full spectral range of the CO$_2$ band in Ganymede spectra.

Infrared spectra of CO$_2$ clathrate hydrate (135 K) show two components at 2338 and 2347 cm$^{-1}$ (4.277 and 4.261 $\mu$m) \citep{Fleyfel1991}. However, the position of the most intense component lies at the edge of the CO$_2$ band in Ganymede spectra (Fig.~\ref{Speciation_CO2}), which shows that clathrates contribution is, at best, weak.

\begin{figure}[h!]
\includegraphics[angle=-90,width=9cm]{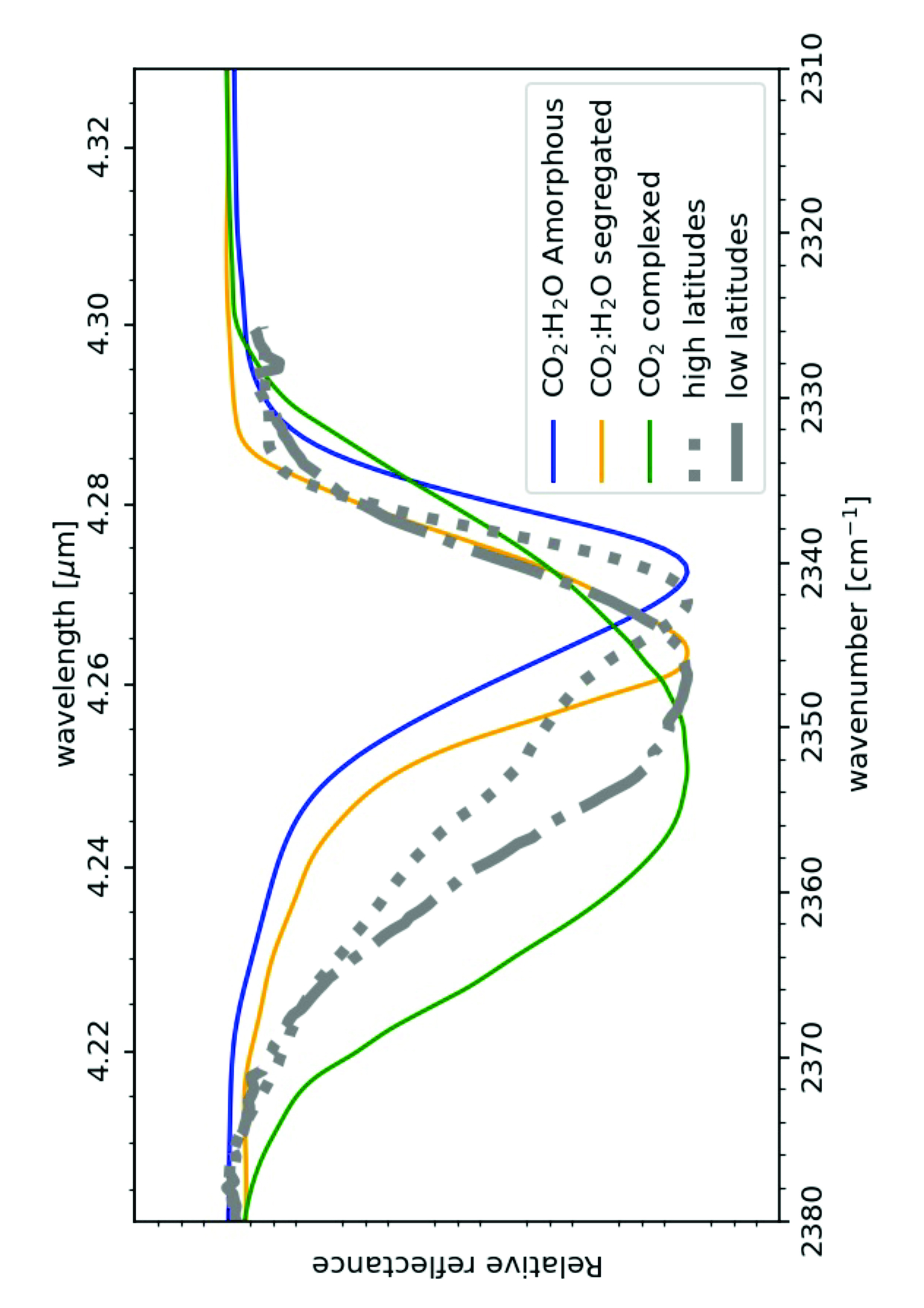}
\caption{CO$_2$ band in Ganymede spectra at the pole (60-75$^{\circ}$ N; dotted grey line) and the equator (0-15$^{\circ}$ N; dot-dashed grey line) of the leading hemisphere compared to several experimental measurements of CO$_2$ mixed with water ice at 170~K \citep{Ehrenfreund1999} (blue line), CO$_2$ segregated in water ice at 138~K \citep{He2018} (yellow line) and CO$_2$ complexed \citep{Jones2014} (green line). 
\label{CO2H2O}}
\end{figure}

\begin{figure}[h!]
\includegraphics[width=9.cm]{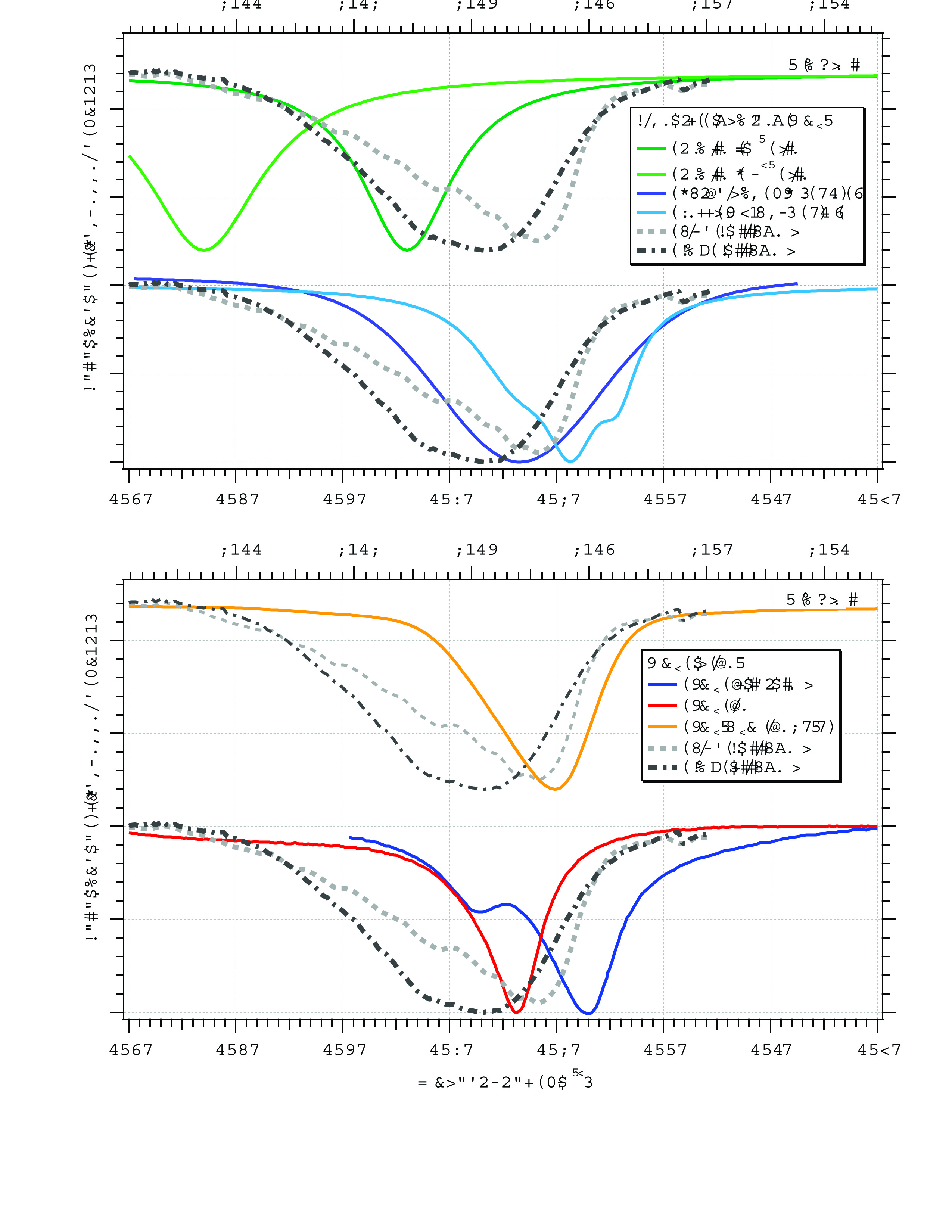}
\caption{CO$_2$ band in Ganymede spectra at the pole (60-75$^{\circ}$ N; dotted light grey lines) and the equator (0-15$^{\circ}$ N; dot-dashed grey lines) of the leading hemisphere compared with transmission spectra from \citet{Berlanga16} (carboneous chondrites), \citet{Bekhti21} (zeolites), \citet{Ehrenfreund1999} (CO$_2$ ice and CO$_2$/H$_2$O ice) and \citet{Fleyfel1991} (CO$_2$ clathrates). Note that some spectra are vertically shifted for the sake of clarity.
\label{Speciation_CO2}}
\end{figure}

Spectra of a H$_2$O:CO$_2$=10:1 ice mixture collected from 10 to 170 K \citep{Ehrenfreund1999} show that (1) carbon dioxide can be stabilised in the solid state when mixed with amorphous and crystalline water ice, over the whole range of temperature at Ganymede's surface; (2) the band is shifted towards high wavelengths (resp. low wavenumbers) when CO$_2$ is mixed with crystalline water ice; (3) even at 10\% concentration, CO$_2$ molecules do not form crystallites of pure CO$_2$, but are more likely distributed in the lattice as multimers. In spectra of H$_2$O:CO$_2$=10:1 mixtures collected at 170 K, the CO$_2$ peak is located at 2337.2 cm$^{-1}$ (4.2786 $\mu$m), at the edge of the Ganymede CO$_2$ band. For amorphous water ice at 80 K, as shown in Fig.~\ref{CO2H2O}, the band is located at 2340.6 cm$^{-1}$ (4.2724 $\mu$m), closer to the Ganymede CO$_2$ band at high latitudes. High latitudes of the leading hemisphere are precisely where amorphous water ice is detected by NIRSpec (Fig. \ref{fig:spectral-ratios}), as discussed in Sect.~\ref{h2o_polar}. 

In the study of \citet{He2018}, a H$_2$O:CO$_2$=50:100 mixture was deposited at 10 K, and subsequently warmed up. The CO$_2$ band lies between 2345 and 2350 cm$^{-1}$ (4.264 and 4.255 $\mu$m) in the range 90-138 K (Fig. \ref{CO2H2O} for 138~K), highlighting CO$_2$ segregated in the water matrix. But in this experiment, pure crystalline CO$_2$ ice is present in the sample (no pure CO$_2$ ice is observed for samples with a CO$_2$ concentration $<$ 23\%). In all these experiments, the wavenumber of the CO$_2$ band remains lower or equal to $\sim$ 2350 cm$^{-1}$ ($\lambda$ $\ge$ 4.255 $\mu$m).

The theoretical calculations of \cite{Chaban2007} show that the position of the CO$_2$ band for CO$_2$-H$_2$O complexes becomes blue-shifted by 8 to 46 cm$^{-1}$, with respect to isolated CO$_2$, and reach values well beyond 2350 cm$^{-1}$ (4.255 $\mu$m). They have been used to suggest that CO$_2$ could be present in the form of complexes at the surface of icy bodies like Iapetus, Hyperion, and Dione \citep{Cruikshank2010-pz}. However, the position of isolated CO$_2$ is predicted at 2361 and 2384 cm$^{-1}$ (4.236 and 4.195 $\mu$m), depending on the base used in DFT calculations. In addition, these positions are far from the position of CO$_2$ in the gas phase, 2349 cm$^{-1}$ (4.257 $\mu$m), which shows that these predictions of the positions of the CO$_2$-H$_2$O complexes are not reliable. Last, \citet{Jones2014} report an experimental study of CO$_2$ produced from the radiolysis of almost pure H$_2$CO$_3$. They observe a feature as broad as the CO$_2$ band in Ganymede's spectra (Fig. \ref{CO2H2O}, shown as CO$_2$ complexed), which is inconsistent with the observed presence of several spectral components.

To sum up, the main conclusion of this section is that CO$_2$ trapped in water ice cannot account for the whole CO$_2$ band observed in Ganymede spectra. Therefore, a contribution from non-ice materials must be examined. 

\subsubsection{CO$_2$ in non-ice materials}
The position of the $\nu_3$ band of CO$_2$ mixed in water ice is at wavenumbers less than 2350 cm$^{-1}$, therefore other spectral components are required to explain the whole CO$_2$ band. We discuss here the issue of CO$_2$ adsorbed or trapped inside microporosity of non-ice materials \citep{McCord1998, Hibbits07}. Linearly adsorbed CO$_2$ preserves its linear geometry (D$_{\infty h}$ symmetry) and leads to a $\nu_3$ CO$_2$ band around 4.26 $\mu$m, with position and width controlled by the nature of the surrounding cations and the local crystalline structure. In the case of zeolites doped with a variety of cations, linearly adsorbed CO$_2$ displays peak positions at around 2349 cm$^{-1}$ (K$^{+}$), 2354 cm$^{-1}$ (Na$^{+}$), 2367 cm$^{-1}$ (Ca$^{2+}$) and 2372 cm$^{-1}$ (K$^{+}$) (resp. 4.257, 4.248, 4.225 and 4.216 $\mu$m) \citep[estimated from data of][see Fig.~\ref{Speciation_CO2}]{F19898501149, Bonelli00, Martra02, Bekhti21, Taifan16}. These values account well for the high wavenumbers region of the CO$_2$ band in Ganymede spectra. We have satisfactorily fitted the CO$_2$ band in Ganymede spectra using a model based on three Lorentzian components, with peak positions covering the ranges 2340--2345, 2348--2351 and 2352--2367 cm$^{-1}$ (resp. 4.274--4.264, 4.259--.4.254 and 4.252-4.225 $\mu$m). However, zeolites are Al-rich tectosilicates and are expected to be scarce on the surface of Ganymede. Therefore, the question arises whether other minerals would display similar infrared features to zeolites. Unfortunately, far fewer studies have been devoted to mafic minerals, oxides, sulfides and salts, which could be present on the surface of Ganymede \citep{Taifan16}. The study of \citet{Berlanga16} reports adsorption measurements on a series of 13 carbonaceous chondrites, spanning a diversity of chemical groups, mineralogical compositions and post-accretional histories. The highest abundance of adsorbed CO$_2$ at 150 K was, unsurprisingly, observed with Ivuna (CI), the most finely-grained, porous and brittle object, mostly devoid of metals and sulfides. For other chondrites, the abundance of adsorbed CO$_2$ is lower by a factor of 7. The range of variations of the band position over the full range of chondrites is 2337.5--2345.2 cm$^{-1}$ (4.278--4.264 $\mu$m) and no correlation with the mineralogical composition nor organic matter abundance was found. Carbonaceous chondrites are therefore not good candidates for accounting for the long wavelength components of the CO$_2$ band. Another plausible candidate is cometary dust \citep{Poppe16}, which is essentially made of organic polyaromatic matter \citep{Bardyn17}, mafic silicates as Mg-rich olivine and pyroxenes, sulfides, metals and possibly glasses \citep{Dobrica09}. However, we lack data about the adsorption properties of these assemblages. Last, one needs to keep in mind that non-ice materials have been continuously irradiated by magnetospheric electrons and ions (H, O and S), leading to modifications of their topography and porosity, and possibly their adsorption properties. To our knowledge, no relevant data are yet available in the literature.

\subsubsection{Consistency with observations}
The correlations between the band centre and the asymmetry parameter of the CO$_2$ band on one hand, and with the bandwidth on the other hand, support the presence of several components (Fig. \ref{fig:correlations-CO2}). The CO$_2$ band in the bright northern polar cap of the leading hemisphere is very red-shifted and narrower compared to other terrains. The spectral ratios shown in Figure \ref{fig:zoom-CO2} suggest a component at 2340 cm$^{-1}$ (4.274 $\mu$m) which is very narrow and essentially present on the leading hemisphere, mostly in the polar regions. This narrow band at 2340 cm$^{-1}$ is very similar to the band of CO$_2$ mixed with amorphous water ice (Fig.~\ref{CO2H2O}). The association of the CO$_2$ responsible for this narrow band with amorphous water ice is further confirmed because the exact same area of the leading north pole has the most redshifted CO$_2$ band and the most blueshifted H$_2$O Fresnel peak position (Fig.~\ref{fig:maps-CO2}, Fig.~\ref{fig:maps-Fresnel}, Fig.~\ref{fig:correlations-CO2}), due to a higher abundance of amorphous H$_2$O ice (see Sect.~\ref{h2o_polar}, Fig.~\ref{fig:spectral-ratios}). Therefore, the CO$_2$ responsible for the redshifted component of the band is probably mixed with H$_2$O ice, and particularly with amorphous H$_2$O ice at the leading pole. Figure \ref{fig:zoom-CO2} also suggests that the other components of the band are broader and that residual CO$_2$ is associated with non-ice material(s) in the polar regions.

At lower latitudes, we observe a fairly linear correlation between the CO$_2$ band centre (expressed in wavelength) and the Bond albedo, and, for the leading side, a correlation with the area of the Fresnel peak area (Fig. \ref{fig:CO2-albedo}B, Fig.~\ref{fig:CO2-H2O}B). The contribution of CO$_2$ associated with dark materials blue-shifts the band, and also broadens it, as observed from the correlation between width and band center (Fig. \ref{fig:correlations-CO2}).

In contrast, excluding the polar regions, the variations of the CO$_2$ abundance (i.e. CO$_2$ band area; Fig.~\ref{fig:depth-CO2}) do not display significant correlations with the Bond albedo, nor with the area of the Fresnel peaks (Sec~\ref{sec:CO2-an}). The CO$_2$ abundance is possibly controlled by sputtering due to surface irradiation. The sputtering yield of CO$_2$ is higher than that of water ice, at least in the case of electronic interactions, which holds for electrons and protons \citep{Brown84,Dartois15,Seperuelo09}. It means that, when CO$_2$ is trapped within water ice, CO$_2$ is preferentially sputtered, resulting in the decrease of its abundance. In the case of CO$_2$ adsorbed on dark material, we expect as well a higher sputtering yield of CO$_2$ compared to atoms or molecular groups forming the dark materials. The flux density of ions and electrons is weaker in the equatorial regions, and, in this respect, we expect a lower sputtering efficiency at low latitudes. This is consistent with the latitudinal variations of the CO$_2$ band depth  (Fig. \ref{fig:depth-CO2}). Note that this mechanism also requires a regolith turnover that brings unirradiated grains to the top of the surface, where sputtering happens.

%%%%%%%%%%%%%%%%%%%%%%%%%%%%%%%%%%%%%%%%%%%%%%%%%%%%%%%%%
%%%%%%%%%%%%%     5.9 mic band               %%%%%%%%%%%%
%%%%%%%%%%%%%%%%%%%%%%%%%%%%%%%%%%%%%%%%%%%%%%%%%%%%%%%%%

\subsection{Origin of the 5.9-$\mu$m band}
\label{sec:5.9mu}

The MIRI spectra show an absorption feature at 5.9~$\mu$m (centred between 5.80 and 5.94 $\mu$m -- respectively 1724 cm$^{-1}$ and 1683 cm$^{-1}$ -- taking into account uncertainties), with a shoulder at 6.5 $\mu$m (1538 cm$^{-1}$) (Sect.~\ref{5umband}).

The band at 5.9 $\mu$m cannot be attributed to water ice. Indeed, the $\nu_2$ water absorption band is located at slightly longer wavelengths ($\sim$ 6.2 $\mu$m, 1613 cm$^{-1}$) and this water band is expected to be broader \citep[][]{Mastrapa2009} than the band reported in our observations (0.8 $\mu$m-width).
This is illustrated in Fig~\ref{fig:Hapkemodel_CO2} where merged $I/F$ NIRSpec/MIRI spectra of the leading and trailing hemispheres for several latitude bins are plotted together with the simulated spectrum of CO$_2$/H$_2$O mixture (Appendix~\ref{appendix:merged} explains how these merged spectra corrected for thermal radiation, including in the NIRSPec wavelength range, were obtained). In addition, the band depth of the 5.9-$\mu$m band corrected for thermal emission shows some variations but without a clear geographical trend (Fig.~\ref{fig:5mu9_band}), while the Fresnel peaks and Bond albedo, which are indicative of the abundance/properties of water ice, vary with latitude, type of terrains, and hemisphere. 
Still, water ice is present all over the surface, so it influences the MIRI spectrum to some extent. Notably, the derived reflectance at 5 $\mu$m, which is the location of a H$_2$O ice inter-band, shows a marked anticorrelation with the Bond albedo (Fig.~\ref{fig:fitted_param}, Sect.~\ref{5umband}). Moreover, the 5.9-$\mu$m band is also present all over Ganymede's surface and is slightly enhanced on the leading hemisphere, following an overall trend similar to that of water ice. Interestingly, the spectra of the polar regions of leading and trailing hemispheres exhibit marked differences in their water ice spectral features and in the shape of the 5.9-$\mu$m band as well (Fig.~\ref{fig:Hapkemodel_CO2}). In addition, on the leading hemisphere the 5.9 $\mu$m band position (Figs~\ref{fig:5mu9_band_only_model2},~\ref{fig:5mu9_band}) seems correlated with H$_2$O ice Fresnel peak area and inter-band position (Fig.~\ref{fig:maps-Fresnel}, Fig.~\ref{fig:maps-interband}). Therefore, although the 5.9-$\mu$m band is not attributed to water ice, it may be influenced by the presence of water ice, and it might be possible that the component responsible for it is associated to H$_2$O ice.

The 5.9-$\mu$m band falls in the 1750--1650 cm$^{-1}$ (5.714--6.061 $\mu$m) region where the carbonyl C=O stretching mode has a strong absorption band. Other double bonds like C=C and C=N have absorptions in lower frequency regions of about 1500--1650 cm$^{-1}$ (6.667--6.061 $\mu$m). The assignment of the 5.9-$\mu$m band to such compounds is challenging as no other unattributed functional groups (e.g., C-H or N-H stretching modes) are seen in NIRSpec spectra.

Organic materials in carbonaceous chondrites and cometary dust, in particular the insoluble fraction, display a C=O band around 1710 cm$^{-1}$ (5.8480 $\mu$m), along with features in the range 2800--3000 cm$^{-1}$ (3.571--3.333 $\mu$m) (CH$_2$ and CH$_3$ stretching modes), at 1600 cm$^{-1}$ (6.250 $\mu$m) (aromatic C=C), 1450 cm$^{-1}$ (6.897 $\mu$m) (CH$_2$ deformation), 1380 cm$^{-1}$ (7.246 $\mu$m) (CH$_3$ deformation) and a broad congested feature centred at $\sim$ 1200 cm$^{-1}$ (8.333 $\mu$m). Carbonaceous chondrites that have experienced short-duration thermal metamorphism, presumably controlled by hypervelocity impacts, have a much lower aliphatic abundance resulting in a weak 2800-3000 cm$^{-1}$ band (\citealt{Flynn03, Kebukawa11, Orthous-Daunay13, Quirico18}). However, they also often display a C=O band of lower intensity, and, in addition, the width of the C=O band is significantly lower than that of the 5.9-$\mu$m band. It is then unlikely that this feature is due to chondritic or cometary insoluble organic matter.

We also investigated whether these features could be attributed to carbonic acid (H$_2$CO$_3$), which has been predicted to be present on icy moons for decades (\citealt{Carlson2005,Johnson2004,Peeters2010,Iopollo2021}), since it is an irradiation product of CO$_2$ and water ice mixtures (\citealt{Moore1991,Zheng2007,Jones2014,Brucato1997,Gerakines2000}). H$_2$CO$_3$ has an absorption band assigned to the C=O stretch at 1719 cm$^{-1}$ (5.817 $\mu$m) and the C-OH asymmetric stretch at 1508 cm$^{-1}$ (6.631  $\mu$m) \citep{Gerakines2000}, i.e. nearby the wavelength of the observed features. However, it presents a strong absorption band around 2600 cm$^{-1}$ (3.846 $\mu$m), which is not observed in NIRSpec spectra. In addition, the band strengths of the above mentioned  mid-IR bands  have been reported to be of 11 $\pm$ 1 and 6.5 $\pm$ 0.6 $\times$ 10$^{-17}$ cm molecule$^{-1}$, respectively \citep{Gerakines2000}. These band strengths are strong and of the order of the band strength of pure CO$_2$ at 2342 cm$^{-1}$ ($4.2699 \mu$m) of 7.6 $\times$ 10$^{-17}$ cm molecule$^{-1}$ (\citet{Gerakines1995}) and of 7.6$\times$  10$^{-17}$-1.1$\times$ 10$^{-16}$ in other studies (\citet{bouilloud2015}). The area of the 5.9-$\mu$m band is tens to hundred times larger than the area of the CO$_2$ band. Since the band strengths are of the same order for CO$_2$ and H$_2$CO$_3$, this would imply that H$_2$CO$_3$ would be a dominating species on the entire surface of Ganymede, and that CO$_2$ would be only a small fraction of H$_2$CO$_3$. H$_2$CO$_3$ is very unstable to radiation \citep{Gerakines2000,Jones2014}, which makes this hypothesis unlikely.

Another possible carrier of the 5.9-$\mu$m bands is non-linearly physisorbed CO$_2$. In that case, the $\nu_3$ band shifts towards low wavenumbers, covering a broad range of values (1780-1340 cm$^{-1}$, resp. 5.618-7.463 $\mu$m) depending on the type of binding arrangement and the nature of the cations. The amplitude of the shift correlates with the number of metal cations to which the molecule is bonded \citep{Taifan16}. As the CO$_2$ molecule is bent, its symmetry changes from D$_{\infty h}$ to C$_{2v}$ and the $\nu_1$ symmetric stretching mode becomes infrared active, with a range of values of 1340-980 cm$^{-1}$ (7.463-10.204~$\mu$m). Complicating this picture, CO$_2$ can also be chemically adsorbed into carbonates, through a variety of binding configurations (free symmetrical, monodentate, bidentate, polydentate). Those different speciations lead to strong spectral shifts and splitting depending both on the nature of cations and binding arrangements, compared to the positions of the $\nu_2$, $\nu_3$ and $\nu_4$ vibration modes of isolated carbonates \citep{farmer_infrared_1974}. Therefore, interestingly, the bands observed in MIRI spectra at $\sim$ 5.9 and 6.5 $\mu$m could be controlled by non-linearly adsorbed or chemically adsorbed CO$_2$. As discussed before, it is not possible to assign them univocally to specific cations and binding arrangements, and dedicated experiments are required. Additional bands due to physisorbed or chemisorbed CO$_2$ are not expected in the MIRI/NIRSPec spectral ranges. 

H$_2$SO$_4$.8H$_2$O has been proposed to be present at the surface of Ganymede, as this compound leads to a better fit of the water bands in the near-infrared region (\citealt{Shirley10, Ligier2019,King2022-is}). The H$_2$SO$_4$/H$_2$O phase diagram displays several sulfuric acid hydrates, depending on H$_2$SO$_4$ concentration: monohydrate, dihydrate, trihydrate, tetrahydrate, heptahydrate and octahydrate \citep{Kinnibrugh22}. The spectra of monohydrate, tetrahydrate and octahydrate have been determined experimentally (\citealt{Nash2001, Beyer03}). The spectra show a group of bands between 800 and 1350 cm$^{-1}$ (12.50--7.41 $\mu$m), associated with S-O bonds; a band in the range 1698--1720 cm$^{-1}$ (5.889--5.814 $\mu$m) assigned to a deformation mode of H$_3$O$^+$ or H$_2$O$_5^+$; and bands in the range 2800--2900 cm$^{-1}$ (3.571--3.448 $\mu$m), due stretching modes of H$_3$O$^+$, H$_2$O$_5^+$ or OH. Except at boundaries between different stability fields of those hydrates, a H$_2$SO$_4$-H$_2$O mixture is made of two different hydrates (except at the edges of the phase diagram, where one hydrate is mixed with either H$_2$O ice or solid H$_2$SO$_4$). That said, the position of the 5.9-$\mu$m band in Ganymede (in the range 1678--1718 cm$^{-1}$ (5.96--5.82 $\mu$m), considering uncertainties and both hemispheres, Sect.~\ref{5umband}) is consistent with the band positions at 1698, 1715 and 1720 cm$^{-1}$ (5.889, 5.831 and 5.814 $\mu$m) for octa (SAO), tetra (SAT) and monohydrates (SAM), respectively. In addition, its width is quite similar to those in experimental spectra (around 170 cm$^{-1}$). However, according to \citet{Nash2001}, features are present between 2800 and 2900 cm$^{-1}$ (resp. 3.571 and 3.448 $\mu$m) due to stretching modes in H$_3$O$^+$, H$_2$O$_5^+$ and OH bonded to the S-O core. In fact, these assignments are unclear. We have reanalyzed the spectra of SAM, SAT and SAO from \citet{Nash2001} and \citet{Beyer03}, and found several arguable interpretations. The spectrum of SAM displays a broad feature centred at 2881 cm$^{-1}$, however, it does not fit with the spectrum of the H$_3$O$^+$-rich HCl ice, which peaks at 3250 cm$^{-1}$. The spectra of SAO and SAT display a broad feature centred at 3250 cm$^{-1}$, but this feature looks more similar to water ice. All in all, we get to the conclusion that (i) the assignment of these broad structures is unclear, (ii) but they are broad and likely share similarities with water ice in terms of position (3250 cm$^{-1}$, i.e., 3.076 $\mu$m) and broadness. As a consequence, we expect them to essentially contribute to the continuum along with water ice. Last, as discussed at the beginning of this section, the component responsible for the 5.9-$\mu$m band might be associated with H$_2$O ice, favouring hydrates, and possibly H$_2$SO$_4$ hydrate(s), over other non-ice candidates.

In conclusion, the 5.9-$\mu$m band in Ganymede spectra can be tentatively assigned to a sulfuric acid hydrate. A firmer assignment would require more robust experimental data.

\subsection{H$_2$O$_2$} 

H$_2$O$_2$ is present all over Ganymede's surface, with a higher abundance in the polar regions (\citealt{Trumbo2023}), which are the regions where the water ice is the most abundant and/or have the smallest optical mean free path (Sect.~\ref{h2o_polar}, Fig.~\ref{fig:maps-interband}). These observations are consistent with H$_2$O$_2$ being a product of water ice radiolysis, enhanced in the open field line regions \citep{Trumbo2023}. However, the H$_2$O$_2$ and CO$_2$ maps are not well correlated (see CO$_2$ map in Fig.~\ref{fig:maps-CO2} and H$_2$O$_2$ map in \citealt{Trumbo2023}), making CO$_2$ abundance unlikely to influence H$_2$O$_2$ production. Moreover, although the CO$_2$ responsible for the red-shifted component of the band, interpreted as CO$_2$ associated with water ice, is enhanced on the polar regions of the leading, like H$_2$O$_2$, there is no exact pixel-to-pixel match between the locations of both species. In addition, on the trailing hemisphere, the red-shifted CO$_2$ component is enhanced on the polar regions, whereas the H$_2$O$_2$ is not.

As described in \citet{Trumbo2023}, the distribution of H$_2$O$_2$ on Ganymede (H$_2$O$_2$ abundance enhanced on the most ice-rich and coldest regions) is at odds with that observed on Europa (H$_2$O$_2$ abundance enhanced on the most ice-poor and warmest regions) (\citealt{spencer1999,rathbun2010,trumbo2018,trumbo2019}). The enhanced abundance of H$_2$O$_2$ on Europa's chaos terrains has been hypothesized to be related to the presence of CO$_2$ or possibly other electron-accepting ice contaminants, which may scavenge destructive electrons created during the irradiation of the ice, thereby slowing down the destruction of H$_2$O$_2$ (\citealt{trumbo2019,Trumbo2023}). Ganymede's NIRSpec observations do not support CO$_2$ to be an electron scavenger enhancing H$_2$O$_2$. However, the production/destruction pathways of H$_2$O$_2$ may be different on Europa and Ganymede, and/or other species might play the role of electron scavengers.

\section{Summary}

Ganymede Leading and Trailing hemispheres were observed with the NIRSpec/IFU (2.9--5.3 $\mu$m) and MIRI/MRS (unsaturated data from 4.9--11.7 $\mu$m) instruments of the JWST, as part of the ERS program 1373 "ERS observations of the Jovian System as a demonstration of JWST's capabilities for Solar System science". The aim was to investigate the composition and thermal properties of the surface, as well as its exosphere. These observations led to the first detection of  H$_2$O$_2$, published by \citet{Trumbo2023}. Results on the exosphere will be reported elsewhere. Although the SNR was the highest ever obtained in the near-IR for Ganymede, minor signatures observed with Galileo/NIMS and Juno/JIRAM at 3.3, 3.4, 3.88, 4.38 and 4.57 $\mu$m were not confirmed. In this paper, we focused on the distribution and spectral properties of CO$_2$ and water ice signatures observed with NIRSpec, and on the analysis of the MIRI data. The main results are:

\begin{itemize}
\item Maps of the H$_2$O Fresnel peak area at $\sim$3.1 $\mu$m are consistent with the distribution of water ice inferred from H$_2$O bands at 4.5 $\mu$m and in the near-IR \citep{Ligier2019,King2022-is}, and correlate with Bond albedo maps. Water ice is the most abundant on the polar regions of the leading hemisphere. On the trailing side, the Fresnel peak area shows a minimum  slightly eastward (by 9\dg) from the central longitude (270\dg), increasing around this minimum with a bull's-eye distribution, with some excess at the poles. An excess of ice is also observed on the trailing morning limb, which may trace the formation of ice crystals during the night, from recondensation of water subliming from the warmer subsurface.  
\item At low to mid latitudes, the central wavelength of the H$_2$O Fresnel peaks shifts with longitude, and this is observed for both the leading and trailing sides. This shift seems to indicate a progressive and cumulative modification of the H$_2$O ice molecular structure (via amorphisation, other phase changes, chemical changes?) and/or nano/micrometre-scale texture (via sublimation/re-condensation/photodesorption?), induced by the accumulation of solar radiation from the morning to the evening, and which is reversible during the night (via re-condensation, metamorphisation, phase change, chemical change?). Dedicated studies are needed to constrain the origin of this longitudinal modification of the surface ice.
\item Amorphous water ice, probably formed by irradiation from Jupiter's magnetospheric energetic particles, is detected in the polar regions, where temperature conditions block the kinetics of water ice recrystallization. The polar regions of the leading side contain more amorphous ice than those of the trailing side, and the boreal region of the leading hemisphere is the richest.
\item As expected, maps of the wavelength position of the H$_2$O inter-band (in the range 3.65-3.95 $\mu$m) correlate with the distribution and abundance of water ice inferred from the band at 4.5 $\mu$m and the Fresnel peak area. The wavelength position of the inter-band in dark terrains seems to be compatible with the presence of Na-/Mg- sulfate salts, sulfuric acid hydrates, and possibly phyllosilicates. The positive (red) spectral slope of Ganymede's dark/equatorial terrain spectrum continuum from 3.9 to 7 $\mu$m is consistent with the mixing of such compounds with fine-grained opaque minerals having a highly porous surficial micro-texture. However, no absorption bands that would enable firm identification of compounds specific of Ganymede's dark terrain are observed.
\item The CO$_2$ 4.26 $\mu$m band displays latitudinal and regional variations in band center and band shape over the two hemispheres, which indicate that CO$_2$ is present in different physical states on the surface. The band center and shape are correlated with the Bond albedo and H$_2$O ice abundance.
\item The CO$_2$ band depth shows a latitudinal gradient, from 8 to 22\% on the trailing side, and from 11 to 21\% on the leading side with the lower values in polar regions. The CO$_2$ band depth does not show significant correlation with Bond albedo and ice content in the equatorial regions. 
\item In the ice-rich polar regions, which are the most exposed to Jupiter's plasma irradiation, the CO$_2$ band is redshifted with respect to other terrains. In the boreal region of the leading hemisphere, the CO$_2$ band is dominated by a high wavelength component at $\sim$ 4.27 $\mu$m, consistent with CO$_2$ trapped in amorphous water ice. In the same region, amorphous water ice is observed. Therefore, the formation and/or trapping of this CO$_2$ may be linked to the formation and/or  structure of the amorphous water ice.
\item At equatorial latitudes (and especially on dark terrains) the observed band broadening and blueshift suggest CO$_2$ adsorbed or trapped on non-icy materials, such as minerals or salts.
\item From Hapke modelling the abundance of CO$_2$ trapped in water ice is roughly 1\% in mass.
\item Clathrate hydrates cannot explain the characteristics of the CO$_2$ band, and are at best a minor contributor.
\item The brightness temperature from 7 to 11 $\mu$m exhibits a monotonic (and quasi-linear) decrease with increasing wavelength, by over 10 K at disk center, which may be due to a spectral emissivity $<$1, and/or an emissivity that decreases with increasing wavelength, and/or a Planck-weighted mixing of a variety of surface temperatures within the PSF.
\item From thermophysical modelling, the distribution of the brightness temperatures at 8.25 $\mu$m indicates a rough surface at scales of 0.1 mm-10 cm, with mean slope angles of 15\dg-25\dg and thermal inertias in the range $\Gamma$ = 20–40 SI, with no obvious difference between leading and trailing sides.
\item The MIRI spectra show an absorption feature at 5.9 $\mu$m, present all over Ganymede's surface, with a depth in the reflected sunlight component ranging typically over 35-54 \% on the leading side, and 30-52 \% on the trailing side, and a width of $\sim$0.8 $\mu$m, with evidence for an absorption shoulder at 6.5 $\mu$m (more pronounced on the leading side). The 5.9~$\mu$m band cannot be attributed to the $\nu_2$ band of water ice, expected to be broader and at a slightly higher wavelength. The features cannot be assigned to carbonic acid H$_2$CO$_3$, predicted to be present on icy moons since decades. These bands could be controlled by non-linearly physisorbed or chemically adsorbed CO$_2$. Sulfuric acid hydrates H$_2$SO$_4$.nH$_2$O appear to be good candidates to explain the 5.9-$\mu$m band.
\item H$_2$O$_2$ is enhanced in Ganymede's polar regions, where it is probably produced by water ice radiolysis. These observations do not support CO$_2$ to be an electron scavenger enhancing H$_2$O$_2$ on Ganymede, contrary to what was hypothesized on Europa.
\end{itemize}

The spectral properties of the polar regions are very different for leading vs. trailing sides. The leading polar caps have less exposed non-ice materials, a smaller optical mean free path in water ice (indicative of a larger external or internal surface such as cracks or pores), more amorphous water ice, as well as deeper CO$_2$ (more red-shifted) and H$_2$O$_2$ absorption bands than trailing poles. The origins of these differences remain to be investigated, in particular considering leading/trailing asymmetries in Jovian plasma precipitation and micro-meteoroid gardening, and how these processes influence the icy surface physico-chemical transformations.

Radiative transfer modelling and experimental measurements are needed to provide deeper data interpretations. From an observational perspective, this JWST investigation has shown that observations designed to investigate diurnal variations of Ganymede's surface properties may unravel unexpected processes. Altogether the results obtained in this study will certainly help in optimizing the observation strategies of the Moons And Jupiter Imaging Spectrometer (MAJIS) onboard the ESA/JUICE mission which will explore Ganymede further.

\begin{acknowledgements} 
This work is based on observations made with the NASA/ESA/CSA James Webb Space Telescope. The data were obtained from the Mikulski Archive for Space Telescopes at the Space Telescope Science Institute, which is operated by the Association of Universities for Research in Astronomy, Inc., under NASA contract NAS 5-03127 for JWST. These observations are associated with program 1373, which is led by co-PIs Imke de Pater and Thierry Fouchet and has a zero-exclusive-access period. IdP, MHW, PMF, MRS are in part supported by the Space Telescope Science Institute grant nr. JWST-ERS-01373. The work of O.P., E.Q. and B.S. was in part supported by the CNES. D.B.-M, E.Q., E.L., T.F, and O.P. acknowledge support from the French Agence Nationale de la Recherche (program PRESSE, ANR-21-CE49-0020-01). L.N.F. was supported by a European Research Council Consolidator Grant (under the European Union’s Horizon 2020 research and innovation programme, grant agreement No 723890) at the University of Leicester. G.S.O. was supported at the Jet Propulsion Laboratory under contract with the National Aeronautics and Space Administration (80NM0018D004).

\end{acknowledgements}

  \bibliographystyle{aa} % style aa.bst
   \bibliography{ganymedebiblio-v2} % your references Yourfile.bib

%-------------- Appendix-------------------------------------------
\begin{appendix}

\section{Bull's eye structure of Fresnel peak area distribution on the trailing hemisphere}
\begin{figure}
\includegraphics[width=9.cm]{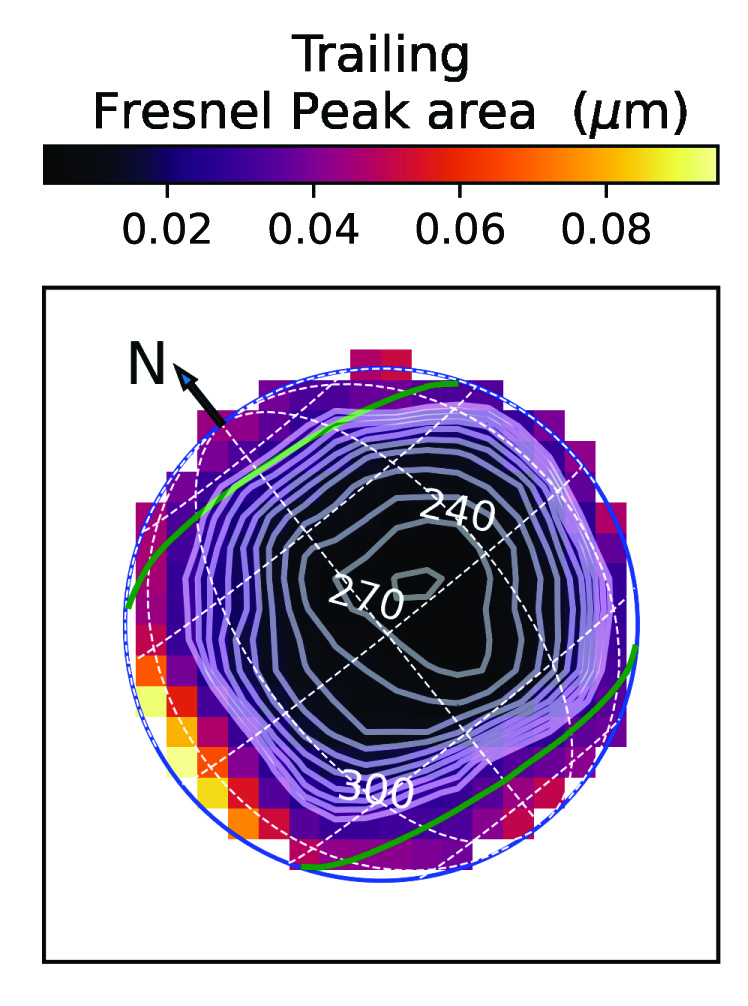}
\caption{The distribution of the Fresnel peak area on the trailing hemisphere with iso-contours (levels spaced by 0.0025 $\mu$m with the central level at 0.0035 $\mu$m). \label{fig:bullseye-trailing}}
\end{figure}

\section{Correlations between CO$_2$ band parameters and Fresnel peak area}
\label{appendix:CO2-H2O}

\begin{figure}
\includegraphics[width=9.cm]{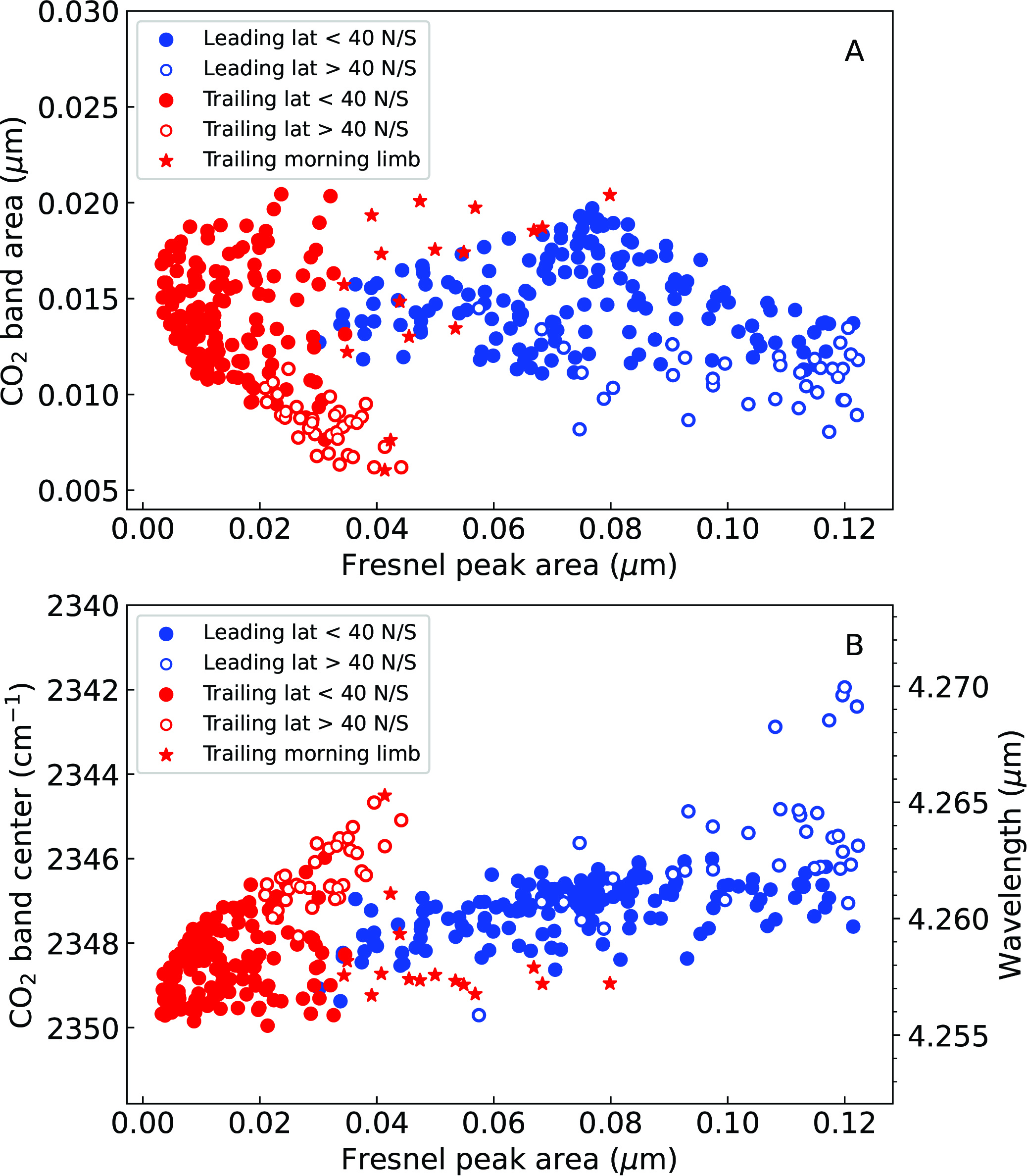}
\caption{CO$_2$ band area (A) and band center (B) as a function of Fresnel-peak area (i.e. $EqW$). Data for the leading and trailing hemispheres are shown with blue and red symbols, respectively. Filled and open dots are for latitudes lower and higher than 40$^{\circ}$ N/S, respectively. The data in the morning limb of the trailing hemisphere (namely longitudes > 320\dg) are shown with the star symbol. The positive correlation between CO$_2$ band center (in wavelength) and Fresnel peak area has a Spearman's rank coefficient $r$ =  0.62 and 0.52 with a significance of 6.6$\sigma$ and 6.9$\sigma$ for leading and trailing hemispheres, respectively. A weak negative correlation between  CO$_2$ band area and Fresnel peak area is present when considering all data ($r$ = -0.31 (4.5$\sigma$) and $r$ = -0.53 (6.6 $\sigma$) for leading/trailing hemispheres respectively), but not when considering only equatorial ($<$ 40\dg) latitudes ($r$ $\sim$ 0 (0$\sigma$) and $r$ = -0.15 (1.8 $\sigma$), for leading/trailing hemispheres, respectively). These calculations of correlation coefficients exclude data on the morning limb of the trailing hemisphere. \label{fig:CO2-H2O}}
\end{figure}

\section{Additional figures describing MIRI data fits}

       \begin{figure*}
\vspace*{-0cm}
   \includegraphics[angle=0,width=18cm]{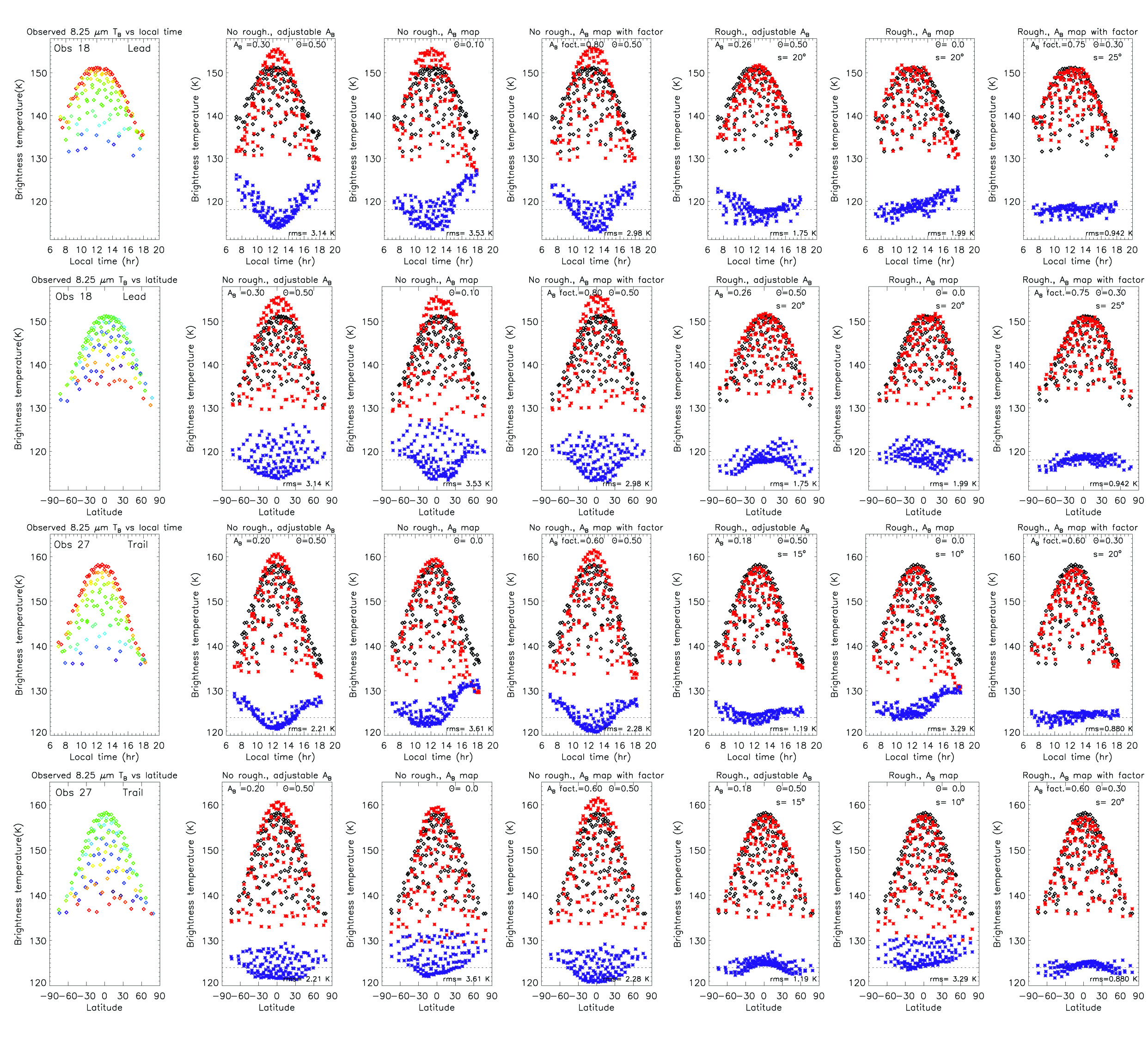}
\vspace*{-0.25cm}
   \caption{8.25 $\mu$m brightness temperature (T$_B$) for Obs. 18 (first two rows) and Obs. 27(last two rows), plotted as a function of local time (first and third rows) or latitude (second and fourth rows). In each row, observational data (first column) are compared to a series of models. In the first column, the data are color-coded according to latitude (i.e. to equal values of the latitude difference from the sub-observer point) in the T$_B$ vs local time plots, and according to local time in the T$_B$ vs latitude plots. The following six column describe the six considered models, with the three options for the Bond albedos and inclusion or not of surface slopes (see text for details). In those panels, data are in black, models in red, and the data minus model difference in blue, with the 0 value indicated by the dashed line. Parameter values and rms of fit are indicated in each panel. }
   \label{fig:thermalfits}
    \end{figure*}

       \begin{figure*}
\vspace*{-3cm}
   \includegraphics[angle=0,width=18cm]{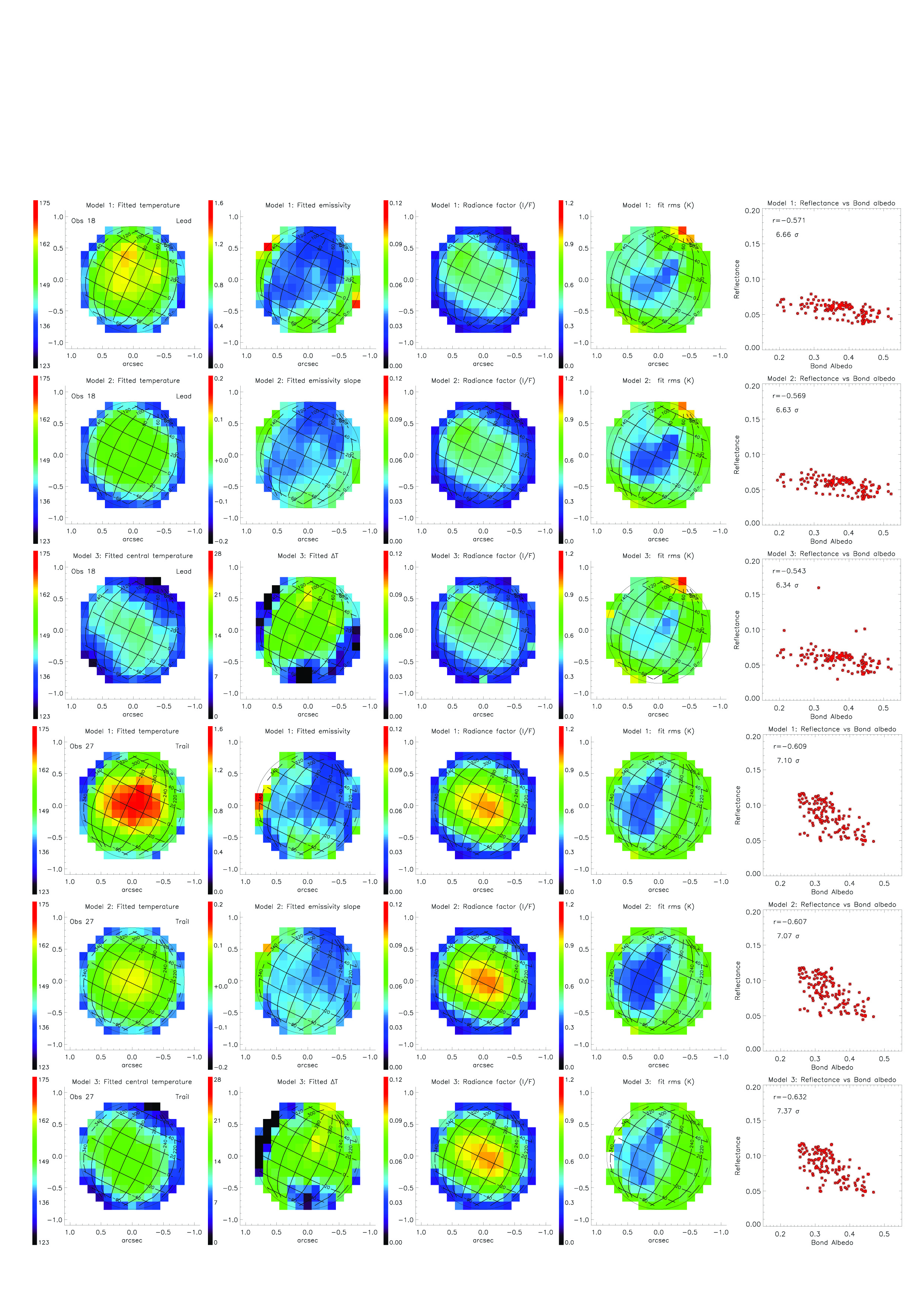}
\vspace*{-0.75cm}
   \caption{Fitted model parameters for Obs. 18 (Leading, first three rows) and Obs. 27 (Trailing, last three rows), for the three models described in Sect.~\ref{5umband}: temperature (T$_1$, T$_2$, or T$_3$), second fit parameter ($\varepsilon_1$, $\varepsilon$$^\prime_2$, or $\Delta$T), $I/F$ radiance factor (Radf$_1$, Radf$_2$ or Radf$_3$) and corresponding fit rms (fourth column). Fifth column: correlation between the reflectance factor deduced from $I/F$ using the Oren-Nayar model (see text) and the Bond albedo. Spearman correlation coefficients and number of standard deviations significance are indicated as insets.}
   \label{fig:fitted_param}
    \end{figure*}
    
       \begin{figure*}
\vspace*{-3cm}
   \includegraphics[angle=0,width=18cm]{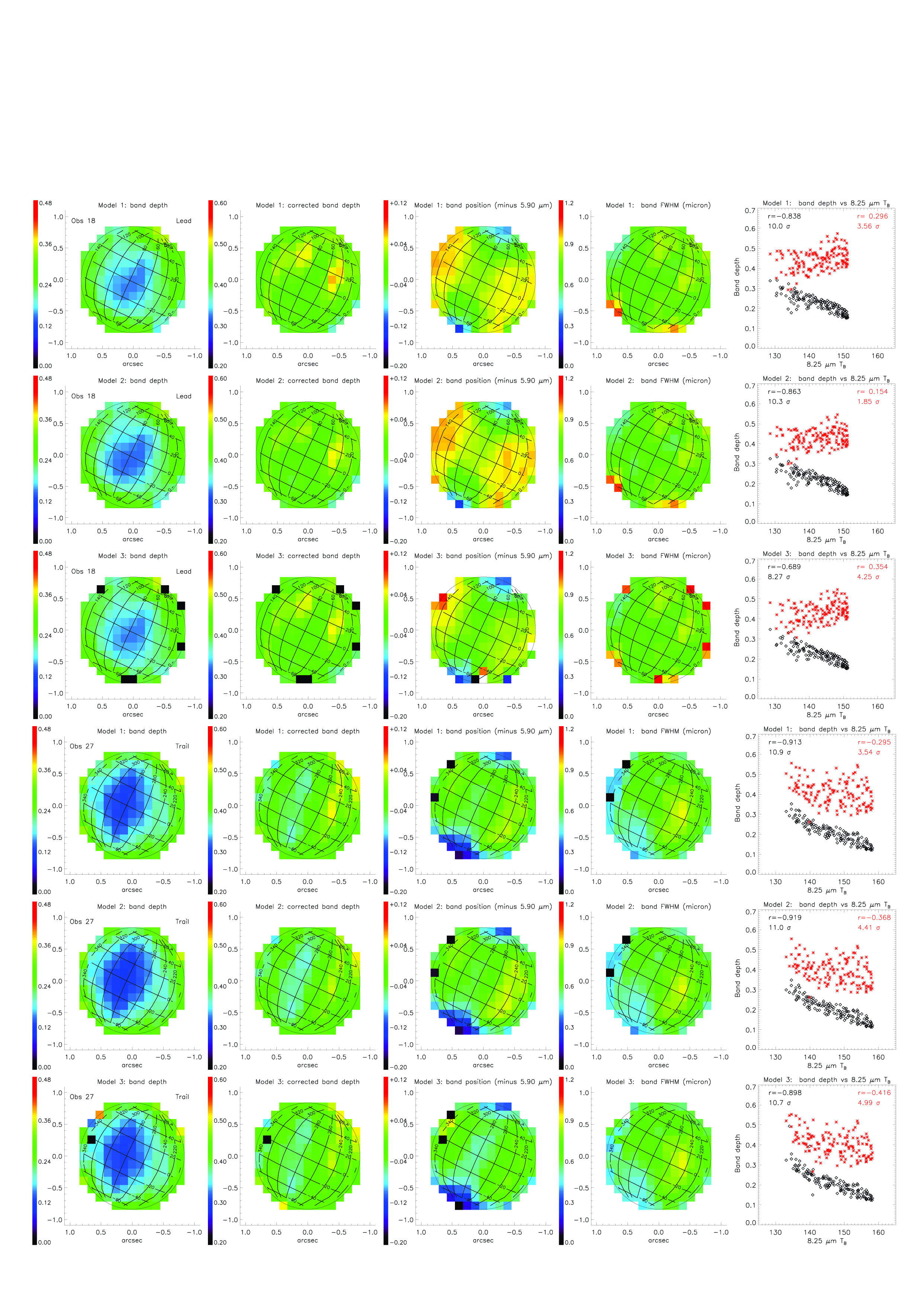}
\vspace*{-0.75cm}
   \caption{Fitted characteristics of the 5.9-$\mu$m band in Obs. 18 (leading side, first three rows) and Obs. 27 (trailing side, last three rows), for the three models described in the text: observed band depth relative to total continuum, band depth corrected for thermal contribution (i.e. relative to model solar reflected component), band position (offset from 5.90 $\mu$m), and band FWHM width. The last column shows the correlation between the observed (black) and corrected (red) band depths with the observed 8.25 $\mu$m T$_B$. Spearman correlation coefficients and number of standard deviations significance are indicated as insets.}
   \label{fig:5mu9_band}
    \end{figure*}

\section{Estimation of polar H$_2$O ice temperature from 3.6-$\mu$m inter-band position}
\label{appendix:h2o_ice_temp}
\begin{figure}
\includegraphics[width=9.cm]{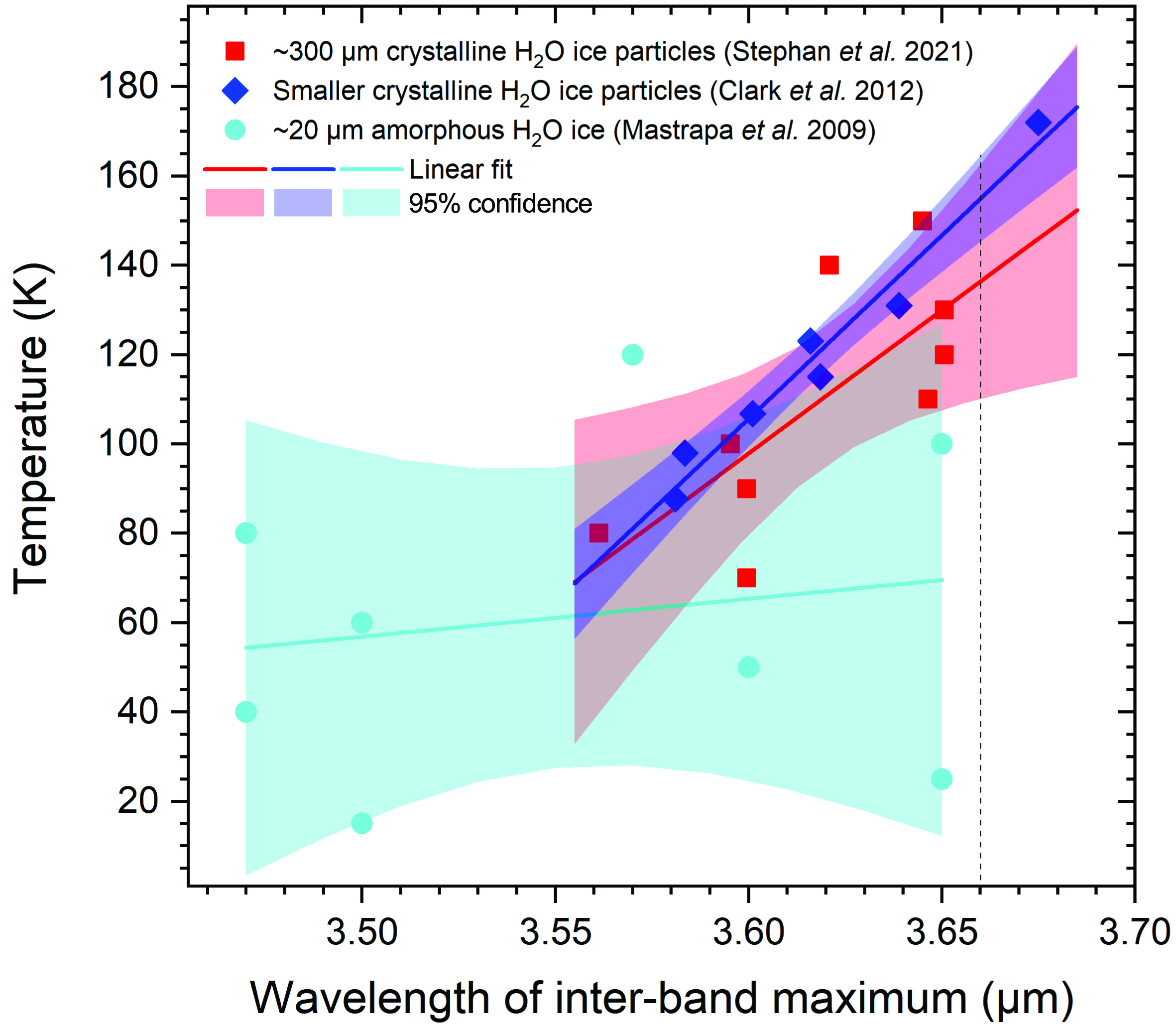}
\caption{Correlation between the temperature of pure H$_2$O ice and the wavelength at which the reflectance bump of the 3.6-$\mu$m H$_2$O ice inter-band is maximum. For pure crystalline H$_2$O ice, the laboratory reflectance spectra are from \citet{Clark2012} and \citet{Stephan2021}. For amorphous ice, points have been estimated by reflectance spectra with \cite{Mastrapa2009} optical constants, and applying a shift of 0.1 $\mu$m.
\label{fig:h2o_ice_temp}}
\end{figure}

The correlation between the temperature of pure H$_2$O ice and the wavelength at which the reflectance bump of the 3.6-$\mu$m H$_2$O ice inter-band is maximum is represented in Fig. \ref{fig:h2o_ice_temp}. For crystalline ice, laboratory reflectance from \citet{Clark2012} and \citet{Stephan2021} are reported. For amorphous H$_2$O ice, the reflectance spectra for ice ranging between 15~K and 120~K are represented. To do so, we used the optical constants of amorphous H$_2$O ice \citep{Mastrapa2009} to predict the inter-band spectra in reflectance \citep{1993tres.book.....H}, and applied a shift similar than the shift observed for crystalline H$_2$O ice between calculated reflectance (with optical constants from \cite{Mastrapa2009}) and laboratory spectra \citep{Clark2012}. By considering that this shift (of 0.1 $\mu$m) is similar for amorphous and crystalline ice, the reflectance peaks simulated for amorphous ice are presented. As shown in the figure, these maxima are mostly located at lower temperatures than for crystalline ice. We here therefore assume that the inter-band maxima for amorphous ice are located at lower wavelengths than those for crystalline ice. 

\section{Preparation of Aguas Zarcas hyperfine and hyperporous sample}
\label{appendix:aguas_zarcas}

Mid-infrared reflectance spectra are extremely dependent on the surface texture (grain size, porosity) (\citealt{salisbury1992}), and we know from recent observations the extreme surface porosity of some Ganymede's dark terrains \citep[85\%,][]{2021PSJ.....2....5D}. 

Therefore, we have tested how such extreme porosity influences the reflectance spectrum of a dark material of composition potentially analogue to Ganymede's dark terrains. We chose the Aguas Zarcas (CM) meteorite, which is a recent fall (2019). It is mainly made of phyllosilicates constituting a fine-grained matrix, hosting anhydrous silicates, calcite, opaque minerals (iron sulfides, magnetite etc.), and salts among other constituents (\citealt{garvie2021}). The fine-grained opaque minerals appear to be responsible for the low albedo of the meteorite, and possibly of many primitive bodies such as comets or P-/D-type asteroids \citep{Quirico2016}. Moreover, we know that the degree of dispersion/agglomeration of sub-micrometer-sized grains of opaques and silicates (hydrated or not) controls their spectroscopic properties from visible to mid-infrared wavelengths (\citealt{sultana2023}). Therefore, we have produced a hyperfine powder (i.e., made of sub-micrometer-sized grains) of the Aguas Zarcas meteorite, and we have made a hyperporous (>80\%) sample of this powder to measure its mid-infrared reflectance spectrum.

The detailed experimental protocol for the production of the hyperfine and hyperporous sample is given in \citet{sultana2021}. Aguas Zarcas was grounded in sub-micrometer-sized grains using a planetary ball mill and following a series of grinding and sieving steps. This hyperfine powder was then mixed in liquid water and produced frozen droplets of this mixture, which were put in a thermal-vacuum chamber for the water to sublimate. After complete sublimation of the water, a hyperporous (>80\%) sublimation lag deposit was formed, made of the Aguas Zarcas sub-$\mu$m grains arranged in a highly porous structure. The spectrum of this lag deposit shown in Figure~\ref{fig:phyllo-zarcas}b exhibits a strong positive (red) spectral slope reminiscent of the spectral slope of the continuum of Ganymede's dark terrains from 3.9 to 7 $\mu$m (dashed-dotted gray line in Fig.~\ref{fig:phyllo-zarcas}b). The radiative transfer mechanism explaining how an increase of porosity induces a spectral reddening is not yet established (it will be the subject of an upcoming study), but it requires the presence of sub-$\mu$m opaque grains such as sulfides.

Aguas Zarcas spectrum exhibits absorption bands centered at 2.8-2.9 $\mu$m due to OH/H$_2$O in phyllosilicates, at 6.09 $\mu$m due to adsorbed H$_2$O and at 6.89 $\mu$m due to carbonates.

\section{Merged NIRSpec/MIRI spectra}
\label{appendix:merged}

Thermal emission affects to some extent the continuum emission in the long wavelength part of NIRSpec NRS2 spectra. We used the models developed for analysing the MIRI observations to estimate the thermal contribution at the longest NIRSpec wavelengths. For that, we extrapolated the calculations of the thermal and reflected components down to 4 $\mu$m, yielding the thermal to solar reflected ($TH$/$SR$) ratio as a function of wavelength and $I/F$. In doing so, we used Model 2 of Sect.~\ref{5umband}, and assumed unit spectral emissivity shortwards of 4.9 $\mu$m. This estimate was done for spectra averaged over the whole disk, equatorial and polar latitudes (Fig.~\ref{fig:Hapkemodel_CO2}). For these various spectra shown in Fig.~\ref{fig:Hapkemodel_CO2} (and also in Figs~\ref{fig:salts} and \ref{fig:phyllo-zarcas}), the $TH/SR$ varies from 0.12--0.25 at 5.3 $\mu$m, and depending on the spectra, falls below 0.01 at 4.4--4.65 $\mu$m. This ratio was then used to iteratively correct the NIRSPec spectra for thermal radiation. A correcting factor from 0.95 to 1.14 was then applied to the MIRI spectra to produce the merged NIRSPec/MIRI spectra shown in Fig.~\ref{fig:Hapkemodel_CO2}.
   
\end{appendix}

\end{document}